\documentclass[12pt]{article}
\usepackage[dvips]{epsfig}
\usepackage{graphicx}
\usepackage{amssymb,amsmath,bm}
\usepackage{verbatim,enumerate}
\usepackage{setspace}
\usepackage{mdframed}
\usepackage[flushleft]{threeparttable}
\usepackage{subfigure}
\usepackage{longtable}
\usepackage{supertabular}
\usepackage{algorithmicx,algorithm}
\usepackage{comment}
\usepackage{CJK}
\usepackage[colorlinks=true,allcolors=blue]{hyperref}%
\usepackage[T1]{fontenc}
\usepackage{tabularx,ragged2e,booktabs,caption}
\usepackage{array}
\newcolumntype{L}[1]{>{\raggedright\let\newline\\\arraybackslash\hspace{0pt}}m{#1}}
\newcolumntype{C}[1]{>{\centering\let\newline\\\arraybackslash\hspace{0pt}}m{#1}}
\newcolumntype{R}[1]{>{\raggedleft\let\newline\\\arraybackslash\hspace{0pt}}m{#1}}
\usepackage{natbib}
\usepackage{multirow}
\usepackage{theorem}
\usepackage{float}
\usepackage{caption}
\newcommand{\RNum}[1]{\uppercase\expandafter{\romannumeral #1\relax}}
\def\00{\mathrm{0}}

\def\mG{\mathcal{G}}

\newtheorem{thm}{Theorem}

\begin{document}
\title{\bf Fast Community Detection in Dynamic and Heterogeneous Networks}
\author{Maoyu Zhang$^\text{1}$, Jingfei Zhang$^\text{2}$ and Wenlin Dai$^\text{1}$
\medskip \\
$^\text{1}$ {\normalsize Institute of Statistics and Big Data, Renmin University of China}\\
$^\text{2}$ {\normalsize Department of Management Science, University of Miami}
}
\date{}
\def\thefootnote{\#}\footnotetext{Maoyu Zhang and Jingfei Zhang are joint first authors.}
\maketitle

\begin{abstract}
\baselineskip=17.5pt
Dynamic heterogeneous networks describe the temporal evolution of interactions among nodes and edges of different types. While there is a rich literature on finding communities in dynamic networks, the application of these methods to dynamic heterogeneous networks can be inappropriate, due to the involvement of different types of nodes and edges and the need to treat them differently.In this paper, we propose a statistical framework for detecting common communities in dynamic and heterogeneous networks. Under this framework, we develop a fast community detection method called DHNet that can efficiently estimate the community label as well as the number of communities. An attractive feature of DHNet is that it does not require the number of communities to be known a priori, a common assumption in community detection methods. While DHNet does not require any parametric assumptions on the underlying network model, we show that the identified label is consistent under a time-varying heterogeneous stochastic block model with a temporal correlation structure and edge sparsity. We further illustrate the utility of DHNet through simulations and an application to review data from Yelp, where DHNet shows improvements both in terms of accuracy and interpretability over existing solutions.
\end{abstract}

\par\medskip\noindent

\newpage
\baselineskip=26.5pt
\section{Introduction}

One of the fundamental problems in network data analysis is community detection that 
aims to divide the network into non-overlapping groups of nodes such that nodes within the same community are densely connected and nodes from different communities are relatively sparsely connected. 
Community detection can provide valuable insights on the organization of a network and greatly facilitate the analysis 
of network characteristics. 
As such, community detection methods have been applied to numerous scientific fields such as social science \citep{moody2003structural}, biology \citep{sorlie2001gene} and business \citep{linden2003amazon}. Over the past few decades, the problem of community detection has been approached from methodological, algorithmic and theoretical perspectives with substantial developments. We refer to \citet{fortunato2010community} and \citet{abbe2017community} for comprehensive reviews on this topic.

\begin{figure}[t!]
		\centering
		\includegraphics[width=0.75\linewidth]{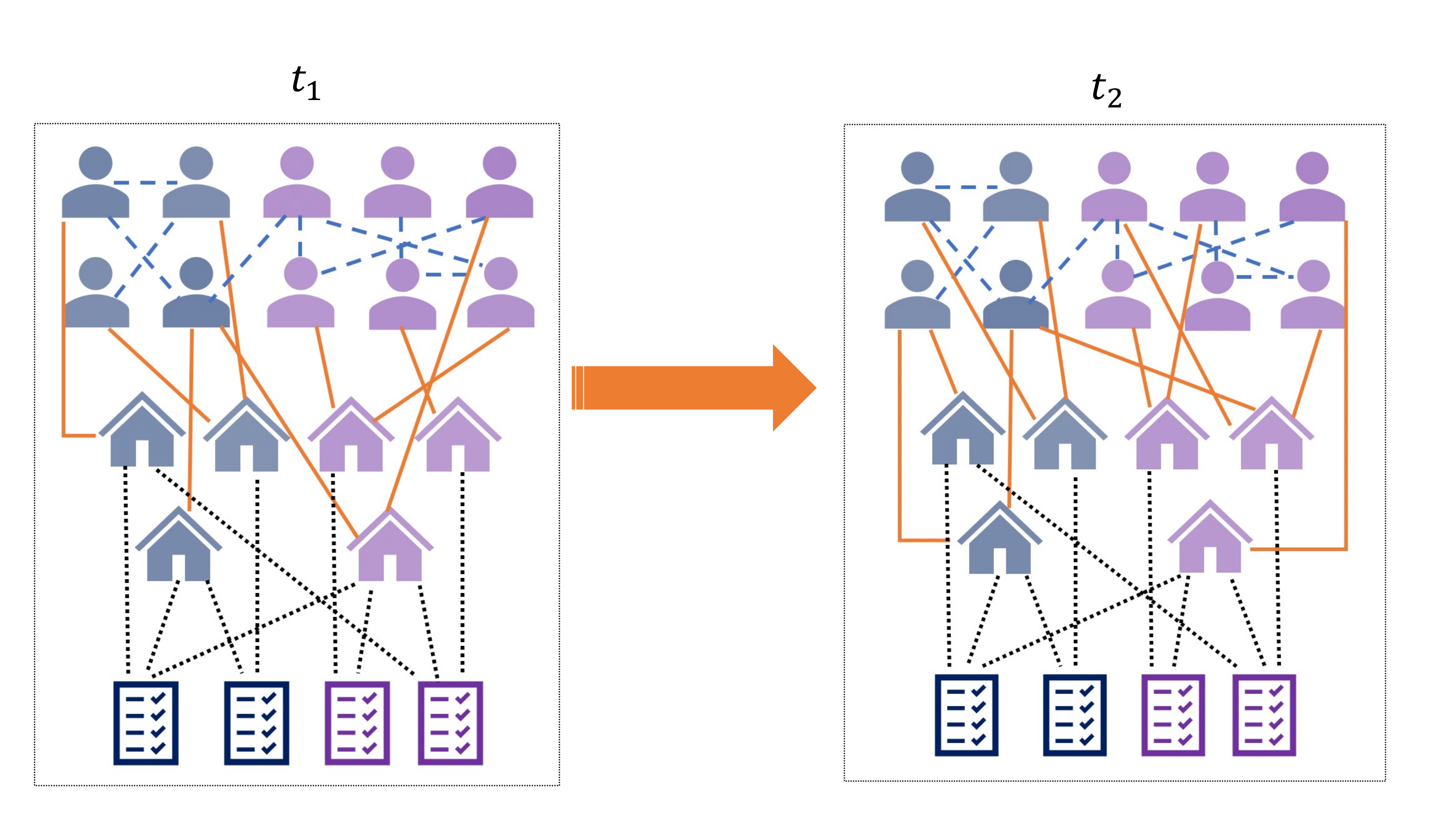}
	\caption{The dynamic heterogeneous Yelp network with two communities, including three types of nodes (user, business, category) and edges (user is friend with user, business is reviewed by user, business is labeled with category).} \label{yelp}
\end{figure}

While the majority of existing community detection methods are developed for a homogeneous network or a dynamic network, networks that are dynamic and heterogeneous are fast emerging in recent years. For example, in a dynamic healthcare
network, nodes can be {\it patients}, {\it diseases}, {\it doctors} and {\it hospitals} and edges can be in
the type of patient-disease (patient treated for disease) and patient-doctor (patient treated by
doctor) and doctor-hospital (doctor works at hospital). 
These edges are expected to evolve with time as patients may develop new diseases that are treated by different doctors at possibly different hospitals.
Figure \ref{yelp} provides an illustration of a dynamic heterogeneous Yelp review network, which is analyzed in Section \ref{sec:real}.  
In this figure, there are three types of nodes including {\it users}, {\it businesses} and {\it categories} and three types of edges including user-user (user is friend with user), user-business (business is reviewed by user) and business-category (business is labeled with category). 
As users review different businesses over time, this network is both heterogeneous and dynamic. 

Due to the rich information embedded in a dynamic heterogeneous network, many methods have been developed recently for its analysis, such as network embedding \citep{wang2020dynamic,zhang2022multi}, representation learning \citep{yin2019dhne} and link prediction \citep{xue2020modeling,jiang2021hints}. 
However, community detection in dynamic heterogeneous networks is less studied. 
One relevant work 
is \cite{sun2010community}, which provides a mixture model-based generative model for estimating the community structure, which is assumed to be time-varying. 
Other works on this topic include \citet{sengupta2015spectral} and \citet{zhang2017finding}, though they only focus on a single heterogeneous network.

In our work, we focus on detecting common communities in a dynamic heterogeneous network, that is, the community assignment does not vary with time but the interactions within and between communities do. 
Finding common communities are useful in many applications. For example, in genetic studies and brain connectivity studies, the common communities represent functional groups of genes or brain regions that are coordinated in biological processes, and identifying them is of keen scientific interests \citep{zhang2017finding,zhang2020mixed}. 
In the Yelp review network, it is plausible that businesses, categories and the majority of users have a common community structure over time, as the service offered by a business and the interests of users (e.g., pets, parks, fine dining) are often stable over a period of time. 
One notable advantage of considering a common community structure is that the networks observed at different time points are allowed to be highly sparse if $S$, the number of time points, increases.
For example, we show in Theorem \ref{the1} that consistent community detection is achievable as long as $\lambda S\rightarrow\infty$, where $\lambda$ is the average degree, while the single network case requires $\lambda\rightarrow\infty$ to achieve community detection consistency.
Moreover, our approach allows the community strength to be highly variable over time. For example, a community needs to be active for only a very short period of time for it to be consistently identified; see more discussions after Theorem \ref{the1}.


In this paper, 
we propose a statistical framework for modularity-based common community detection in the dynamic heterogeneous network, where no parametric assumptions are made on the model underlying the observed networks. Under this framework, we develop a fast community detection method called \texttt{DHNet} that can efficiently estimate the community label as well as the number of
communities. An attractive feature of \texttt{DHNet} is that it does not require the number
of communities to be known a priori, a common assumption in community detection methods. 
Although \texttt{DHNet} does not rely on parametric assumptions on the underlying network model, we propose a new dynamic heterogeneous stochastic block model with a temporal correlation structure and edge sparsity, and show that \texttt{DHNet} can consistently estimate the community label under this model. This provides theoretical justifications of the proposed method and also sheds lights on how different network properties (e.g., sparsity, size, community strength) affect its performance.
The consistency property of our method when applied to dynamic bi-partite or multi-partite networks follows as special cases. 

The remainder of the article is organized as follows. 
Section \ref{sec:mod} describes a community detection framework and 
proposes a modularity function for finding common communities in a dynamic heterogeneous network. 
Section \ref{sec:max} describes a fast community detection method called \texttt{DHNet} that can efficiently estimate the community label as well as the number of communities. 
Section \ref{sec:theory} shows the consistency property of \texttt{DHNet} under a dynamic heterogeneous stochastic block model. Section \ref{sec:sim} demonstrates the efficacy of \texttt{DHNet} through simulation studies and Section \ref{sec:real} applies the proposed method to review data from Yelp. The paper is concluded with a short discussion section.

\section{Community Detection with Modularity}\label{sec:mod}
\subsection{Notation}\label{sec:not}
\begin{figure}[t!]
\centering
\includegraphics[width=\linewidth]{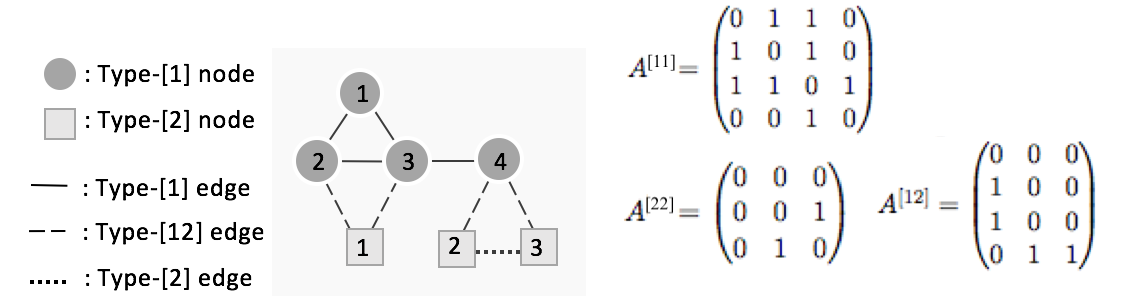}
\caption{An illustrative example of a heterogeneous network with two types of nodes.} 
\label{example}
\end{figure}

We write $[m]=\{1,\ldots,m\}$ for an integer $m>0$. To ease notation, we start the introduction with a \textit{single} heterogeneous networks with $L$ types of nodes. Let $V^{[l]}=(v^{[l]}_1,\ldots,v^{[l]}_{n_l})$ be the set containing the $l$-th type of nodes for $l\in[L]$, where $n_{l}$ is the number of $l$-th type nodes.
Denote the heterogeneous network as $\mG=(\cup_{l=1}^L V^{[l]}, \mathcal{E}\cup\mathcal{E}^+)$, where set $\mathcal{E}$ contains edges between nodes of the same type and set $\mathcal{E}^+$ contains edges between nodes of different types. 
When $\mathcal{E}=\emptyset$, $\mG$ forms a multi-partite network, i.e., edges are only established between different types of nodes. 
Let $G^{[l]}$ denote the homogeneous network formed within node set $V^{[l]}$ with an $n_l\times n_l$ adjacency matrix $A^{[l]}$, and $G^{[l_1l_2]}=(V^{[l_1]}\cup V^{[l_2]}, E^{[l_1l_2]})$ denote the bi-partite network formed between node sets $V^{[l_1]}$ and $V^{[l_2]}$ with an $n_{l_1}\times n_{l_2}$ bi-adjacency matrix $A^{[l_1l_2]}$, $l_1,l_2\in[L]$. See Figure~\ref{example} for an example of a heterogeneous network with $L=2$.

Consider a dynamic heterogeneous network $\{\mG(t), t \in \mathcal{T}\}$ with $L$ types of nodes, where $\mG(t)=(\bigcup_{l=1}^{L} V^{[l]}, \mathcal{E}(t)\cup \mathcal{E}^{+}(t))$ is a heterogeneous network at time $t$ defined as above.
The network $\mG(t)$ at time $t$ can be uniquely represented by its adjacency matrix $\mathcal{A}(t)$ defined as 
$$
\mathcal{A}(t)=\left(\begin{array}{ccc}A^{[11]}(t)  & \ldots & A^{[1 L]}(t) \\ \vdots & \ddots & \vdots \\ A^{[L 1]}(t) & \ldots & A^{[LL]}(t) \end{array}\right),
$$
where $A^{[l_1l_2]}(t)\in\mathbb{R}^{n_{l_1}\times n_{l_2}}$ is defined as in Figure \ref{example}.
Define
$\mathbf{d}^{[l]}(t)=(d^{[l]}_1(t),\ldots,d^{[l]}_{n_1}(t))$, where $d^{[l]}_i(t)$ is the number of links incident to $v^{[l]}_i$ from $V^{[l]}$ at time $t$, and $\mathbf{d}^{[l_1l_2]}(t)=(d^{[l_1l_2]}_1(t),\ldots,d^{[l_1l_2]}_{n_1}(t))$, where $d^{[l_1l_2]}_i$ is the number of links incident to $v^{[l_1]}_i$ from $V^{[l_2]}$ at time $t$.
Write the number of edges in $A^{[l_1l_2]}(t)$ as $m^{[l_1l_2]}(t)=\sum_{i,j}A_{ij}^{l_1l_2}(t)$ for $l_1,l_2\in[L]$ . 


\subsection{Modularity function}
The modularity function measures the strength of division of a network into communities,
and the maximum modularity function value is a metric frequently used for quantifying the strength of community structure within a network \citep{fortunato2010community}. The function was first defined in \citet{newman2004finding} for a simple network $G(V,E)$ with $n$ nodes, $m$ edges, adjacency matrix $A_{n\times n}$ and a community assignment $\bm{e}=(e_1,\ldots,e_n)$, where $e_i\in[K]$, as
\begin{equation}\label{eq0}
Q(\bm{e},G)=\frac{1}{2m}\sum_{1\le i<j\le n}\left[A_{ij}-\mathbb{E}(A_{ij})\right]1(e_i=e_j),
\end{equation}
where $1(\cdot)$ is the indicator function. 
In \eqref{eq0}, the expectation $\mathbb{E}(A_{ij})$ is calculated under a null model for random networks with no community structure. 
The most common choice for the null model is the Chung-Lu model \citep{newman2004finding,newman2006finding}. In a Chung-Lu model \citep{chung2006complex}, given the expected degrees for the nodes, the probability of having an edge between nodes $i$ and $j$ depends only on the their expected degrees. 
As noted by \citet{newman2006finding}, the Chung–Lu model is the only random graph model where the probability of having an edge between nodes $i$ and $j$ is the product of separate
functions of the expected degrees of nodes $i$ and $j$, written as $f(d_i)f(d_j)$, where the functions must be the same since the adjacency matrix is symmetric. 
Under the Chung–Lu model, it has been shown that every network in the null space occurs with the same probability and there is no preference for any particular graph configuration \citep{zhang2017hypothesis}, which makes the model a desirable choice as the null.
It is seen that the modularity function in \eqref{eq0} measures the difference between the observed number of intra-community edges and the expected number of intra-community edges under the null with no community structure. Correspondingly, the community label of a network is identified by maximizing the modularity function with respect to $\bm{e}$.

To define the modularity function in a dynamic heterogeneous network, we first describe the corresponding null model that characterizes a dynamic heterogeneous network with no community structure. 
Consider the heterogeneous network at time $t$, $\mG(t)=\left(\cup_{l=1}^L V^{[l]}, \mathcal{E}(t) \cup \mathcal{E}^{+}(t)\right)$ with degree sequence
$\boldsymbol{D}(t)=\{\mathbf{d}^{[l_1l_2]}(t),l_1,l_2\in[L]\}$.  
We define a heterogeneous Chung-Lu model as the null.
Specifically, under the null, we assume that a heterogeneous network at time $t$ is generated with
\begin{equation}\label{eq2}
A_{i j}^{[l_{1} l_{2}]}(t)\sim\text{Bernoulli}\left(\frac{d_{i}^{\left[l_{1} l_{2}\right]}(t) d_{j}^{\left[l_{2} l_{1}\right]}(t)}{m^{\left[l_{1} l_{2}\right]}(t)}\right), \quad l_1,l_2\in[L],
\end{equation}
where all edges in $\mathcal{G}(t)$ are independent.
Under \eqref{eq2}, it is easy to show that the expected degree sequence under the null is the same as the observed degree sequence $\boldsymbol{D}(t)$. 
Following the same argument as in \citet{zhang2017hypothesis}, it can be shown that under \eqref{eq2}, every heterogeneous network in the null space occurs with the same probability.

Next, we move to define the modularity matrix.
At time $t\in\mathcal{T}$ and given $\mathcal{A}(t)$, we write the $(n_1 +\dots
+n_L)\times (n_1 + \dots +n_L)$ modularity matrix $\mathcal{M}(t)$ as
$$\mathcal{M}(t)=\left(\begin{array}{ccc}\frac{M^{[11]}(t)}{m^{[11]}(t) }& \ldots & \frac{M^{[1L]}(t)}{ m^{[1L]}(t)}\\ \vdots & \ddots & \vdots \\ 	\frac{M^{[L1]}(t)}{ m^{[L1]}(t)} & \ldots & 	\frac{M^{[LL]}(t)}{ m^{[LL]}(t)}\end{array}\right),
$$
where $M^{[l_1l_2]}(t)=A^{[l_1l_2]}(t)-\mathbb{E}\left(A^{[l_1l_2]}(t)\right)$. 
The modularity matrix $\mathcal{M}(t)$ measures the distance between the observed network and the expected network under the null model at time $t$. Given the dynamic heterogeneous networks $\{\mathcal{A}(t),\,t\in\mathcal{T}\}$, the integrated modularity matrix $\mathcal{M}$ is defined as
$$
\mathcal{M}=\left(\begin{array}{ccc}\mathcal{M}^{[11]}& \ldots & \mathcal{M}^{[1L]}\\ \vdots & \ddots & \vdots \\ \mathcal{M}^{[L 1]} & \ldots & \mathcal{M}^{[LL]}\end{array}\right),\,\,
\text{where}\,\,
\mathcal{M}^{[l_1l_2]}=\frac{\int_{t \in \mathcal{T}}M^{[l_1l_2]}(t)}{\bar{m}^{[l_1l_2]}},
$$
$\bar{m}^{[l_1l_2]}=\int_{t \in \mathcal{T}} m^{[l_1l_2]}(t) $. The integrated modularity matrix $\mathcal{M}$ measures the distance between the observed network and the expected network under the null model over all $t\in\mathcal{T}$.

We are now ready to define the modularity function. Write the community assignment label as  $\boldsymbol{e}=\left(\mathbf{e}^{[1]},\dots, \mathbf{e}^{[L]}\right)$ with $\mathbf{e}^{[l]}=\left(e_{1}^{[l]}, \ldots, e_{n_l}^{[l]}\right), l\in [L]$, the modularity function of the dynamic heterogeneous network is defined as
\begin{equation}\label{eq3}
\begin{aligned}
	Q(\boldsymbol{e}, \{\mG(t)\}_{t \in \mathcal{T}}) &=\frac{1}{L^2}\sum_{1\le l_1, l_2\le L} \sum_{i, j} \mathcal{M}^{[l_1l_2]}_{i j} 1(e^{[l_1]}_{i}=e^{[l_2]}_{j}).
\end{aligned}
\end{equation}
From the above definitions, it can be shown that $Q(\boldsymbol{e}, \{\mG(t)\}_{t \in \mathcal{T}})\in[-1,1]$. 
This modularity function measures the overall difference between the observed number of intra-community edges and the expected number of intra-community edges under the null model.
When $Q(\boldsymbol{e}, \{\mG(t)\}_{t \in \mathcal{T}})$ approaches 1, the observed number of intra-community edges is greater than the expected values, which indicates a strong community structure. In contrast, when $Q(\boldsymbol{e}, \{\mG(t)\}_{t \in \mathcal{T}})$ approaches 0, the observed number of intra-community edges is close to the expected values under the null, which indicates no or weak community structure.

In practice, the networks are often only observed on a number of time points $\mathcal{T}=\left\{t_{1}, t_{2}, \ldots, t_{S}\right\}$, where $S$ is the total number of observations or snapshots. 
In this case, we can define
\begin{equation}\label{eq:mod}
\mathcal{M}=\left(\begin{array}{ccc}\sum_{s=1}^SM^{[11]}(t_s) /  \bar{m}^{[1]} & \ldots & \sum_{s=1}^SM^{[1 L]}(t_s) / \bar{m}^{[1 L]} \\ \vdots & \ddots & \vdots \\ \sum_{s=1}^SM^{[L 1]}(t_s)/ \bar{m}^{[L 1]} & \ldots & \sum_{s=1}^SM^{[LL]}(t_s) /  \bar{m}^{[L]}\end{array}\right),
\end{equation}
where $\bar{m}^{[l_1l_2]}=\sum_{s=1}^S m^{[l_1l_2]}(t_s) $, $l_1, l_2\in[L]$, and write the modularity function as
$$
Q(\boldsymbol{e}, \{\mG(t_s)\}_{s\in[S]})=\frac{\sum_{s=1}^{S} m^{[l_1l_2]}\left(t_{s}\right) Q^{[l_1l_2]}\left(\boldsymbol{e},\mG\left(t_{s}\right)\right)}{\sum_{s=1}^{S} m^{[l_1l_2]}\left(t_{s}\right)},
$$
where 
$Q^{[l_1l_2]}\left(\boldsymbol{e},\mG\left(t_{s}\right)\right)=\frac{1}{m^{[l_1l_2]}(t_s)L^2}\sum_{l_1, l_2=1}^L \sum_{i, j} M^{[l_1l_2]}_{i j}(t_s) 1\left(e^{[l_1]}_{i}=e^{[l_2]}_{j}\right)$.
The above modularity function can be considered as an averaged version of the modularity in each graph $\mG(t_s)$, $s\in[S]$.

\section{Modularity maximization}\label{sec:max}
We aim to find the community assignment that maximizes the modularity
function (\ref{eq3}), that is, 
\begin{equation}\label{eq4}
\hat{\boldsymbol{c}}=\arg \max \limits_{\substack{\boldsymbol{e}=(\mathbf{e}^{[1]},\ldots,\mathbf{e}^{[L]}), \\ e_{i}^{[l]} \in\{1, \ldots, K\}}}{Q}(\boldsymbol{e},\{\mG(t)\}_{t \in \mathcal{T}}).
\end{equation}
Finding the exact maximizer of (\ref{eq3}) is challenging due to the combinatorial nature of the problem and the fact that the number of communities $K$ is generally unknown. 
\cite{brandes2007modularity} showed that finding the partition that maximizes the modularity function for a simple graph is NP-hard. 
There are a number of existing heuristic algorithmic solutions to maximizing the modularity function,
some of which are fast and hence feasible for very large 
networks \citep{clauset2004finding,wakita2007finding,blondel2008fast}, while some others could be more precise though restricted to graphs of moderate sizes \citep{guimera2004modularity,massen2005identifying}.

In our approach, we adopt a fast Louvain-type maximization method.
The Louvain method was first proposed by \cite{blondel2008fast} for modularity maximization in simple graphs. In the Louvain method, small communities are first identified by optimizing the modularity function locally on all nodes. Then each small community is grouped into one ``meta'' node and the first step is repeated.
The Louvain method is fast to compute and enjoys a good empirical performance. It has been successful applied to network analyses from various scientific fields, permitting up to 100 million nodes and billions of edges. Notably, the modularity maximum found by the Louvain method often compares favorably with those found by alternative methods such as \cite{clauset2004finding} and \cite{wakita2007finding}; see \cite{fortunato2010community}.


Motivated by the Louvian algorithm, we propose a {\bf d}ynamic {\bf h}eterogeneous {\bf net}work modularity maximization algorithm, referred to \texttt{DHNet}.
To do so, we first define a \textit{unit}, which is set of nodes with at most one from each node type.
For example, a unit may contain one node of any type or $L$ nodes of different types. 
A unit serves as the building block of a community in a heterogeneous network.
Next, given a heterogeneous $n\times n$ modularity matrix $\mathcal{M}$ as in \eqref{eq:mod}, we define a \textit{modularity network}, which is a network of $n$ nodes and the edge between nodes $(i,j)$ is $\mathcal{M}_{i,j}$. From \eqref{eq3} and \eqref{eq4}, it is easy to see that our optimization task is to find a partition of the modularity network such that the within-community sum of edges from $\mathcal{M}$ is maximized.

The algorithm \texttt{DHNet} starts with assigning each node to its own unit and then each unit to its own community, leading to $n$ communities at the start of the algorithm with each community containing only one node (or unit). 
The optimization procedure is then carried out in two phases that are repeated iteratively.
In the first phase and for each unit $i$, \texttt{DHNet} removes this unit from its current community and assigns it to its neighboring community (communities to which unit $i$ is linked to), such that it leads to the largest increase of the modularity in \eqref{eq3}. 
If no move increases the modularity, then unit $i$ remains in its current community. 
In the second phase, the algorithm merges nodes of the same type in each community, such that each community contains at most one node from each node type, and builds a new modularity network. 
In the new modularity network, the units are communities from the first phase and the edge between two nodes are given by summing the edge weights connecting two corresponding sets of nodes from the first phase. 
These steps are repeated iteratively until the modularity value no longer increases. 
The algorithm can be summarized as Algorithm \ref{lv}.


\begin{algorithm}[!t]
\setstretch{1.35}
	\caption{{\bf D}ynamic {\bf H}eterogeneous {\bf Net}work Modularity Maximization (\texttt{DHNet})}
	\hspace*{0.02in}{\bf Input}: Dynamic heterogeneous networks $A_1$,\dots, $A_S$.
	\begin{algorithmic}
		\State \textbf{Step 1}: Calculate the modularity matrix using \eqref{eq:mod}.
		\State \textbf{Step 2}: Assign each node to its own unit and assign each unit to its own community.
		\State \textbf{Step 3}: Repeat Steps 3.1-3.4 until the modularity value no longer increase.
		\State \hspace{0.15in}\textbf{Step 3.1}: For each unit, place it into the neighboring community that leads to the\\ 
		\hspace{0.5in}largest modularity increase in \eqref{eq3}. If no such move is possible, then this unit stays\\
		\hspace{0.5in}in its present community.
		\State \hspace{0.15in}\textbf{Step 3.2}: Repeated apply Step 3.1 to all units until none can be moved.
		\State \hspace{0.15in}\textbf{Step 3.3}: If the modularity is higher than that from the previous iteration, merge\\
		\hspace{0.5in}nodes of the same type in each community such that each community is regarded\\ \hspace{0.5in}as a unit and go to Step 3.4. If not, exit with the assignment from the previous \\ 
		\hspace{0.5in}iteration.
        \State \hspace{0.15in}\textbf{Step 3.4}: Calculate the modularity matrix of the merged network.\\
	\hspace*{0.02in}{\bf Output}: Community assignment and the corresponding modularity value. 
	\end{algorithmic}\label{lv}
\end{algorithm}

\begin{figure}[h!]
		\centering
		\includegraphics[width=\linewidth]{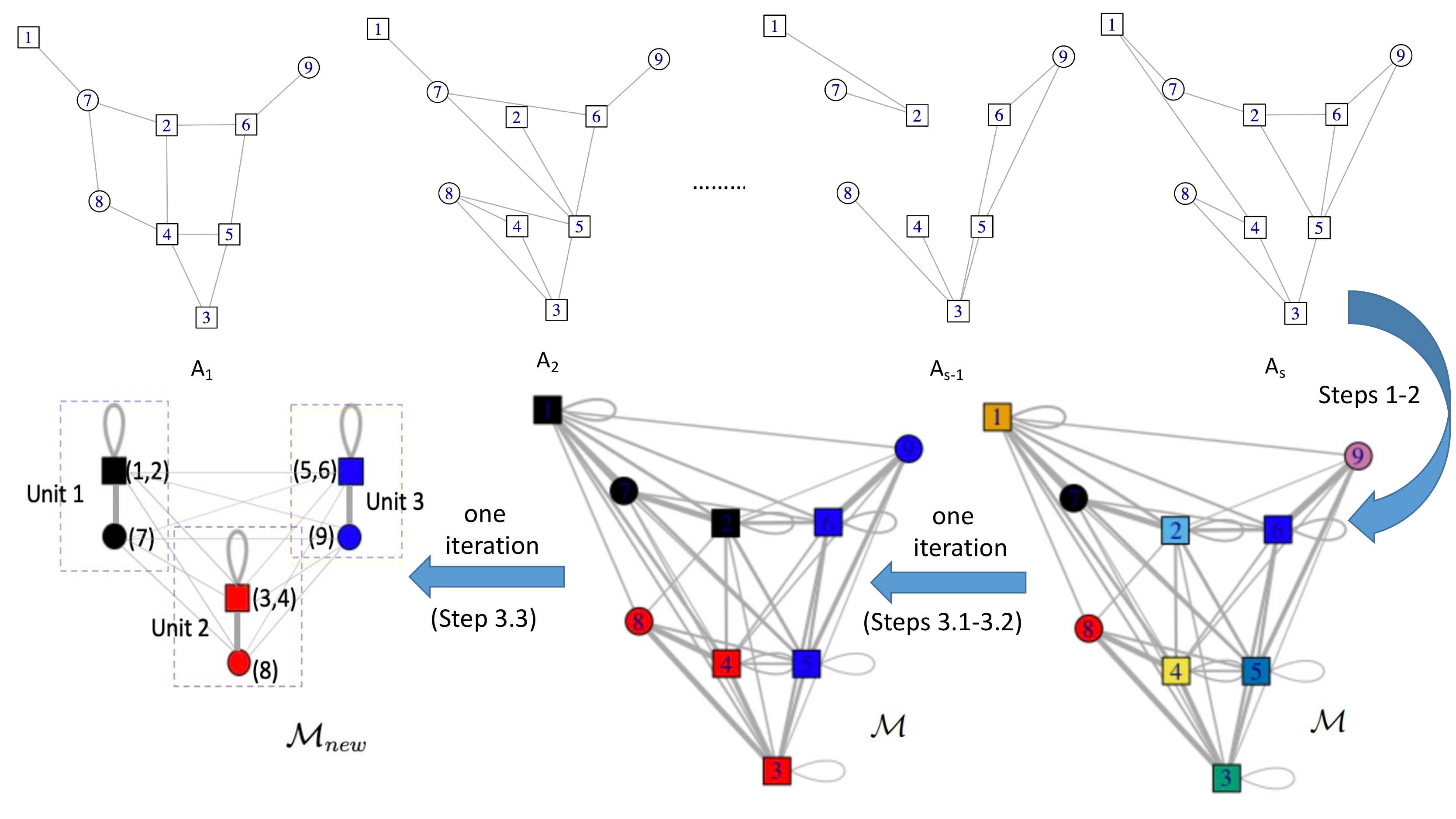}
	\caption{A simple illustration of \texttt{DHNet}. Nodes of the same type are marked using the same shape and nodes of the same color are in the same community.}\label{vis} 
\end{figure}
	

Figure \ref{vis} shows an example of 
applying \texttt{DHNet} to a dynamic heterogeneous network with two types of nodes. 
First, Step 1 calculates the modularity matrix $\mathcal{M}$ from $S$ heterogeneous networks $A_1$,\dots, $A_S$ as in \eqref{eq:mod} and Step 2 assigns each node to its own unit and assign each unit to its own community. Then Step 3 is implemented once and the algorithm reaches convergence. 
Specifically, in Figure \ref{vis} and after Steps 3.1-3.2, nodes $\{1,2,7\}$, $\{3,4,8\}$ and $\{5,6,9\}$ are placed into three communities, colored black, red and blue, respectively. 
Step 3.3 then produces a new modularity network with 6 nodes and 3 units. 
The first unit has nodes $\{1,2\}$ and 7, where nodes $1$ and $2$ are merged as they are of the same type. Similarly, the second unit has $\{3,4\}$ and 8 and the third unit has $\{5,6\}$ and 9.
After this step, merging any of the three units cannot further increase the modularity and thus \texttt{DHNet} returns three communities with nodes $\{1,2,7\}$, $\{3,4,8\}$ and $\{5,6,9\}$, respectively. 

\textbf{Remark 1 (initialization).} In Step 3.1, if there are multiple communities that lead to the same maximum modularity increase,
\texttt{DHNet} randomly selects a community to assign the unit to. Hence, the result of \texttt{DHNet} may differ each time the algorithm is implemented. Moreover, the result of the algorithm may differ depending the node ordering in Step 2. That is, a node ordering of $\{1,2,3\}$ or $\{3,1,2\}$ may give different results.
We recommend applying the Louvain method $\kappa$ times with random node orderings and using the assignment with the largest modularity function value as the final output.
In our simulation studies and real data analysis, we set $\kappa=100$ and notice that the output from \texttt{DHNet} is not sensitive to node orderings. 
Generally, it is recommended that $\kappa$ should increase with the size of the network. 

\begin{figure}[t!]
		\centering
		\includegraphics[width=0.35\linewidth]{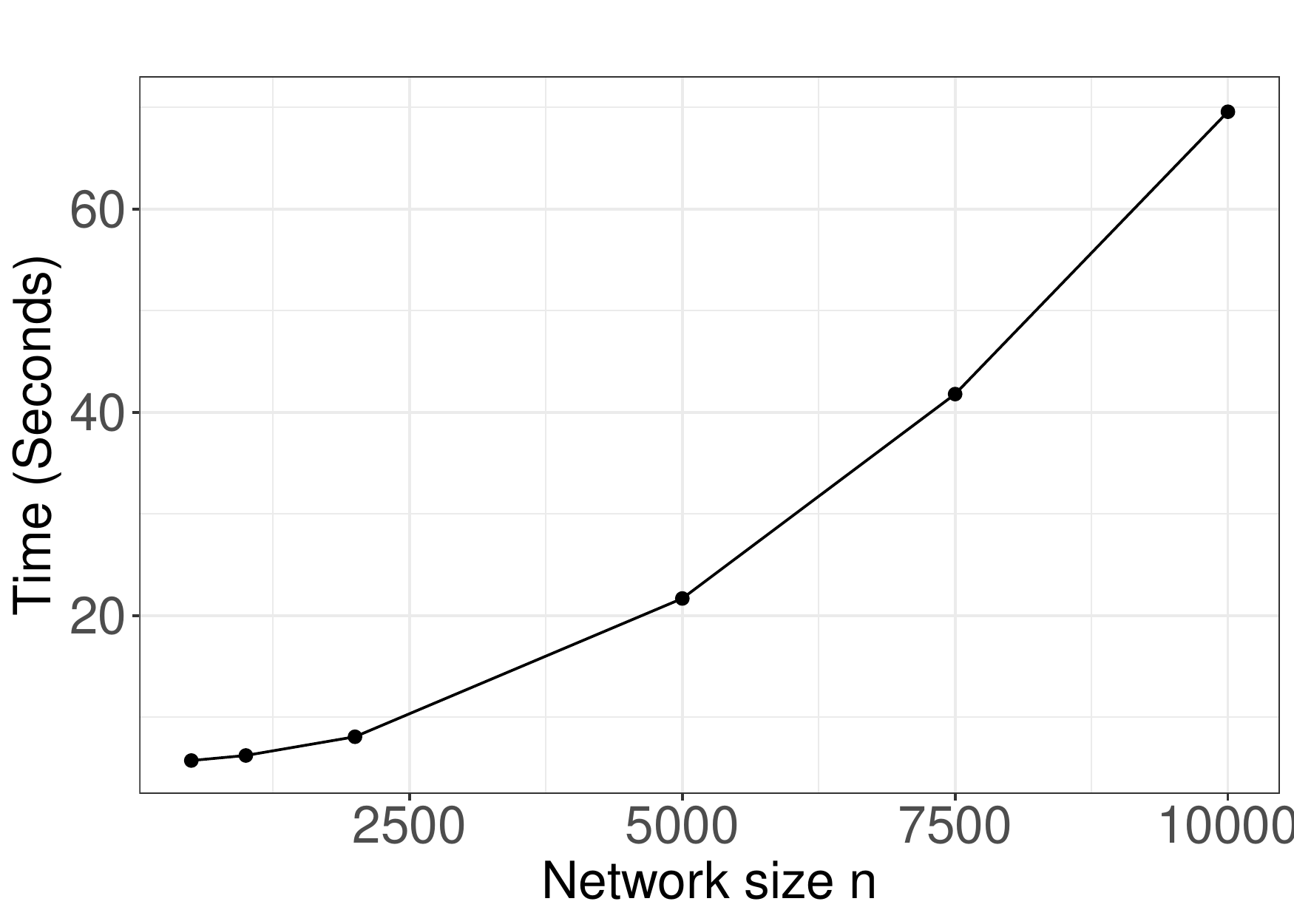}\qquad
		\centering
		\includegraphics[width=0.35\linewidth]{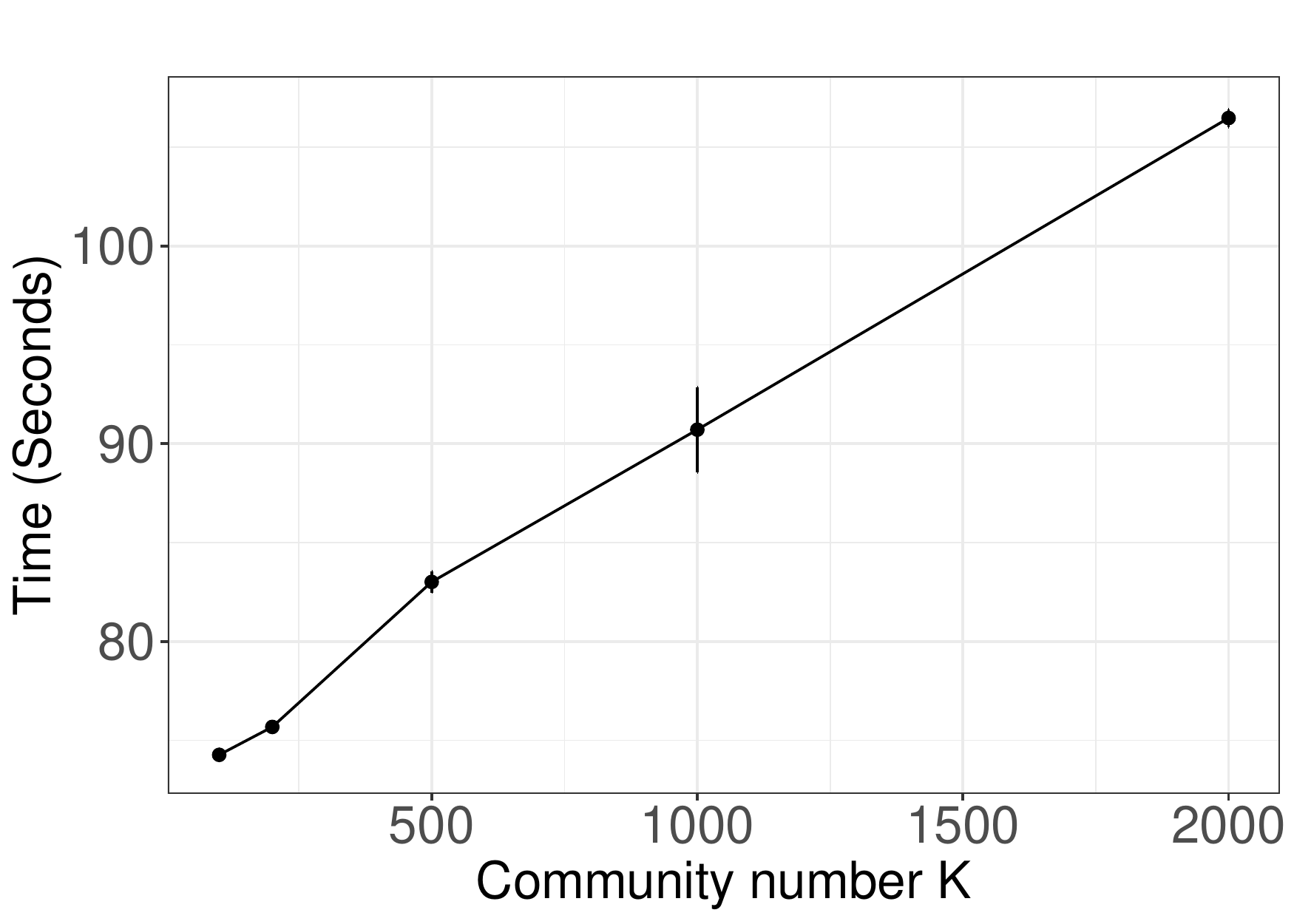}
	\caption{
	The computation time of \texttt{DHNet} with a varying network size $n$ and number of communities $K$. 
	In the left panel, we set the number of communities $K=10$, and in the right panel, we set the network size $n=10,000$. } \label{time}
\end{figure}

\textbf{Remark 2 (time complexity).} 
In \texttt{DHNet}, computing whether and where to move 
each unit based on modularity changes 
is of time complexity $O(1)$. Hence, the computation time at each iteration is roughly linear in the number of units, which is less than or equal to the total number of nodes.  
Figure \ref{time} provides the computation time of \texttt{DHNet} with 
varying network size $n$ and number of communities $K$. We set $S=20$ and generate these networks from DHSBM where the inter- and intra-community connecting probabilities are 0.1 and 0.15, respectively, in the homogeneous networks, and the inter- and intra-community connecting probabilities are, respectively, 0.05 and 0.1 in the multi-partite networks.
All experiments are ran on an Intel(R) Xeon(R) with $3.10 \mathrm{GHz}$ and $192 \mathrm{~GB}$ memory processor.

\section{Consistency}\label{sec:theory}
In this section, we investigate the theoretical properties of \texttt{DHNet} for finding common communities in a dynamic heterogeneous network. To do so, we first propose a discrete-time heterogeneous stochastic block model with a temporal correlation structure.

\noindent
\textbf{Dynamic Heterogeneous Stochastic Block Model (DHSBM)}
\begin{enumerate}
\item Dynamic heterogeneous network $\{\mG(t_s),\,s\in[S]\}$ with $L$ node types has a latent community label $\boldsymbol{c}=$ $\left(\boldsymbol{c}^{[1]}, \ldots, \boldsymbol{c}^{[L]}\right)$, where $\boldsymbol{c}^{[l]}=(c_1^{[l]}, \ldots, c_{n_l}^{[l]})$ and $c_i^{[l]} \in \{1,\dots, K\}$ denotes the community that node $i$ of type-$[l]$ belongs to, $l\in[L]$. 

\item The label $\boldsymbol{c}^{[l]}$ follows a multinomial distribution with $n_l$ trials and probability $\bm\pi^{[l]}=\left(\pi_{1}^{[l]}, \ldots, \pi_{K}^{[l]}\right)$, $l\in[L]$. 

\item 
Define the time-varying probability matrix $\bm\Theta(t_s)$ such that 
$$
\bm\Theta(t_s)=\left(\begin{array}{ccc}\Theta^{[11]}(t_s)  & \ldots & \Theta^{[1 L]}(t_s) \\ \vdots & \ddots & \vdots \\ \Theta^{[L 1]}(t_s) & \ldots & \Theta^{[LL]}(t_s) \end{array}\right),\,\,\text{where}\,\,
\Theta^{[l_1l_2]}(t_s)=\left(\begin{array}{ccc}
\theta_{11}^{[l_1l_2]}(t_s) & \cdots & \theta_{1 K}^{[l_1l_2]}(t_s)\\
\vdots & \ddots & \vdots\\ 
\theta_{K 1}^{[l_1l_2]}(t_s) & \cdots & \theta_{K K}^{[l_1l_2]}(t_s)
\end{array}\right)
$$
and $\theta_{k_1k_2}^{[l_1l_2]}(t_s)$ is a function of $t_s$, $l_1,l_2\in[L]$, and $k_1,k_2 \in [K]$.


\item Given 
$\boldsymbol{c}$, we treat $A_{i j}^{[l_1l_2]}\left(t_{s}\right)$'s as independent Bernoulli random variables satisfying 
$$
A_{i j}^{[l_1l_2]}\left(t_{s}\right)=uA^{[l_1l_2]}_{i j}\left(t_{s-1}\right)+(1-u) v^{[l_1l_2]},
$$
where $u\overset{iid} \sim \operatorname{Bernoulli}(\alpha)$, 
and given $c^{[l_1]}_{i}=k_1$ and $c^{[l_2]}_{j}=k_2$,
$$
v^{[l_1l_2]} \overset{iid}\sim \text { Bernoulli }\left(\frac{\theta^{[l_1l_2]}_{k_1k_2}\left(t_{s}\right)-\alpha\theta^{[l_1l_2]}_{k_1k_2}\left(t_{s-1}\right)}{1-\alpha}\right),\quad l_1, l_2\in[L].
$$

\end{enumerate}
In Assumption 4, it is possible to let $u\overset{iid} \sim \operatorname{Bernoulli}(\alpha^{[l_1l_2]})$, though we assume $\alpha^{[l_1l_2]}=\alpha$ to simplify notation. In our theoretical analysis, we allow $\alpha^{[l_1l_2]}$ to vary with $l_1,l_2\in[L]$. Next, we require that $0 \leq \alpha<1$, $\alpha \theta^{[l_1l_2]}_{k_1k_2}\left(t_{s-1}\right)\le \theta^{[l_1l_2]}_{k_1k_2}\left(t_{s}\right)$, and $\alpha\left(1-\theta^{[l_1l_2]}_{k_1k_2}\left(t_{s-1}\right)\right)\le 1-\theta^{[l_1l_2]}_{k_1k_2}\left(t_{s}\right)$, so that the above Bernoulli distribution is valid with the probability parameter in $[0,1]$.
Based on Assumption 4, 
some algebra shows that 
$$
\mathbb{P}\left(A^{[l_1l_2]}_{i j}(t_s)=1\right)=\theta^{[l_1l_2]}_{k_1k_2}(t_s), 
$$
which shows that the marginal distribution of $A_{i j}^{[l_1l_2]}(t_s)$ is $\operatorname{Bernoulli}\left(\theta_{k_1k_2}^{[l_1l_2]}(t_s)\right)$. 
Hence, for a fixed $t_s$, $A^{[l_1l_2]}(t_s)$ follows a stochastic block model with a probability matrix $\Theta^{[l_1l_2]}(t_s)$. Additionally, under our DHSBM model, we have 
$$
\operatorname{corr}\left(A^{[l_1l_2]}_{i j}\left(t_{s}\right), A^{[l_1l_2]}_{i j}\left(t_{s-1}\right)\right)=\alpha^{[l_1l_2]} \sqrt{\frac{\theta^{[l_1l_2]}_{k_1k_2}\left(t_{s-1}\right)\left(1-\theta^{[l_1l_2]}_{k_1k_2}\left(t_{s-1}\right)\right)}{\theta^{[l_1l_2]}_{k_1k_2}\left(t_{s}\right)\left(1-\theta^{[l_1l_2]}_{k_1k_2}\left(t_{s}\right)\right)}}.
$$
And for the special case $\alpha^{[l_1l_2]}=0$, $A_{ij}(t_s), s=1,\dots S$, are independent. If $\Theta(t_s)$ is constant over time, 
then 
$\operatorname{corr}\left(A^{[l_1l_2]}_{i j}\left(t_{s}\right), A^{[l_1l_2]}_{i j}\left(t_{s-k}\right)\right)=(\alpha^{[l_1l_2]})^{k}$ for $k=1,2,\ldots$.

Next, we show the consistency property of the estimated assignment vector $\boldsymbol{\hat c}$ under the DHSBM model when the network size $n$ and the number of time points increases in that $nS \rightarrow \infty$.
This regime is more general and includes the results from \citet{zhang2017finding} and \citet{zhang2018modularity} as special cases.
We say a label $\boldsymbol{e}=\left(\boldsymbol{e}^{[1]}, \ldots, \boldsymbol{e}^{[L]}\right)$ is consistent if it satisfies
$$
\forall \epsilon>0, \quad P\left[\frac{1}{n} \sum_{l=1}^{L} \sum_{i=1}^{n_{l}} I\left(e_{i}^{[l]} \neq c_{i}^{[l]}\right)<\epsilon\right] \rightarrow 1 \text { as } nS \rightarrow \infty,
$$
which stipulates that the misclassification ratio tends to zero. Here $\hat{c}_{i}^{[l]}=c_{i}^{[l]}$ means that they belong to the same equivalent class of label permutations. To allow sparsity, we reparameterize $\bm\Theta(t_s)$ as $\tilde{\bm\Theta}(t_s)=\rho_{n,S}\bm\Theta(t_s)$, where $\bm\Theta(t_s)$ is fixed as $nS\rightarrow\infty$. This reparameterization allows us to separate $\rho_{n,S}$, the sparsity parameter, from the structure of the network.


\begin{thm}\label{the1}
Consider a dynamic heterogeneous network $\mG\left(\bigcup_{i=1}^{L} V^{[i]}, \mathcal{E}(t_s) \cup \mathcal{E}^{+}(t_s)\right)$ from the {\rm DHSBM} with $\boldsymbol{c}$, $\boldsymbol{\pi}^{[l]}$'s, $\alpha$ and $\bm\Theta(t_s)$'s, and further assume that the community sizes are balanced, i.e., $\min _{l} n_{l} / n$ is bounded away from zero. Define a $K \times K$ matrix
$$
T_{a b}^{\left[l_{1} l_{2}\right]}(t_s)=\frac{\pi_{a}^{\left[l_{1}\right]} \pi_{b}^{\left[l_{2}\right]} \theta_{a b}^{\left[l_{1} l_{2}\right]}(t_s)}{\sum_{a b} \pi_{a}^{\left[l_{1}\right]} \pi_{b}^{\left[l_{2}\right]} \theta_{a b}^{\left[l_{1} l_{2}\right]}(t_s)} .
$$
Let $W_{[ab]}^{\left[l_{1} l_{2}\right]}(t_s)=T_{ab}^{\left[l_{1} l_{2}\right]}(t_s)-T_{a.}^{\left[l_{1} l_{2}\right]}(t_s)T_{b.}^{\left[l_{1} l_{2}\right]}(t_s)$ with $T_{a.}^{\left[l_{1} l_{2}\right]}(t_s)=\sum_{q=1}^KT_{aq}^{\left[l_{1} l_{2}\right]}(t_s)$. 
If the following assumptions hold
\begin{equation}\label{con1}
\sum_{s=1}^{S}\sum_{l_{1}, l_{2}}^{L} W_{a a}^{\left[l_{1} l_{2}\right]}(t_s)>0 \text{ and } \sum_{s=1}^{S}\sum_{l_{1}, l_{2}}^{L} W_{a b}^{\left[l_{l} l_{2}\right]}(t_s)<0 \quad \text { for all } \quad a \neq b \in [K]
\end{equation}
and $nS\rho_{n,S}\rightarrow\infty$, then we have
$$
\forall \epsilon>0, \quad P\left[\frac{1}{n} \sum_{l=1}^{L} \sum_{i=1}^{n_{l}} I\left(\hat{c}_{i}^{[l]} \neq c_{i}^{[l]}\right)<\epsilon\right] \rightarrow 1 \text { as } nS \rightarrow \infty,
$$
where $\hat{\boldsymbol{c}}$ is the maximizer of \eqref{eq3}.
\end{thm}

It is seen that the network is allowed to be highly sparse at each time point $s$, e.g., the probability of forming an edge can be $O\left(\frac{\log(nS)}{nS}\right)$. 
Denoting the average degree as $\lambda=n\rho_{n,S}$, it is seen that consistency is achievable when $\lambda S\rightarrow\infty$, while the single network case requires $\lambda\rightarrow\infty$ to achieve community detection consistency \citep{zhang2018modularity}.
When $L=1$, the above result reduces to that in \cite{zhang2017finding} and when $S=1$, the above result reduces to that in \cite{zhang2018modularity}. 
We note that \citet{zhang2017finding} only considered the case where the network size $n$ is fixed and their results require $\rho_{n,S}=O(1)$. In comparison, our result in Theorem \ref{the1} allows $n$ and/or $S$ to diverge and only requires $nS\rho_{n,S}\rightarrow\infty$ as $nS\rightarrow\infty$.

The condition in \eqref{con1} requires that edges are on average more likely to be established within communities than they are between communities, though communities may not exist for all types of edges or at all time points. For example, in the simulation setting in Section \ref{sec:sim2}, the edges within type-[1] nodes and/or type-[2] nodes have no community structure, while the edges linking type-[1] and type-[2] nodes do at some time points. 
This type of assortative condition, requiring more edges within communities than between communities, is often required for algorithm-based community detection such as modularity maximization. 
For the special case of $L=1$, $K=2$, and $\bm\Theta(t_s)$ is time homogeneous,
the condition (\ref{con1}) can be simplified as
$$
\theta_{11}^{[11]} \theta_{22}^{[11]}>\left(\theta_{12}^{[11]}
\right)^{2}.
$$
When $L=2$, $K=2$ and $\bm\Theta(t_s)$ is time-varying, the condition (\ref{con1}) is satisfied if
$$
\sum_{s=1}^{S}\left(\theta_{11}^{[11]}(t_s)+\theta_{11}^{[22]}(t_s)+\theta_{11}^{[12]}(t_s)+\theta_{11}^{[21]}(t_s)\right)>\sum_{s=1}^{S}\left(\theta_{12}^{[11]}(t_s)+\theta_{12}^{[22]}(t_s)+\theta_{12}^{[12]}(t_s)+\theta_{12}^{[21]}(t_s)\right),
$$
$$
\sum_{s=1}^{S}\left(\theta_{22}^{[11]}(t_s)+\theta_{22}^{[22]}(t_s)+\theta_{22}^{[12]}(t_s)+\theta_{22}^{[21]}(t_s)\right)>\sum_{s=1}^{S}\left((\theta_{12}^{[11]}(t_s)+\theta_{12}^{[22]}(t_s)+\theta_{12}^{[12]}(t_s)+\theta_{12}^{[21]}(t_s)\right),
$$
which indicate that edges are more likely to form within communities than between communities. 

\section{Simulation}\label{sec:sim}


In this section, we evaluate the clustering accuracy of \texttt{DHNet} and compare it with several alternative solutions including:

\textbf{Method 1}: treat the dynamic heterogeneous network as a dynamic homogeneous network without distinguishing the different node and edge types and apply a dynamic network community detection method \citep{zhang2017finding}. 

\textbf{Method 2}: apply a heterogeneous community detection method \citep{zhang2018modularity} to an aggregated matrix 
$
\mathcal{\bar A}=\left(\begin{array}{ccc}\bar A^{[11]}  & \ldots & \bar A^{[1 L]} \\ \vdots & \ddots & \vdots \\ \bar A^{[L 1]} & \ldots & \bar A^{[LL]} \end{array}\right),
$
where ${\bar A_{i j}^{[l_1l_2]}}=\max_{t} A_{i j}^{[l_1l_2]}(t)$, that is, detect community based on a static summary heterogeneous graph. 

\textbf{Method 3}: infer the community label from $\mG\left(t_s\right)$ for a randomly selected time point $t_s$ in $\{t_{1}, \ldots, t_{S}\}$.
That is, community detection based on a single snapshot of the dynamic heterogeneous network, which is the same as 
\cite{zhang2018modularity}.

\textbf{Method 4}: decompose the dynamic heterogeneous network with $L$ different types of nodes into $L$ dynamic homogeneous networks and apply a dynamic network community detection method \citep{zhang2017finding} to each separately, i.e., discard information from the edges linking different types of nodes.

We generate networks from the DHSBM proposed in Section \ref{sec:theory} with $L$ types of nodes, $K$ communities and $S$ equal-spaced observations within the time interval $[0, 1]$. 
We consider three different settings in our experiments including a time-homogeneous DHSBM with independently sampled networks in Section \ref{sec:sim1}, a DHSBM with independently sampled networks in Section \ref{sec:sim2} and a DHSBM with temporally correlated networks in Section \ref{sec:sim3}. In each setting, we consider dense and sparse networks.
We set $L=2$, $K=3$, $n_1=300$, $n_2=150$ and $\bm\pi^{[1]}=\bm\pi^{[2]}=(1/3,1/3,1/3)$.
To evaluate the clustering accuracy, we adopt the normalized mutual information (NMI) \citep{danon2005comparing}, a commonly used metric in community detection experiments to quantifies the difference between two 
clustering labels. 


\subsection{Simulation setting 1}\label{sec:sim1}
We consider networks independently sampled from a DHSBM 
with a time-homogeneous probability matrix defined as
$$
\bm\Theta(t)=\left(\begin{array}{ccc|ccc}
	\theta_1+r_1& \theta_1 &\theta_1&\theta_3+r_3&\theta_3&\theta_3\\
	\theta_1& \theta_1+r_1&\theta_1&\theta_3& \theta_3+ r_3&\theta_3\\
	\theta_1& \theta_1&\theta_1+r_1&\theta_3&\theta_3&\theta_3+r_3\\
	\hline
	\theta_3+r_3& \theta_3 &\theta_3&\theta_2+r_2&\theta_2&\theta_2 \\
	\theta_3& \theta_3+r_3&\theta_3&\theta_2& \theta_2+ r_2&\theta_2\\
	\theta_3& \theta_3&\theta_3+r_3&\theta_2&\theta_2&\theta_2+r_2\\
\end{array}\right).
$$
In the type-$[1]$ (type-$[2]$) homogeneous network $G^{[1]}$ ($G^{[2]}$), the parameter $\theta_{1}\left(\theta_{2}\right)$ represents the inter-community connecting probability and $\theta_{1}+r_{1}\left(\theta_{2}+r_{2}\right)$ represents the intra-community connecting probability. In the type-$[12]$ bi-partite network, $\theta_{3}$ describes the inter-community connecting probability and $\theta_{3}+r_{3}$ describes the intra-community connecting probability. The strength of the community structure is 
regulated by $r_{1}, r_{2}$ and $r_{3}$. 
We consider both dense and sparse networks in this setting with scenarios 1 and 2 on dense and sparse networks, respectively. Specifically, we consider


\textbf{Scenario 1}: 
$\theta_{1}=0.5$, $\theta_{2}=0.6$, $\theta_{3}=0.3$, $r_{1}=0$, $r_{2}=0$,


\textbf{Scenario 2}: $\theta_{1}=0.1$, $\theta_{2}=0.2$, $\theta_{3}=0.05$, $r_{1}=0$, $r_{2}=0$.

In Scenarios 1 and 2, neither $G^{[1]}$ or $G^{[2]}$ has a community structures. 
We have also considered the case where $G^{[1]}$ has a weak community structure while $G^{[2]}$ has no community structure. The results are similar to those from Scenarios 1 and 2 and delayed to the supplement.
We set $S=20$ and vary $r_3$, i.e., the strength of the community structure in $G^{[12]}$, from $0.05$ to $0.15$. Figure \ref{set1} summarizes the community detection results averaged over 100 data replicates for Scenarios 1-2, respectively.

\begin{figure}[h!]
	\centering
	\subfigure[Type 1 nodes, Scenario 1]{
		\centering
		\includegraphics[width=0.35\linewidth]{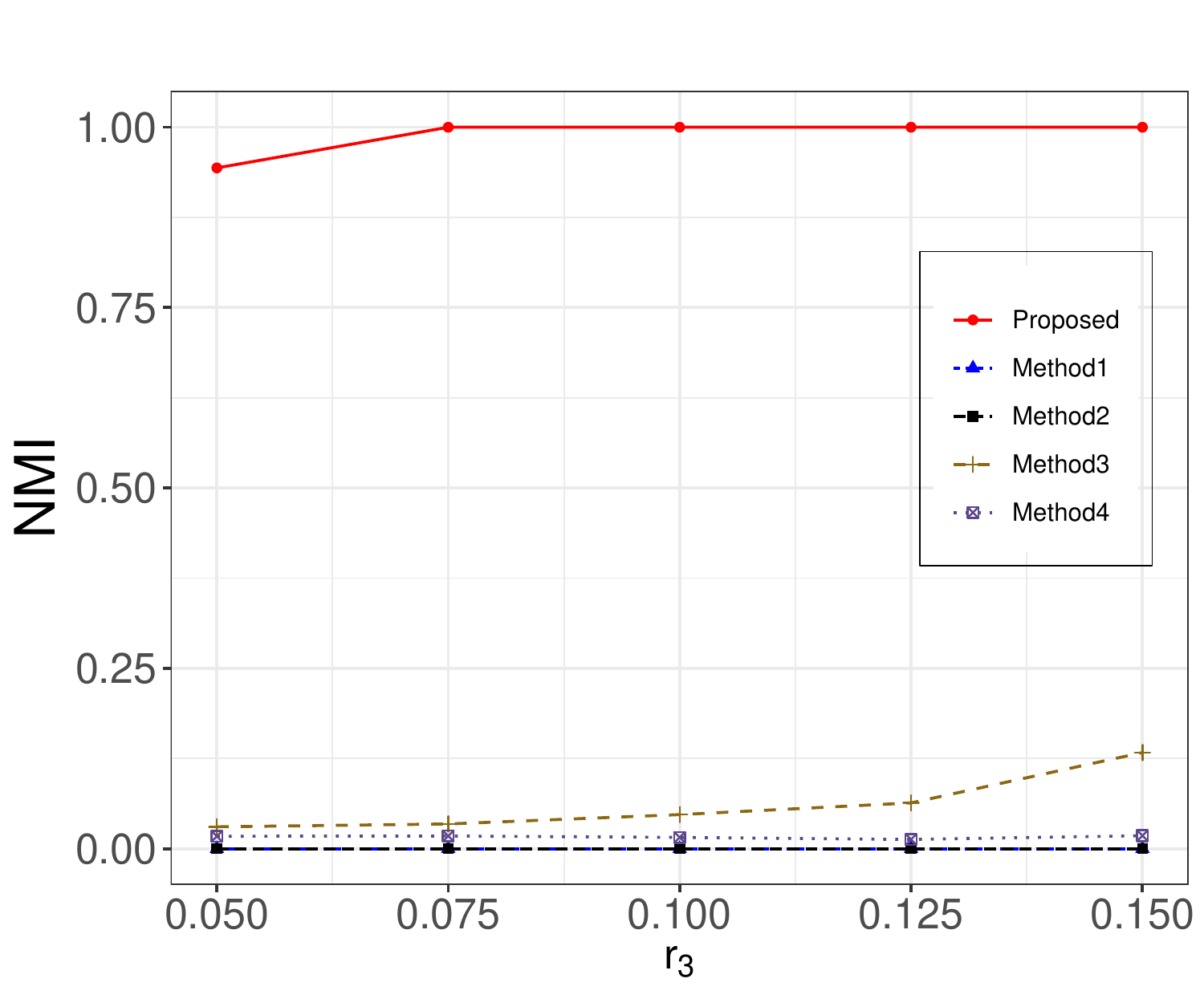}}
	\subfigure[Type 1 nodes, Scenario 2]{
		\centering
		\includegraphics[width=0.35\linewidth]{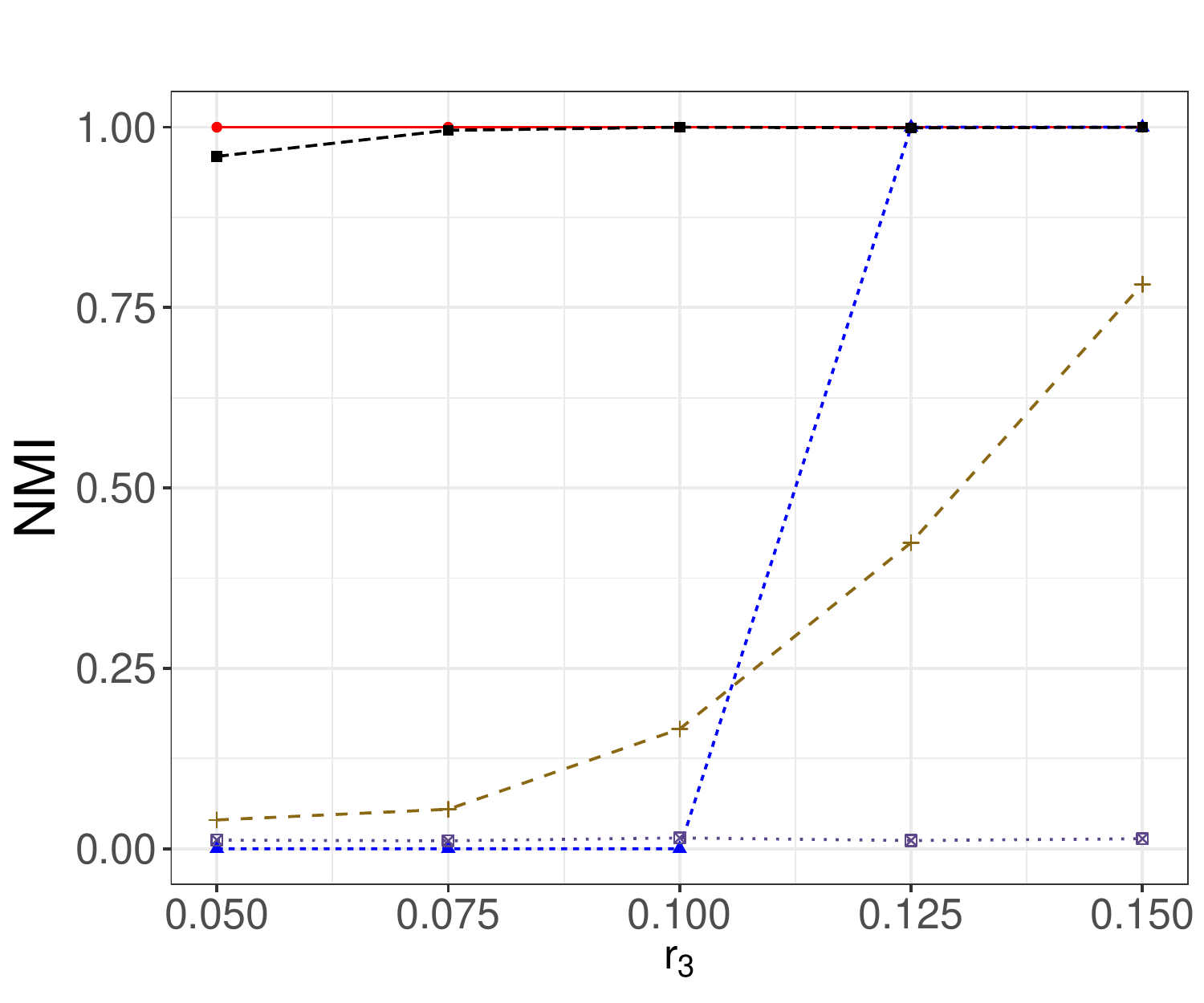}}
	\centering
	\subfigure[Type 2 nodes, Scenario 1]{
		\centering
		\includegraphics[width=0.35\linewidth]{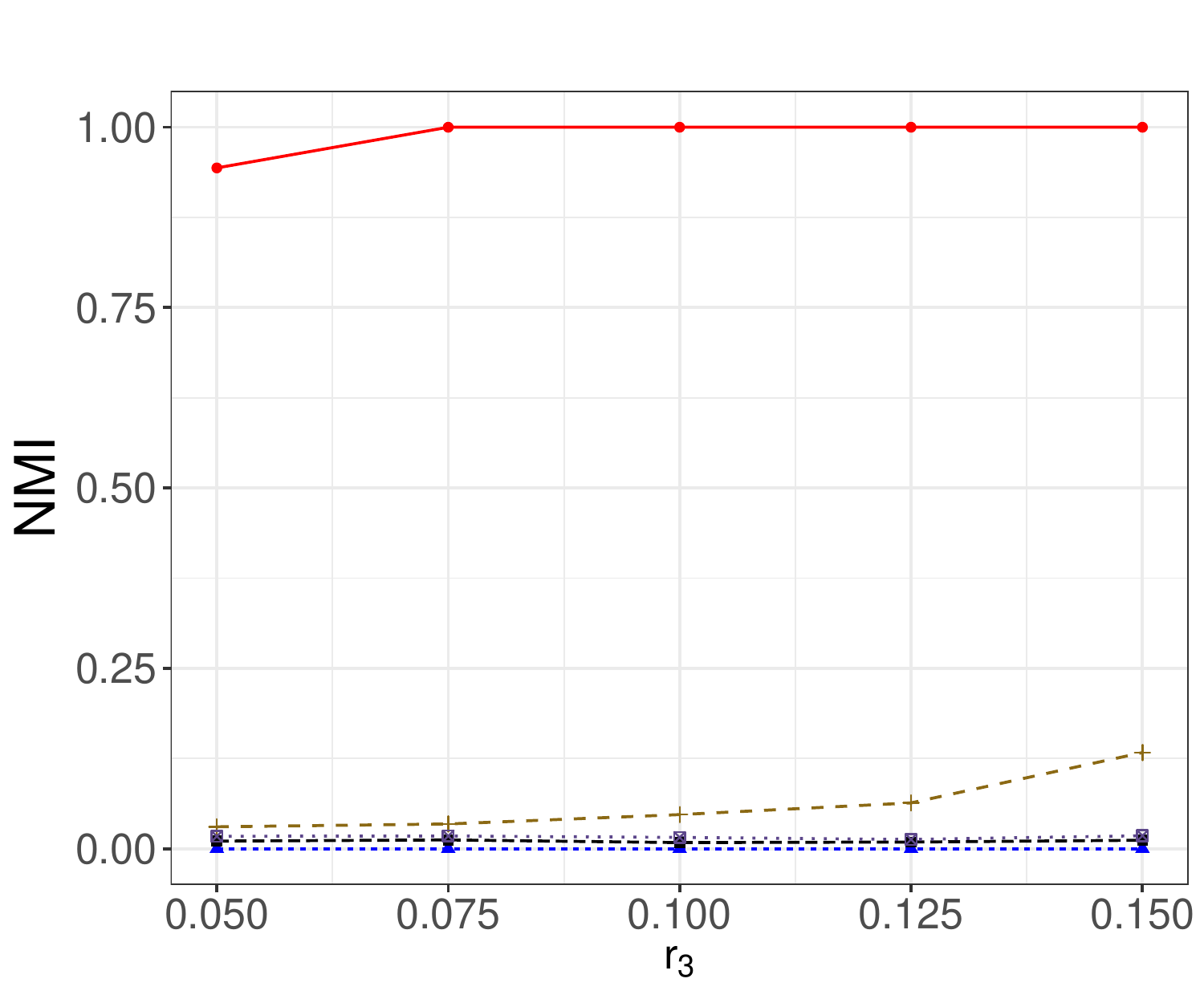}}
	\subfigure[Type 2 nodes, Scenario 2]{
		\centering
		\includegraphics[width=0.35\linewidth]{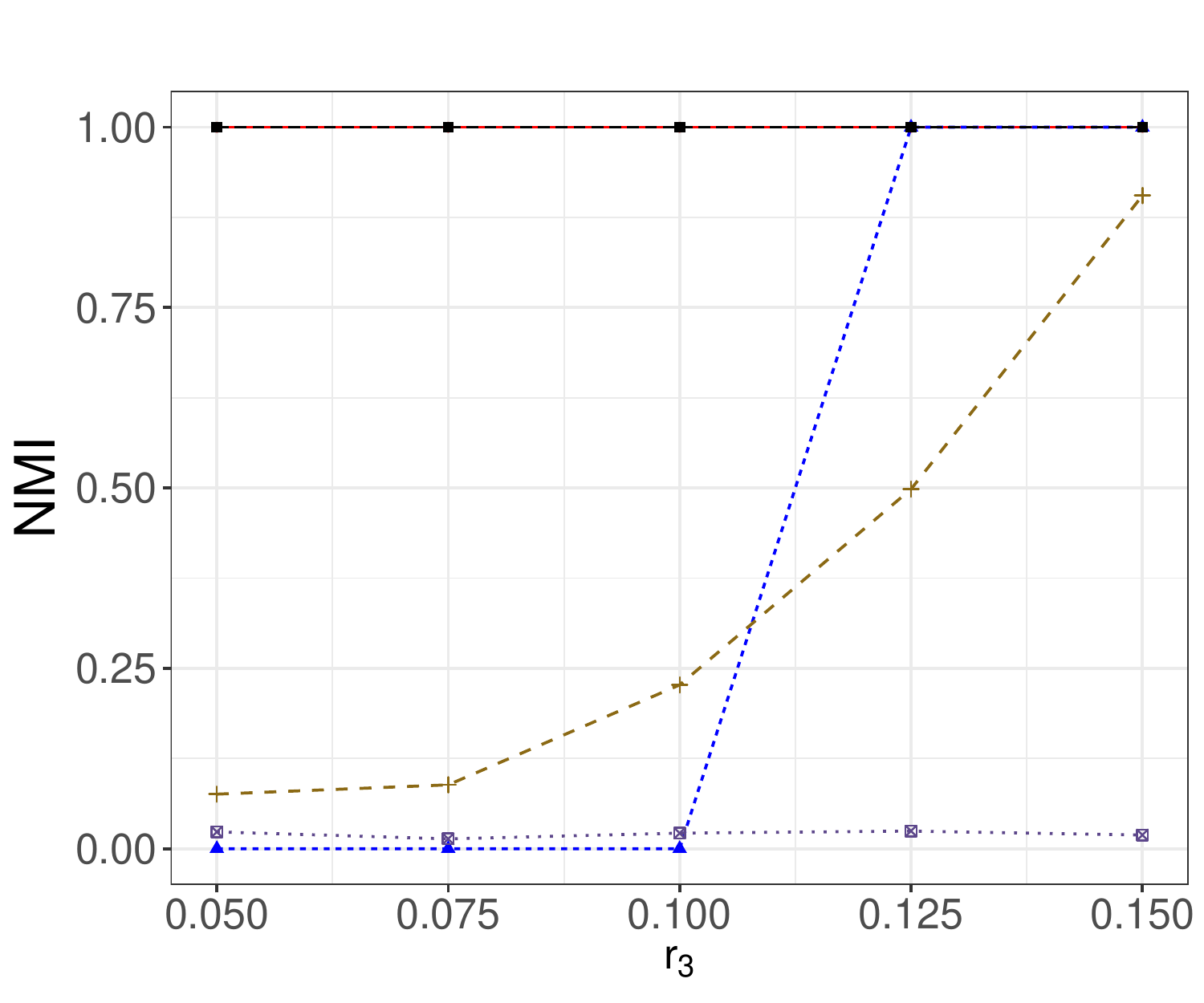}}
	\caption{Average NMIs against the value of $r_3$ for different methods in Setting 1 under Scenarios 1-2.} \label{set1}
\end{figure}

For dense networks in Scenario 1, it is seen from the left panel in Figure \ref{set1} that  \texttt{DHNet} outperforms the other methods on all values of $r_{3}$.  The NMIs from Methods 1-3 are below 0.25 for both types of nodes. 
For Method 1, the clustering output places nodes of the same type in the same community, leading to an NMI close to zero.
For Method 2, the aggregated network becomes very dense and 
the number of inter-community edges are very similar to that of the intra-community edges for each edge type, leading an NMI close to zero.
Method 3 detects community based on a random snapshot of network, which contains relatively weak structural information, 
leading to a lower NMI. 
In addition, Method 4 ignores the edges linking different types of nodes and hence perform well only when a strong 
community structure exists among the investigated type of nodes. 

For sparse networks in Scenario 2, it is seen from the right panel in Figure \ref{set1} that the performance of Methods 1 and 3 increases notably with $r_3$,
as the community structure strength (i.e., the difference between the inter- and intra- community connecting probability) is high in this scenario. 
Due to this reason, Method 2 also performs better in the sparse case as 
the community structure signal is strong in the aggregated network, with many more inter-community edges than intra-community edges.
Our method still 
outperforms most of the other methods when the signal is weak, e.g., $r_3\le 0.1$. 

\subsection{Simulation setting 2}\label{sec:sim2}
\begin{figure}[!t]
	\centering
	\subfigure[$r_{31}(t)$]{
		\centering
		\includegraphics[width=0.3\linewidth]{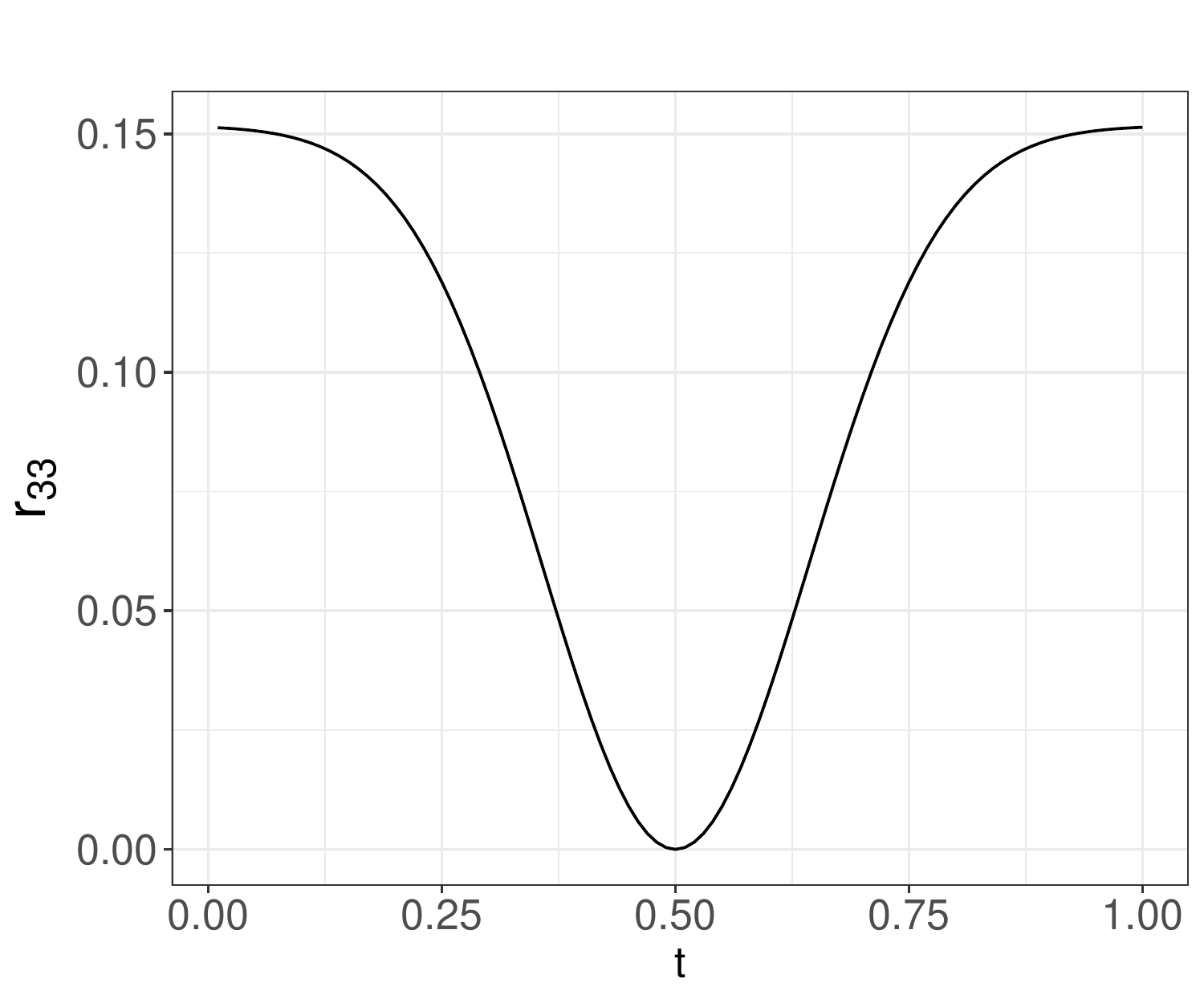}}
	\subfigure[$r_{32}(t)$]{
		\centering
		\includegraphics[width=0.3\linewidth]{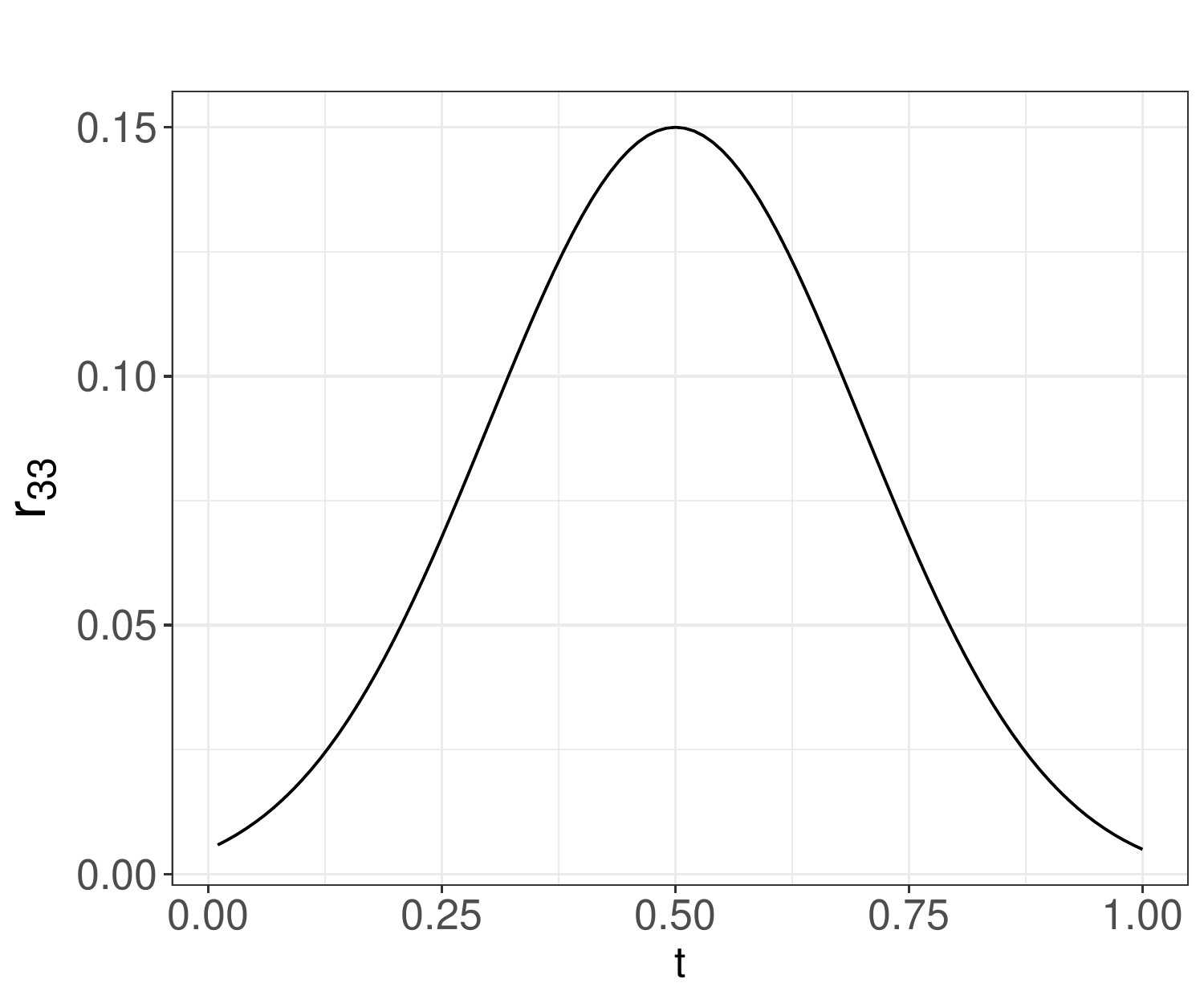}}
	\centering
	\subfigure[$r_{33}(t)$]{
		\centering
		\includegraphics[width=0.3\linewidth]{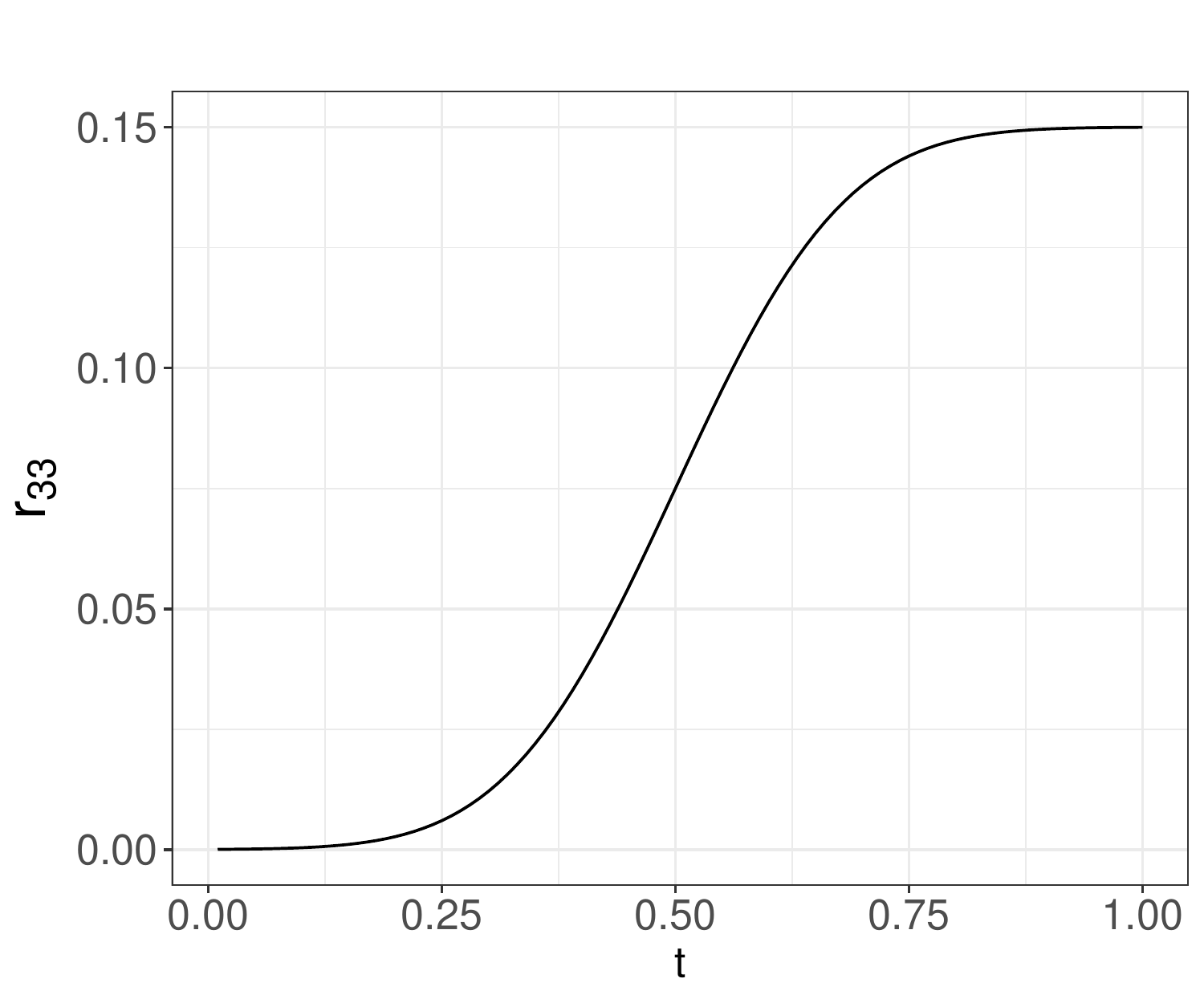}}
	\caption{\label{set2} The three time-varying functions $r_{31}(t)$, $r_{32}(t)$ and $r_{33}(t)$.} 
\end{figure}

We consider networks independently sampled from a DHSBM with a time-varying
probability matrix defined as
$$
\bm\Theta(t)=\left(\begin{array}{ccc|ccc}
	\theta_{1}+r_1& \theta_{1}&\theta_{1}&\theta_3+r_{31}(t)&\theta_3&\theta_3\\
	\theta_{1}& \theta_{1}+r_1&\theta_{1}&\theta_3& \theta_3+ r_{32}(t)&\theta_3\\
	\theta_{1} & \theta_{1}&\theta_{1}+r_1&\theta_3&\theta_3&\theta_3+r_{33}(t)\\
	\hline
	\theta_3+r_{31}(t)& \theta_3&\theta_3&\theta_2+r_2&\theta_2&\theta_2\\
	\theta_3& \theta_3+r_{32}(t)&\theta_3&\theta_2 & \theta_2 + r_2&\theta_2\\
	\theta_3& \theta_3&\theta_3+r_{33}(t)&\theta_2&\theta_2&\theta_2+r_2\\
\end{array}\right).
$$
We set $\theta_1$, $\theta_2$, $\theta_3$, $r_1$ and $r_2$ the same as those in the two scenarios in Simulation 1 and $r_{31}(t)$, $r_{32}(t)$ and $r_{33}(t)$ as plotted in Figure \ref{set2}. 
In this setting, at time $t=0$, community 1 in $G^{[12]}$ is active while communities 2-3 are inactive; at time $t=0.5$, community 1 in $G^{[12]}$ becomes inactive while communities 2-3 are active; 
at time $t=1$, community 2 in $G^{[12]}$ becomes inactive while communities 1 and 3 are both active. 
We consider $S$ in $[20, 100]$, and Figure \ref{set2_dense} summarizes the community detection results averaged over 100 data replicates for Scenarios 1-2, respectively.


\begin{figure}[t!]
	\centering
	\subfigure[Type 1 nodes, Scenario 1]{
		\centering
		\includegraphics[width=0.35\linewidth]{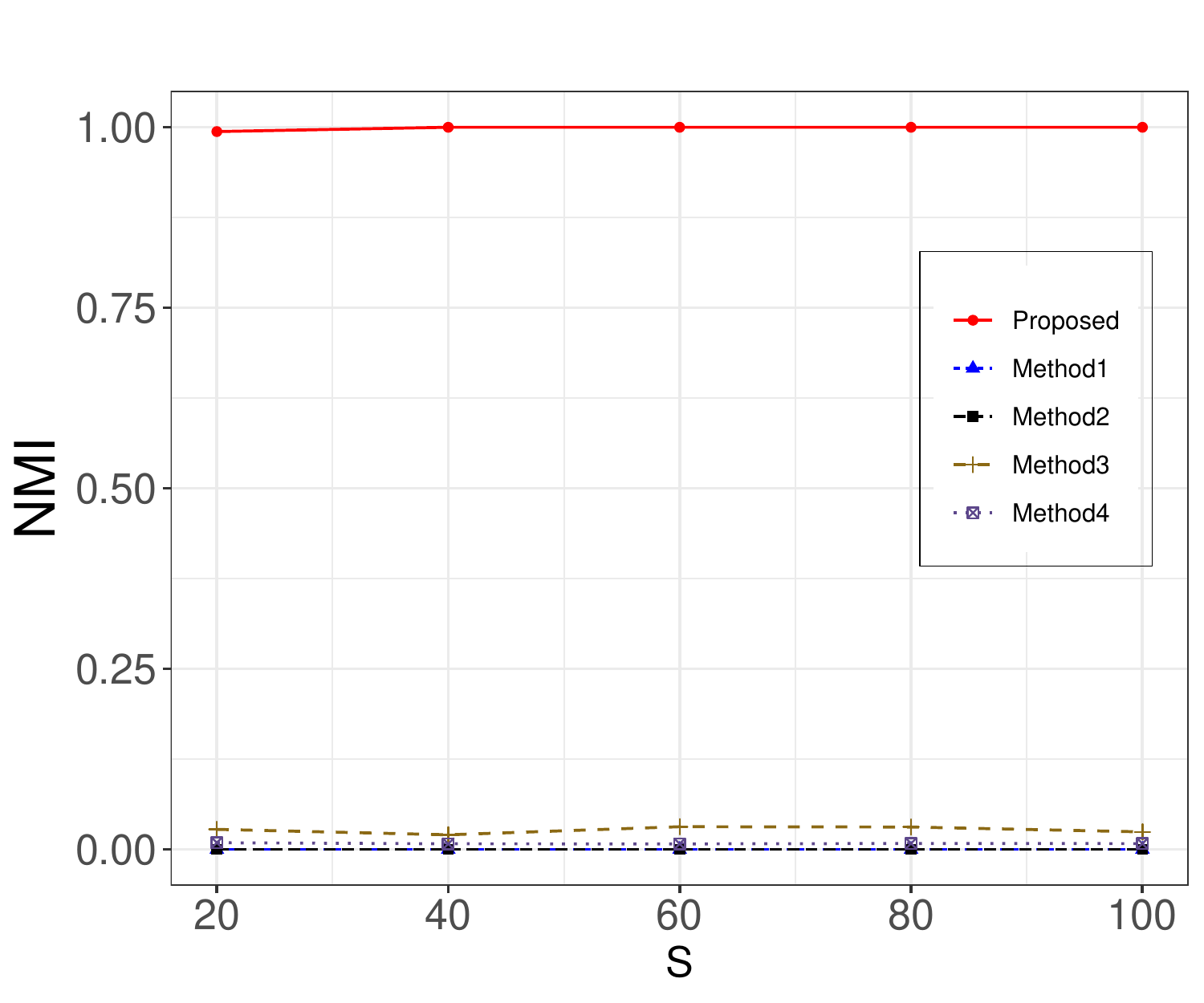}}
	\subfigure[Type 1 nodes, Scenario 2]{
		\centering
		\includegraphics[width=0.35\linewidth]{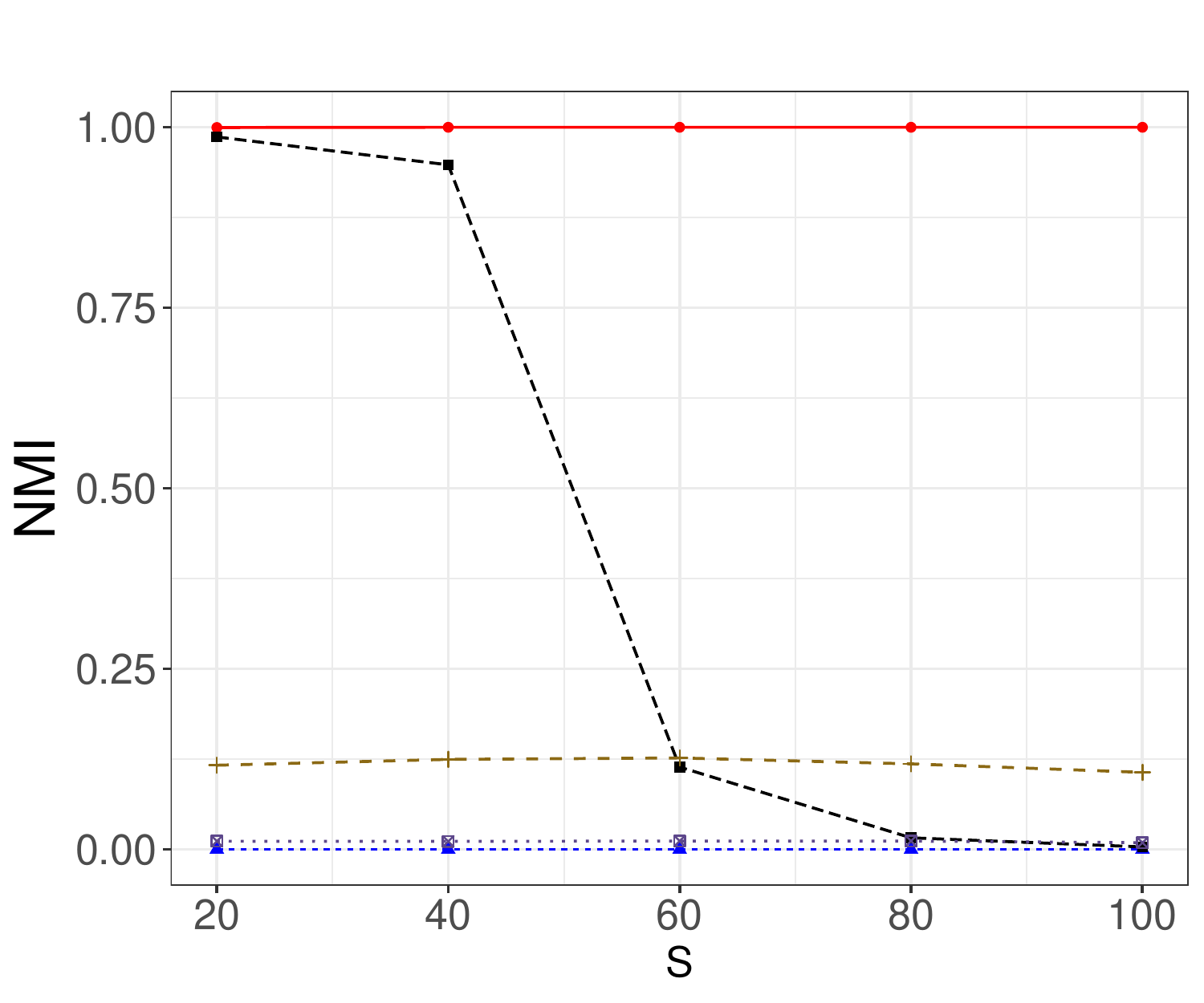}}
	\centering
	\subfigure[Type 2 nodes, Scenario 1]{
		\centering
		\includegraphics[width=0.35\linewidth]{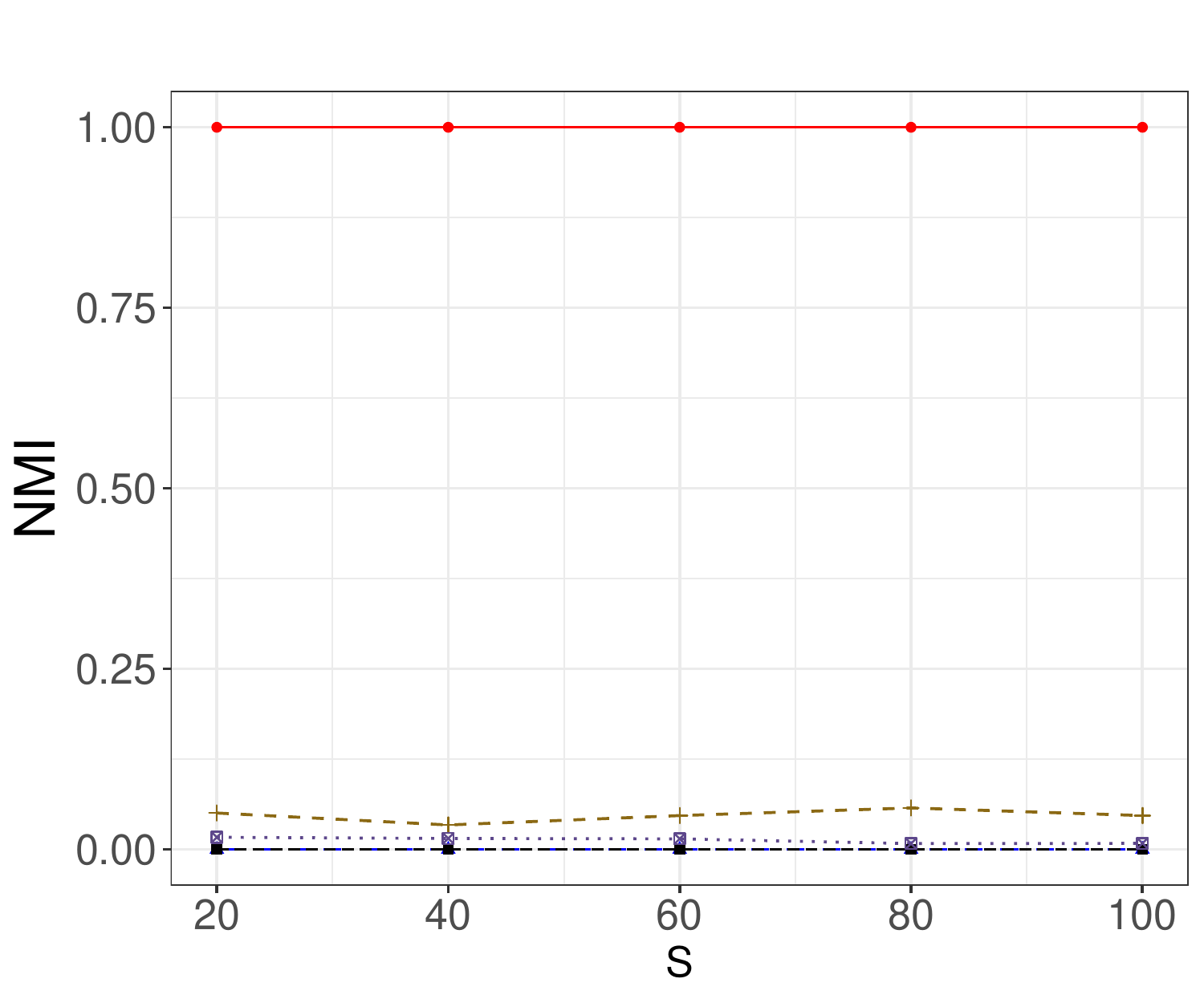}}
	\subfigure[Type 2 nodes, Scenario 2]{
		\centering
		\includegraphics[width=0.35\linewidth]{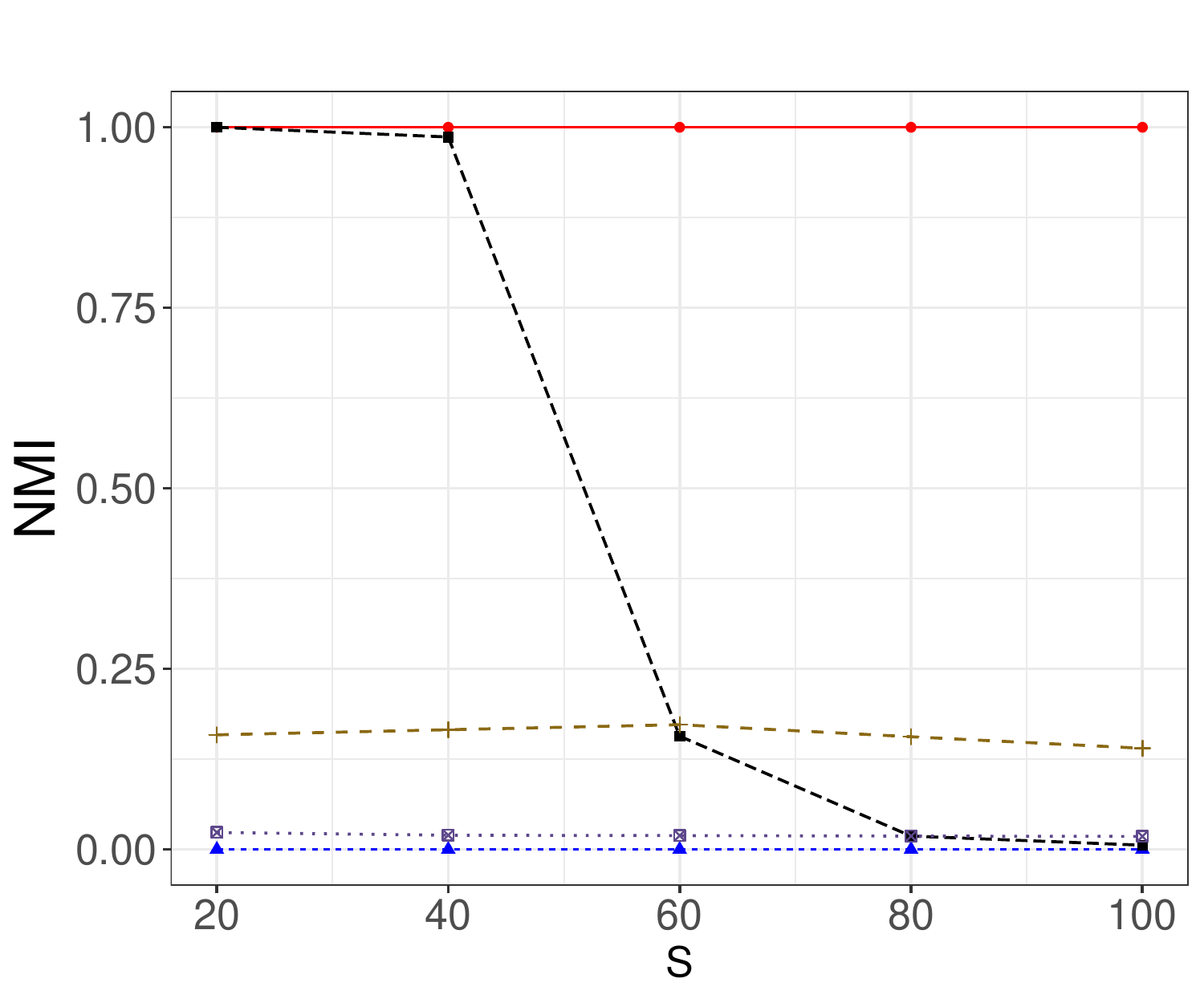}}
	\caption{Average NMIs against the value of $S$ for different methods in Setting 2 under Scenarios 1-2.} \label{set2_dense}
\end{figure}

It is seen that \texttt{DHNet} performs better than Methods 1-4 for all values of $S$, regardless the sparsity of the networks. 
Interestingly, the performance of Method 2 in Figure \ref{set2_dense} is much worse than that in Figure \ref{set1} from Simulation 1 when $S$ is large. This is because the connecting probability is time varying in Simulation 2, and the signal from communities that are active at different time points may get ablated in an aggregated picture when $S$ is large.

\subsection{Simulation setting 3}\label{sec:sim3}

We consider temporally correlated network samples from a DHSBM with a time-varying connecting
probability matrix. Specifically, we adopt the time-varying connecting probability $\bm\Theta(t)$ from Simulation setting 2.
At time $ t_{s} $, the edge $A^{[l_1l_2]}_{i j}\left(t_{s}\right)$ is a Bernoulli random variable with
$$
A_{i j}^{[l_1l_2]}\left(t_{s}\right)=u A^{[l_1l_2]}_{i j}\left(t_{s-1}\right)+(1-u) v^{[l_1l_2]},
$$
where $u \sim$ iid $\operatorname{Bernoulli}(\alpha)$ and
$$
v^{[l_1l_2]} \overset{iid}\sim \text { Bernoulli }\left(\frac{\theta^{[l_1l_2]}_{k_1k_2}\left(t_{s}\right)-\alpha \theta^{[l_1l_2]}_{k_1k_2}\left(t_{s-1}\right)}{1-\alpha}\right),\quad l_1, l_2 \in [L].
$$
Given $\bm\Theta(t)$, a larger $\alpha$ leads to a higher correlation between the networks at two adjacent time points. 
We set $S=100$,
$\alpha\in[0, 0.4]$ for Scenario 1 and $\alpha\in[0, 0.7]$ for Scenario 2, as to keep the probability parameter $\frac{\theta^{[l_1l_2]}_{k_1k_2}\left(t_{s}\right)-\alpha \theta^{[l_1l_2]}_{k_1k_2}\left(t_{s-1}\right)}{1-\alpha}>0$. 
Figure \ref{set3_dense} summarizes the community detection results averaged over 100 data replicates for Scenarios 1-2, respectively.

\begin{figure}[!h]
	\centering
	\subfigure[Type 1 nodes, Scenario 1]{
		\centering
		\includegraphics[width=0.35\linewidth]{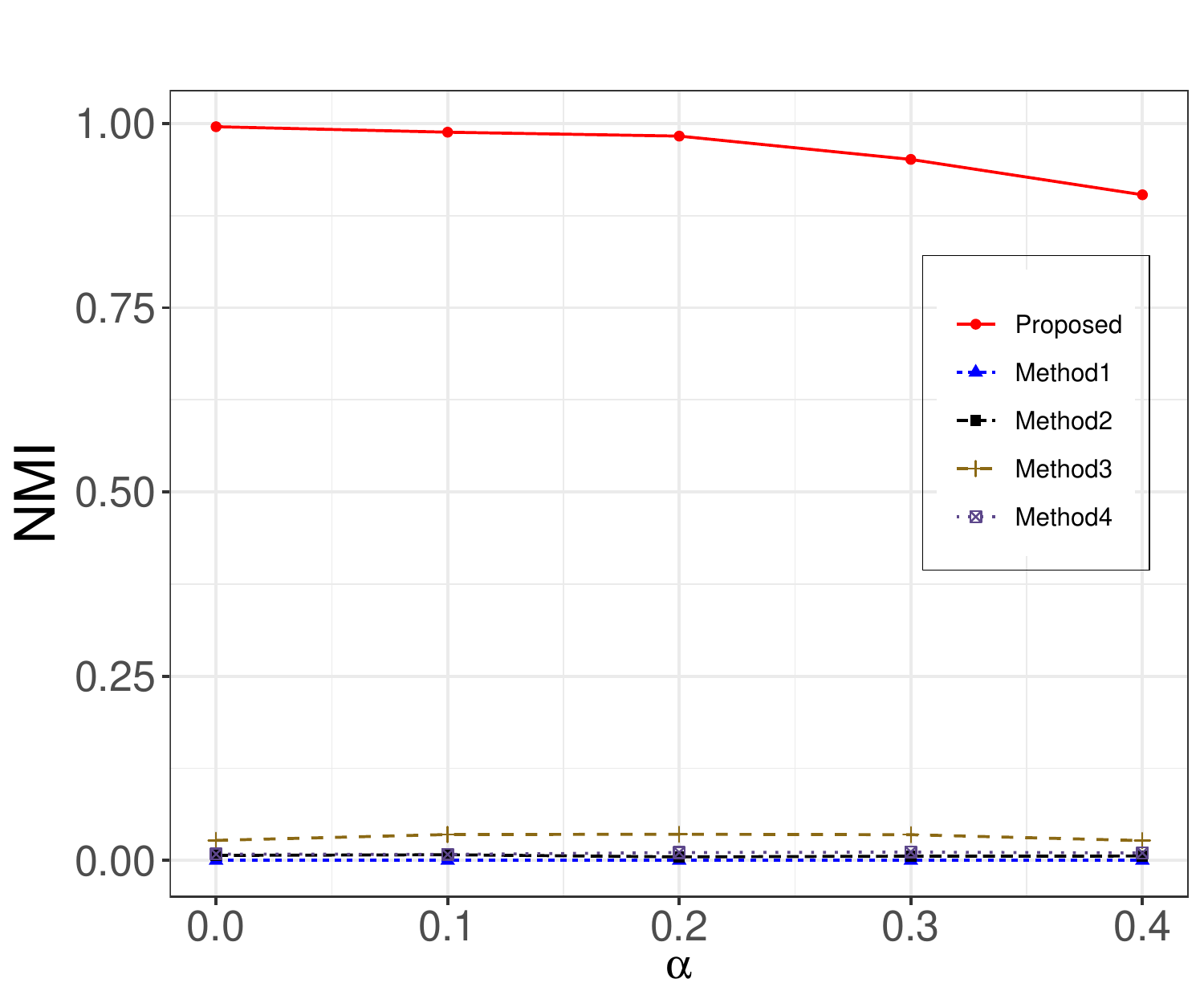}}
	\subfigure[Type 1 nodes, Scenario 2]{
		\centering
		\includegraphics[width=0.35\linewidth]{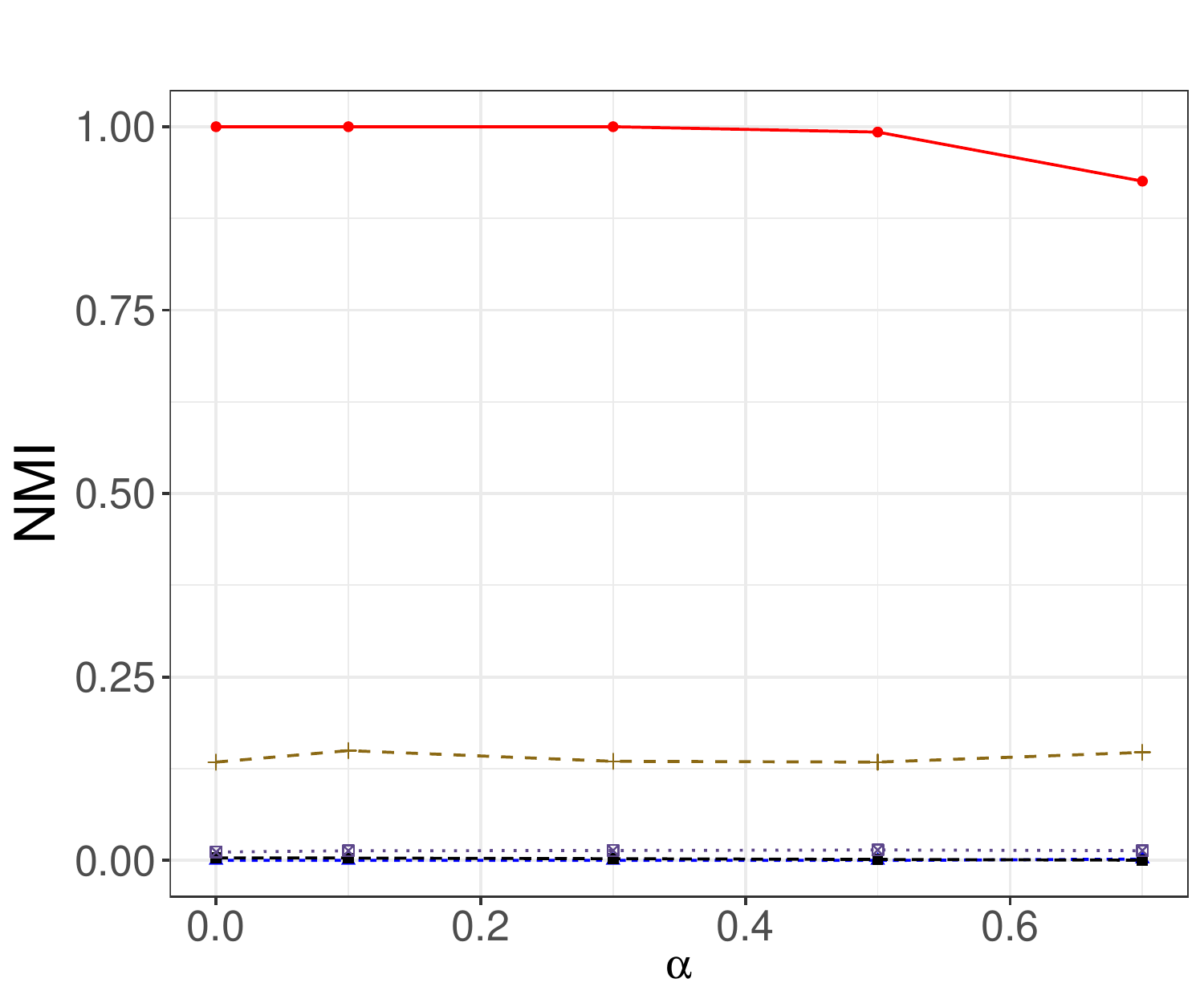}}
	\centering
	\subfigure[Type 2 nodes, Scenario 1]{
		\centering
		\includegraphics[width=0.35\linewidth]{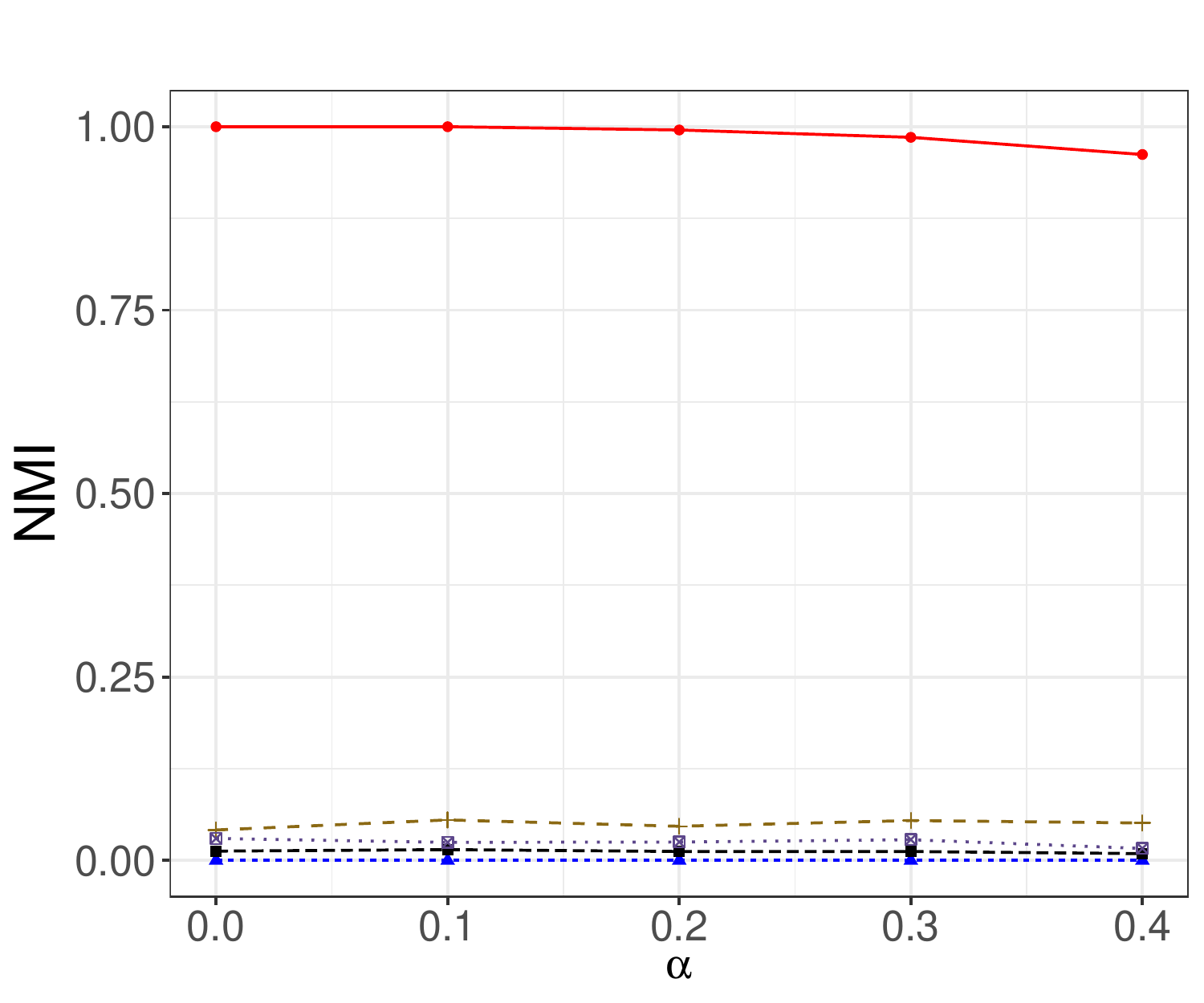}}
	\subfigure[Type 2 nodes, Scenario 2]{
		\centering
		\includegraphics[width=0.35\linewidth]{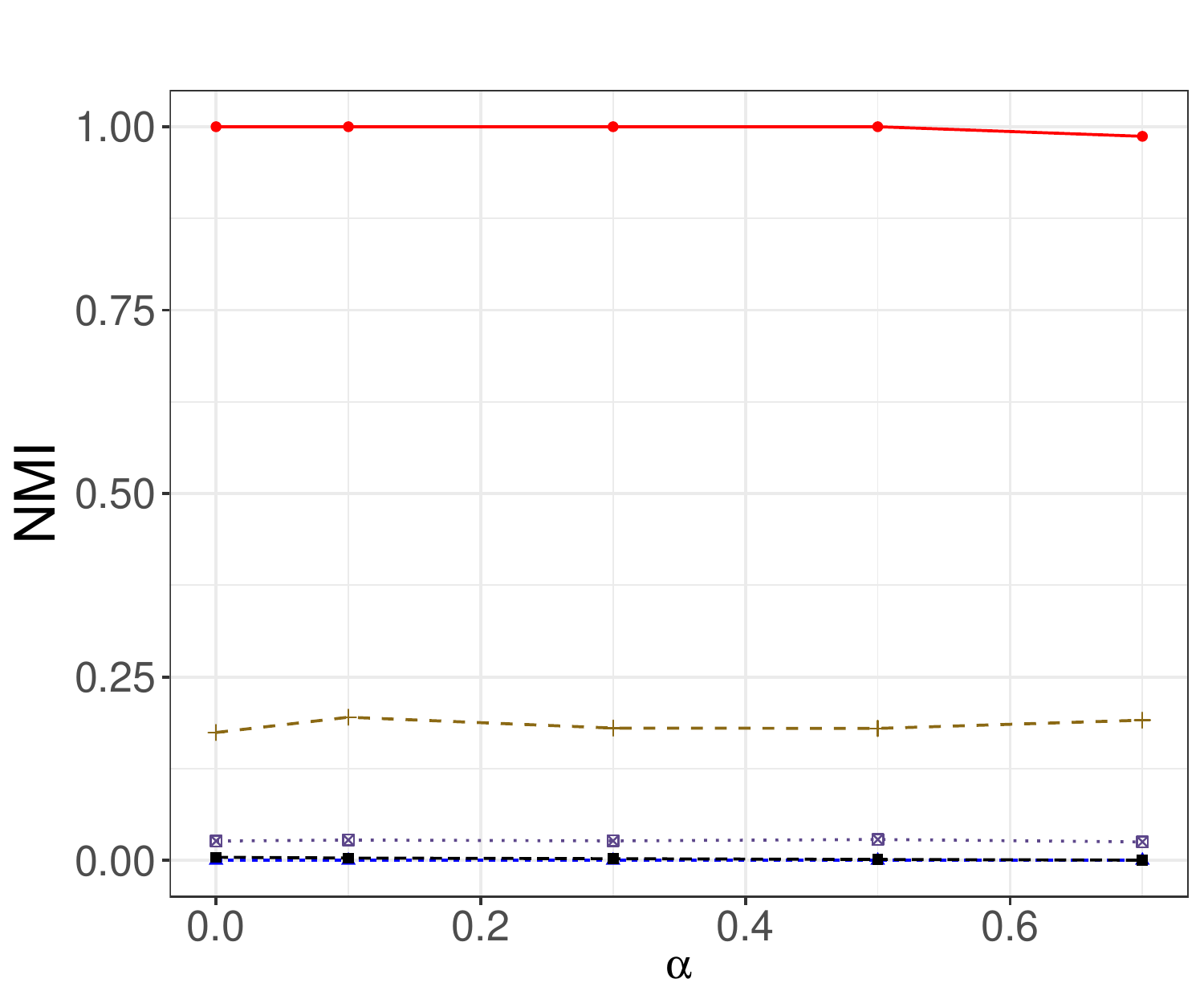}}
	\caption{Average NMIs against the value of $\alpha$ for different methods in Setting 3 under Scenarios 1-2.} \label{set3_dense}
\end{figure}

Similar conclusions as before can be drawn for all methods shown in Figure \ref{set3_dense}. 
\texttt{DHNet} has the best performance out of the five methods. When $\alpha = 0$, $A_{i j}(t)$ is uncorrelated with the past observations, and the model is equivalent to the model used in Simulation setting 2. In fact, as $\alpha$ increases, the effective sample size decreases, leading to a deteriorated performance of \texttt{DHNet}. Method 3 relies only on a random snapshot of network and as such, it is insensitive to changes in $\alpha$.

\section{Yelp review network}\label{sec:real}

Yelp is a well-known review website, founded in 2004 in the United States. It collects reviews on a wide range of businesses such as restaurants, bars and shops from many countries. 
On the Yelp platform, users can rate businesses, submit reviews, and share experiences. 
We analyze 
the review data from the Yelp Challenge (\url{https://www.kaggle.com/yelp-dataset/yelp-dataset}) during the period from January 1st, 2006 to December 31, 2017. This dataset contains a set of businesses and the category labels of each business (a business usually has several labels), a set of users and the friendship information among these users, and the reviews of these businesses by these users. The businesses and users are anonymized and labeled with numerical identifiers. In this dataset, the business-category (business is labeled with category) and user-user (user is friend with user) information are not labeled by time (i.e., not time-varying) while the user-business (business reviewed by user) interactions are labeled by time, and a user may review a business several times. 

Our analysis focuses on finding heterogeneous communities in the Yelp review network 
and predicting interests for new users.
Our results show improvements both in terms of accuracy and interpretability over existing solutions, and demonstrate the need to consider network heterogeneity and dynamics in community detection.


\subsection{Finding heterogeneous communities} 
To get a comprehensive view of the Yelp review network, we consider a heterogeneous network with three types of nodes including business, user and category, connected via three types of edges, user-user (user is friend with user), user-business (business is reviewed by user), and business-category (business is labeled with category); 
see Figure \ref{yelp} for a simple illustration. 
As discussed earlier, the user-user and business-category edges are not time-varying but the user-business edges are. 
We focus on businesses that operated continuously in the study period, business categories that had at least 10 occurrences and users that reviewed at least 20 times in the study period. This gives a total of 3,566 businesses, 207 categories and 5,116 users, with 141,744 user-user and 17,280 business-category and 194,712 user-business edges.
Due to the high sparsity of user-business edges, we use \textit{year} as the time unit when constructing the dynamic network, that is, the network at time $t$ summarizes the review activity between users and businesses in the $t$-th year of the study period, and correspondingly $S=12$. 


We applied \texttt{DHNet} to the constructed dynamic heterogeneous network with $\kappa =200$ and identified 11 communities with a maximized modularity value of 0.237. 
Table \ref{tabc} shows the representative categories, number of users and number of businesses in each identified community, along with a summarizing theme. The complete list of categories in each community can be found in the supplement. 
We found that each community identified by \texttt{DHNet} contains a distinctive type of businesses. 
For example, Communities 3-4 are on dining and Community 7 is mostly on Beauty \& Medical.
Users in Community 1 prefer activities related to pets, users in Community 6 prefer bars and entertainment 
and users in Community 9 show interests in traveling and sports. 
Community 11 is mostly on Auto and we did not identify users whose main review activity and interests are in this type of businesses. These insights can help us understand the life styles and interests of users in each community. 


	\begin{table}[!t] 
	\renewcommand\arraystretch{1.25}
	\centering  
	\caption{Summary of the 11 communities identified by \texttt{DHNet}.} 
	\vspace{0.5em}
	\begin{tabular}{|c|c|c|c|c|}
			\hline
		\multirow{1}{*}{} & \multirow{1}{*}{categories} & 
		\begin{tabular}[c]{@{}c@{}}\# of \\ users \end{tabular}  
		& \begin{tabular}[c]{@{}c@{}}\# of \\ businesses \end{tabular}  
		&\multirow{1}{*}{theme}\\
		\hline
		\multirow{2}{*}{1} & \textit{Animal Shelters},  \textit{Pet Groomers}   & \multirow{2}{*}{719} & \multirow{2}{*}{331}& \multirow{2}{*}{Pets} \\
		&\textit{Pet Services}, \textit{Veterinarians} &&&\\
		\hline
	2&  \textit{Tex-Mex}, \textit{Southern}  & 939&408 & Tex-Mex\\
			\hline
		\multirow{2}{*}{3} & \textit{Tea}, \textit{Fast Food}  & \multirow{2}{*}{2629} & \multirow{2}{*}{1036}& \multirow{2}{*}{Casual Dining} \\
	&\textit{Diners}, \textit{Pizza}, \textit{Restaurants} &&&\\
			\hline
				\multirow{2}{*}{4} & \textit{French}, \textit{Pasta Shops}, \textit{Steak House}     & \multirow{2}{*}{679} & \multirow{2}{*}{337} & \multirow{2}{*}{Fine Dining}\\
			&\textit{Professional Services}, \textit{Seafood} &&&\\
		\hline
			\multirow{2}{*}{5} &  \textit{Candy Stores}, \textit{Farmers Market}    & \multirow{2}{*}{13} & \multirow{2}{*}{239}& \multirow{1}{*}{Stores}\\
		&\textit{Chocolatiers \& Shops}, \textit{Grocery} &&& \& Markets\\
		\hline
			\multirow{2}{*}{6} &  \textit{Adult Entertainment}, \textit{Bars}  & \multirow{2}{*}{46} & \multirow{2}{*}{225} & \multirow{1}{*}{Bars,} \\
		& \textit{Dance Clubs},  \textit{Beer Bar} &&& Entertainment\\
		\hline
			\multirow{2}{*}{7} &  \textit{Beauty \& Spas}, \textit{Doctors}    & \multirow{2}{*}{79} & \multirow{2}{*}{441} & \multirow{1}{*}{Beauty}\\
		&  \textit{Hair Salons}, \textit{Health \& Medical}  &&& \& Medical\\
		\hline
			\multirow{2}{*}{8} &   \textit{Home Services}, \textit{Laundry Services}     & \multirow{2}{*}{5} & \multirow{2}{*}{289} & \multirow{1}{*}{Shopping}\\
		&  \textit{Music \& Video}, \textit{Shopping Centers} &&& \& Life\\
		\hline
			\multirow{2}{*}{9} & \textit{Hotels \& Travel}, \textit{Venues \& Event Spaces}    & \multirow{2}{*}{4} & \multirow{2}{*}{123} & \multirow{1}{*}{Leisure} \\
		&  \textit{Landmarks \& Historical Buildings}, \textit{Tours} &&& \& Travel\\
		\hline
		\multirow{1}{*}{10}&\multirow{1}{*}{\textit{Buffets}, \textit{Indian}, \textit{Pakistani}} &\multirow{1}{*}{3} &\multirow{1}{*}{103}&Asian Fusion \\
		\hline
			\multirow{2}{*}{11}&\textit{Auto Parts \& Supplies}, \textit{Auto Repair} &\multirow{2}{*}{0} &\multirow{2}{*}{34}&\multirow{2}{*}{Auto}\\
			&\textit{Automotive}, \textit{Gas Stations}, \textit{Tires}&&&\\
		\hline
	\end{tabular}
	\label{tabc}
\end{table}

We had also applied Methods 1-3 from Section \ref{sec:sim}, though we did not implement Method 4, which considers each homogeneous networks separately and discards information from the edges linking different types of nodes, as there are no business-business or category-category edges and the user-user edges are not time-varying. 
The results from Method 1, which does not distinguish the different node and edge types, are very difficult to interpret. For example, one community contains only businesses and one community contains only users. 
Method 2, which considers an aggregated heterogeneous network over time, also identified 11 communities (see details of the communities in the supplement). 
The community detection results from Method 2 are less interpretable compared to \texttt{DHNet} and several communities contain mixed businesses themes. 
For example, \textit{Hobby Shops} is placed into Community 4 that is on fine dining, and 
\textit{Colleges \& Universities} and \textit{Education} are placed into Community 9 that is on leisure and travel.
Method 3, which considers a snapshot of the dynamic work, does not perform well,
as the network at each time point is highly sparse with a large number of isolated nodes. 

\subsection{Prediction interests for new users}

In this section, we aim to predict the interests, in terms of business categories, for a new Yelp user based on Yelp activities of his/her friends, a practically useful task in making recommendations and placing advertisements. 
We focus on predicting interests in business categories as opposed to individual businesses, as the number of businesses is large and user-business interactions are highly sparse.
For a new Yelp user, the platform can often collect his/her friendship information with other existing Yelp users, by accessing phone contacts, email contacts and Facebook friendship. In terms of make recommendations, the Yelp activities of friends of a new user can help to ease the ``cold start'' problem, the issue where personalized recommendations cannot be made before a user interacts with the system (e.g., reviewing businesses). 


Consider training and testing datasets taken from two different time periods (e.g., data from years 2006-2015 as training and years 2015-2017 as testing). 
We are interested in making predictions for the new users in the testing set, which are user accounts that did not exist in the training data.
Specifically, for a new user in the testing set, based on Yelp activities of his/her friends in the training data, we predict the his/her interests over the business categories and compare the prediction with the ``true'' measure calculated from the testing set. 

We compare two different prediction strategies. 
The first strategy utilizes community detection results from \texttt{DHNet} in making the prediction and the second strategy directly averages interests from the new user's friends without using any community information, referred to as the naive strategy. The naive strategy is a commonly adopted practice in recommender systems \citep{tey2021accuracy}.
Specifically, let $g_i$ denote the interest measure of the $i$-th new user in the testing set, which is a probability distribution over all categories; it is calculated using the appearance frequency of each category in the businesses reviewed by this user.
In the first strategy, we apply \texttt{DHNet} to the training network data and find
the category distribution of each community, denoted as $f_j$ for the $j$-th community, calculated using the appearance frequency of each category from the businesses in this community. 
We then make prediction $g_i^{\texttt{DHNet}}$ using the weighted average of $f_j$'s as below
$$g_i^{\texttt{DHNet}}=\sum_{j} \frac{n_{ij}}{\sum_j n_{ij}}f_j,$$
where $n_{ij}$ denotes the number of friends that the $i$-th user has in the $j$-th community. 
In this strategy, the prediction is a weighted average of measures from all communities where the weight reflect the number of connections the new user has to each community. 
In the second strategy,
we directly calculate the category distribution $g_i^{Naive}$ based on the businesses that $i$-th user’s friends visited during the training period. 
This strategy only focuses on the ego-centric network of the new user and does not taken into the rich information in the network communities. 
To assess the prediction accuracy, we use the Jensen–Shannon divergence (JSD) to compare the estimated and observed category distributions, that is,
$$\operatorname{JSD}(\hat{g_i}\|g_i )=\frac{1}{2} D(\hat{g_i}\| m)+\frac{1}{2} D(g_i \| m),$$
where $\hat{g}_i$ refers to the estimated category distribution, 
$m=\frac{1}{2}(\hat{g}_i+g_i$) and
$D(\hat{g}_i\| m)$ is the Kullback–Leibler divergence between distributions $\hat{g}_i$ and $m$. 

\begin{table}[!t] 
  \renewcommand\arraystretch{1.5}
  \centering  
  \caption{Jensen–Shannon divergence of the \texttt{DHNet} and naive methods across the 6 moving windows.} 
  \begin{tabular}{|c|c|c|c|c|c|c|}
  	\hline
  	training years &2006-2010 &2007-2011& 2008-2012&2009-2013&2010-2014&2011-2015\\
  	\hline
  	testing years &2011-2012 & 2012-2013 & 2013-2014 & 2014-2015 & 2015-2016 & 2016-2017\\
  	\hline
  	$\operatorname{JSD}^{\texttt{DHNet}}$ & 0.122 & 0.133 & 0.132 & 0.136 & 0.135& 0.138 \\
  	\hline
  	$\operatorname{JSD}^{\texttt{Naive}}$ & 0.252& 0.196& 0.199& 0.196& 0.182& 0.196 \\
  	\hline
  \end{tabular}
  \label{tab:naive}
\end{table}


Table \ref{tab:naive} compares the performance of the two strategies in 6 different sets of training and testing periods, where $$
\operatorname{JSD}^{\texttt{DHNet}}=\frac{1}{n_0}\sum_{i=1}^{n_0}\operatorname{JSD}({g}_i^{\texttt{DHNet}}\|g_i ),\quad \operatorname{JSD}^{\texttt{Naive}}=\frac{1}{n_0}\sum_{i=1}^{n_0}\operatorname{JSD}({g}_i^{\texttt{Naive}}\|g_i )
$$ 
and $n_0$ denotes the number of new users in the testing period.
It is seen that \texttt{DHNet} outperforms the naive strategy in terms of predicting accuracy in all training and testing datasets, demonstrating the advantage of utilizing community structures when predicting user interests.


\section{Discussion}


Maximizing the modularity function as in \eqref{eq4} is not limited to the
Louvain-type method considered in \texttt{DHNet}. 
Other modularity maximization techniques developed for a homogeneous network may be applied
to \eqref{eq4} with some modifications, such as the spectral method based on the eigen
decomposition of the modularity matrix or the stochastic optimization method in \citet{massen2005identifying}. As noted in modularity maximization for other types of networks \citep{fortunato2010community,zhang2018modularity}, we find that the Louvain-type method is computationally much more efficient and yields a good performance in our setting.

While the modularity function value increases at each step of \texttt{DHNet} and the algorithm is guaranteed to converge, there is no guarantee that it will converge to the global optimum. 
Since the modularity maximization problem is NP-hard, most existing methods are heuristic methods that may only find local optima and are not guaranteed to find the global optimum. 
A thorough theoretical investigation of the local convergence of \texttt{DHNet} can be helpful and we leave it as future work.
Finally, our proposed method can be extended to weighted and/or directed networks. 
To incorporate weighted and/or directed edges into our framework, we need to define a null model for a weighted and/or directed heterogeneous dynamic network, followed by calculating the expectations under the null model. This is an interesting topic to investigate next.


\bibliographystyle{asa}

\begin{thebibliography}{27}
\newcommand{\enquote}[1]{``#1''}
\expandafter\ifx\csname natexlab\endcsname\relax\def\natexlab#1{#1}\fi

\bibitem[{Abbe(2017)}]{abbe2017community}
Abbe, E. (2017), \enquote{Community detection and stochastic block models:
  recent developments,} \textit{The Journal of Machine Learning Research}, 18,
  6446--6531.

\bibitem[{Blondel et~al.(2008)Blondel, Guillaume, Lambiotte, and
  Lefebvre}]{blondel2008fast}
Blondel, V.~D., Guillaume, J.-L., Lambiotte, R., and Lefebvre, E. (2008),
  \enquote{Fast unfolding of communities in large networks,} \textit{Journal of
  Statistical Mechanics: Theory and Experiment}, 2008, P10008.

\bibitem[{Brandes et~al.(2008)Brandes, Delling, Gaertler, Gorke, Hoefer,
  Nikoloski, and Wagner}]{brandes2007modularity}
Brandes, U., Delling, D., Gaertler, M., Gorke, R., Hoefer, M., Nikoloski, Z.,
  and Wagner, D. (2008), \enquote{On modularity clustering,} \textit{IEEE
  Transactions on Knowledge and Data Engineering}, 20, 172--188.

\bibitem[{Chung et~al.(2006)Chung, Fan, Chung, Graham, Lu, Chung,
  et~al.}]{chung2006complex}
Chung, F., Fan, R., Chung, F.~R., Graham, F.~C., Lu, L., Chung, K.~F., et~al.
  (2006), \textit{Complex Graphs and Networks}, no. 107, American Mathematical
  Soc.

\bibitem[{Clauset et~al.(2004)Clauset, Newman, and Moore}]{clauset2004finding}
Clauset, A., Newman, M.~E., and Moore, C. (2004), \enquote{Finding community
  structure in very large networks,} \textit{Physical Review E}, 70, 066111.

\bibitem[{Danon et~al.(2005)Danon, Diaz-Guilera, Duch, and
  Arenas}]{danon2005comparing}
Danon, L., Diaz-Guilera, A., Duch, J., and Arenas, A. (2005),
  \enquote{Comparing community structure identification,} \textit{Journal of
  Statistical Mechanics: Theory and Experiment}, 2005, P09008.

\bibitem[{Fortunato(2010)}]{fortunato2010community}
Fortunato, S. (2010), \enquote{Community detection in graphs,} \textit{Physics
  Reports}, 486, 75--174.

\bibitem[{Guimera et~al.(2004)Guimera, Sales-Pardo, and
  Amaral}]{guimera2004modularity}
Guimera, R., Sales-Pardo, M., and Amaral, L. A.~N. (2004), \enquote{Modularity
  from fluctuations in random graphs and complex networks,} \textit{Physical
  Review E}, 70, 025101.

\bibitem[{Jiang et~al.(2021)Jiang, Koch, and Sun}]{jiang2021hints}
Jiang, S., Koch, B., and Sun, Y. (2021), \enquote{HINTS: Citation time series
  prediction for new publications via dynamic heterogeneous information network
  embedding,} in \textit{Proceedings of the Web Conference 2021}, pp.
  3158--3167.

\bibitem[{Linden et~al.(2003)Linden, Smith, and York}]{linden2003amazon}
Linden, G., Smith, B., and York, J. (2003), \enquote{Amazon. com
  recommendations: Item-to-item collaborative filtering,} \textit{IEEE Internet
  Computing}, 7, 76--80.

\bibitem[{Massen and Doye(2005)}]{massen2005identifying}
Massen, C.~P. and Doye, J.~P. (2005), \enquote{Identifying communities within
  energy landscapes,} \textit{Physical Review E}, 71, 046101.

\bibitem[{Moody and White(2003)}]{moody2003structural}
Moody, J. and White, D.~R. (2003), \enquote{Structural cohesion and
  embeddedness: A hierarchical concept of social groups,} \textit{American
  Sociological Review}, 68, 103--127.

\bibitem[{Newman(2006)}]{newman2006finding}
Newman, M.~E. (2006), \enquote{Finding community structure in networks using
  the eigenvectors of matrices,} \textit{Physical Review E}, 74, 036104.

\bibitem[{Newman and Girvan(2004)}]{newman2004finding}
Newman, M.~E. and Girvan, M. (2004), \enquote{Finding and evaluating community
  structure in networks,} \textit{Physical Review E}, 69, 026113.

\bibitem[{Sengupta and Chen(2015)}]{sengupta2015spectral}
Sengupta, S. and Chen, Y. (2015), \enquote{Spectral clustering in heterogeneous
  networks,} \textit{Statistica Sinica}, 25, 1081--1106.

\bibitem[{S{\o}rlie et~al.(2001)S{\o}rlie, Perou, Tibshirani, Aas, Geisler,
  Johnsen, Hastie, Eisen, Van De~Rijn, Jeffrey, et~al.}]{sorlie2001gene}
S{\o}rlie, T., Perou, C.~M., Tibshirani, R., Aas, T., Geisler, S., Johnsen, H.,
  Hastie, T., Eisen, M.~B., Van De~Rijn, M., Jeffrey, S.~S., et~al. (2001),
  \enquote{Gene expression patterns of breast carcinomas distinguish tumor
  subclasses with clinical implications,} \textit{Proceedings of the National
  Academy of Sciences}, 98, 10869--10874.

\bibitem[{Sun et~al.(2010)Sun, Tang, Han, Gupta, and Zhao}]{sun2010community}
Sun, Y., Tang, J., Han, J., Gupta, M., and Zhao, B. (2010), \enquote{Community
  evolution detection in dynamic heterogeneous information networks,} in
  \textit{Proceedings of the Eighth Workshop on Mining and Learning with
  Graphs}, pp. 137--146.

\bibitem[{Tey et~al.(2021)Tey, Wu, Lin, and Chen}]{tey2021accuracy}
Tey, F.~J., Wu, T.-Y., Lin, C.-L., and Chen, J.-L. (2021), \enquote{Accuracy
  improvements for cold-start recommendation problem using indirect relations
  in social networks,} \textit{Journal of Big Data}, 8, 1--18.

\bibitem[{Wakita and Tsurumi(2007)}]{wakita2007finding}
Wakita, K. and Tsurumi, T. (2007), \enquote{Finding community structure in
  mega-scale social networks,} in \textit{Proceedings of the 16th International
  Conference on World Wide Web}, pp. 1275--1276.

\bibitem[{Wang et~al.(2022)Wang, Lu, Shi, Wang, Cui, and Mou}]{wang2020dynamic}
Wang, X., Lu, Y., Shi, C., Wang, R., Cui, P., and Mou, S. (2022),
  \enquote{Dynamic heterogeneous information network embedding with meta-path
  based proximity,} \textit{IEEE Transactions on Knowledge and Data
  Engineering}, 34, 1117 -- 1132.

\bibitem[{Xue et~al.(2020)Xue, Yang, Jiang, Wei, Hu, and Lin}]{xue2020modeling}
Xue, H., Yang, L., Jiang, W., Wei, Y., Hu, Y., and Lin, Y. (2020),
  \enquote{Modeling dynamic heterogeneous network for link prediction using
  hierarchical attention with temporal RNN,} \textit{arXiv preprint
  arXiv:2004.01024}.

\bibitem[{Yin et~al.(2019)Yin, Ji, Zhang, and Pei}]{yin2019dhne}
Yin, Y., Ji, L.-X., Zhang, J.-P., and Pei, Y.-L. (2019), \enquote{DHNE: Network
  representation learning method for dynamic heterogeneous networks,}
  \textit{IEEE Access}, 7, 134782--134792.

\bibitem[{Zhang and Cao(2017)}]{zhang2017finding}
Zhang, J. and Cao, J. (2017), \enquote{Finding common modules in a time-varying
  network with application to the Drosophila Melanogaster gene regulation
  network,} \textit{Journal of the American Statistical Association}, 112,
  994--1008.

\bibitem[{Zhang and Chen(2017)}]{zhang2017hypothesis}
Zhang, J. and Chen, Y. (2017), \enquote{A hypothesis testing framework for
  modularity based network community detection,} \textit{Statistica Sinica},
  27, 437--456.

\bibitem[{Z{}hang and Chen(2020)}]{zhang2018modularity}
Z{}hang, J. and Chen, Y. (2020), \enquote{Modularity based community detection
  in heterogeneous networks,} \textit{Statistica Sinica}, 30, 601--629.

\bibitem[{Zhang et~al.(2020)Zhang, Sun, and Li}]{zhang2020mixed}
Zhang, J., Sun, W.~W., and Li, L. (2020), \enquote{Mixed-effect time-varying
  network model and application in brain connectivity analysis,}
  \textit{Journal of the American Statistical Association}, 115, 2022--2036.

\bibitem[{Zhang et~al.(2022)Zhang, Huang, and Tan}]{zhang2022multi}
Zhang, Z., Huang, J., and Tan, Q. (2022), \enquote{Multi-view Dynamic
  heterogeneous information network embedding,} \textit{The Computer Journal},
  65, 2016--2033.

\end{thebibliography}

\begin{center}
{\Large\bf Supplementary Materials for ``Fast Community Detection in Dynamic and Heterogeneous Networks"} 

\bigskip
{\bf Maoyu Zhang, Jingfei Zhang and Wenlin Dai}
\end{center}

\baselineskip=26.5pt

\setcounter{table}{0}
\renewcommand{\thetable}{S\arabic{table}}%
\setcounter{figure}{0}
\renewcommand{\thefigure}{S\arabic{figure}}%
\setcounter{section}{0}
\renewcommand{\thesection}{S\arabic{section}}%
\setcounter{thm}{0}
\renewcommand{\thethm}{S\arabic{thm}}%
\setcounter{lemma}{0}
\renewcommand{\thelemma}{S\arabic{lemma}}%
\setcounter{equation}{0}
\renewcommand{\theequation}{S\arabic{equation}}%

This supplementary material gives the proof of Theorem 1 in Section \ref{proof}, additional simulation results in Section \ref{sim} and additional real data analysis results in Section \ref{real}.

\section{Proof of Theorem 1}\label{proof}

First, we formalize the notations that will be used in the proof. Consider a dynamic heterogeneous network $\mG\left(\bigcup_{i=1}^{L} V^{[i]}, \mathcal{E}(t) \cup \mathcal{E}^{+}(t)\right)$, let $G^{[l]}(t)$ denote the homogeneous network formed within node set $V^{[l]}$ with an $n_l\times n_l$ adjacency matrix $A^{[l]}(t)$ and $G^{[l_1l_2]}(t)=(V^{[l_1]}\cup V^{[l_2]}, E^{[l_1l_2]}(t))$ denote the bi-partite network formed between node sets $V^{[l_1]}$ and $V^{[l_2]}$ with an $n_{l_1}\times n_{l_2}$ bi-adjacency matrix $A^{[l_1l_2]}(t)$ at time $t$, $l_1,l_2\in[L]$. Write the number of edges in $A^{[l]}(t)$ and $A^{[l_1l_2]}(t)$ as $m^{[l]}(t)=\sum_{i,j}A_{ij}^{l}(t)/2$ and  $m^{[l_1l_2]}(t)=\sum_{i,j}A_{ij}^{l_1l_2}(t)$, respectively. 
For a dynamic heterogeneous network $\{\mG(t_s), s\in \mathcal{S}\}$
from the DHSBM model, each $A_{i j}^{[l]}(t_s)$ and $A_{i j}^{\left[l_1 l_2\right]}(t_s)$ are independent Bernoulli random variables with
$$
E\left(A_{i j}^{[l]}(t_s) \mid c_i^{[l]}=k, c_j^{[l]}=h\right)=\theta_{kh}^{[l]}(t_s),\text{ and } E\left(A_{i j}^{\left[l_1 l_2\right]}(t_s) \mid c_i^{\left[l_1\right]}=k, c_j^{\left[l_2\right]}=h\right)=\theta_{k h}^{\left[l_1 l_2\right]}(t_s).
$$

For a community assignment label $\boldsymbol{e}=\left(\mathbf{e}^{[1]}, \ldots, \mathbf{e}^{[L]}\right)$ with $\mathbf{e}^{[l]}=\left(e_{1}^{[l]}, \ldots, e_{n_l}^{[l]}\right), l\in[L]$, define $K \times K$ matrices $O^{[l],s}, l\in[L]$, and $O^{\left[l_{1} l_{2}\right],s}$ $1 \leq l_{1} \neq l_{2} \leq L$, such that
$$
\begin{gathered}
	O_{k h}^{[l],s}(\boldsymbol{e})=\sum_{i j} A_{i j}^{[l]}(t_s) I\left(e_{i}^{[l]}=k, e_{j}^{[l]}=h\right), \\
	O_{k h}^{\left[l_{1} l_{2}\right],s}(\boldsymbol{e})=\sum_{i j} A_{i j}^{\left[l_{1} l_{2}\right]}(t_s) I\left(e_{i}^{\left[l_{1}\right]}=k, e_{j}^{\left[l_{2}\right]}=h\right).
\end{gathered}
$$
Define $O_{k}^{[l],s}=\sum_{h} O_{k h}^{[l],s}$ and $O_{k}^{\left[l_{1} l_{2}\right],s}=\sum_{h} O_{k h}^{\left[l_{l} l_{2}\right],s}, l\in[L], 1 \leq l_{1} \neq l_{2} \leq L .$ Define $K \times K$ matrices $R^{[l]}(\boldsymbol{e}), V^{[l]}(\boldsymbol{e}), l\in[L]$, such that
$$
	R_{a b}^{[l]}(\boldsymbol{e})=\frac{1}{n} \sum_{i=1}^{n_{l}} I\left(e_{i}^{[l]}=a, c_{i}^{[l]}=b\right), \text{ and }
	V_{a b}^{[l]}(\boldsymbol{e})=\frac{\sum_{i=1}^{n_{l}} I\left(e_{i}^{[l]}=a, c_{i}^{[l]}=b\right)}{\sum_{i=1}^{n_{l}} I\left(c_{i}^{[l]}=b\right)}.
$$
Write $\mathbb{O}(\boldsymbol{e})=\left(\mathcal{O}^{s}(\boldsymbol{e}), s=1, \ldots, S\right)$, where $\mathcal{O}^{s}(\boldsymbol{e})=\left\{O^{{[l]},s}, O^{{\left[l_{1} l_{2}\right]},s}, l\in[L], 1 \leq l_{1} \neq l_{2} \leq L\right\}$, and $\mathcal{R}=\left\{R^{[1]}, \ldots, R^{[L]}\right\}$.

The modularity function $Q(\boldsymbol{e}, \{\mG(t_s)\}_{s\in[S]})$ can be expressed as
$$
\frac{1}{2\bar{m}^{[l]}L^{2}}\sum_{s=1}^S\sum_{l=1}^{L} \sum_{k=1}^{K}\left(O_{k k}^{[l],s}-\frac{(O_{k}^{[l],s})^{2}}{\sum_{k h} O_{kh}^{[l],s}}\right)+\frac{1}{\bar{m}^{[l_1l_2]}L^{2}}\sum_{s=1}^S\sum_{l_{1} \neq l_{2}}^{L} \sum_{k=1}^{K}\left(O_{k k}^{{\left[l_{1} l_{2}\right]},s}-\frac{O_{k}^{{\left[l_{1} l_{2}\right]},s} O_{k}^{{\left[l_{2} l_{1}\right]},s}}{\sum_{k h} O_{k h}^{{\left[l_{1} l_{2}\right]},s}}\right),
$$
where $\bar{m}^{[l]}=\sum_{s=1}^S m^{[l]}(t_s) $ and $\bar{m}^{[l_1l_2]}=\sum_{s=1}^S m^{[l_1l_2]}(t_s) $.

Define $\mu_{n,S}=n^2S \rho_{n,S}$, we have:
$$
\begin{aligned}
	\frac{1}{\mu_{n,S}}E\left(O_{k h}^{{\left[l_{1} l_{2}\right]},s}(\boldsymbol{e}) \mid \boldsymbol{c}\right) &= \frac{1}{\mu_{n,S}}E\left(\sum_{i j} A_{i j}^{\left[l_{1} l_{2}\right]}(t_s) I\left(e_{i}^{\left[l_{1}\right]}=k, e_{j}^{\left[l_{2}\right]}=h\right) \mid \boldsymbol{c}\right) \\
	&=\frac{1}{n^2S}\sum_{i j} \sum_{a b} \theta_{a b}^{\left[l_{1} l_{2}\right]}(t_s) I\left(e_{i}^{\left[l_{1}\right]}=k, c_{i}^{\left[l_{1}\right]}=a\right) I\left(e_{j}^{\left[l_{2}\right]}=h, e_{j}^{\left[l_{2}\right]}=b\right).
\end{aligned}
$$
Define $H^{{\left[l_{1} l_{2}\right]},s}(\mathcal{R}(\boldsymbol{e}))= \frac{1}{\mu_{n,S}}E\left(O^{{\left[l_{1} l_{2}\right]},s}(\boldsymbol{e}) \mid \boldsymbol{c}\right)$, we have
$$
H^{{\left[l_{1} l_{2}\right]},s}(\mathcal{R}(\boldsymbol{e}))=\frac{1}{S}R^{\left[l_{1}\right]}(\boldsymbol{e}) \theta^{\left[l_{1} l_{2}\right]}(t_s) R^{\left[l_{2}\right]}(\boldsymbol{e})^{\prime}, \quad 1 \leq l_{1} \neq l_{2} \leq L.
$$
Similarly, we can define $H^{{[l]},s}(\mathcal{R}(\boldsymbol{e}))= \frac{1}{\mu_{n,S}}E\left(O^{[l],s}(\boldsymbol{e}) \mid \boldsymbol{c}\right)$ and write
$$
H^{{[l]},s}(\mathcal{R}(\boldsymbol{e}))=\frac{1}{S}R^{[l]}(\boldsymbol{e}) \theta^{[l]}(t_s) R^{[l]}(\boldsymbol{e})^{\prime}, \quad l\in[L], s\in[S].
$$
Write $\mathcal{H}=\left\{H^{{[l]},1}, H^{{\left[l_{1} l_{2}\right]},S},\dots,H^{{[l]},1}, H^{{\left[l_{1} l_{2}\right]},S} l\in[L], 1 \leq l_{1} \neq l_{2} \leq L\right\} $.

Consider a community label $\boldsymbol{e}=\left(\mathbf{e}^{[1]}, \ldots, \mathbf{e}^{[L]}\right)$ with $\mathbf{e}^{[l]}=\left(e_{1}^{[l]}, \ldots, e_{n_l}^{[l]}\right), l\in[L]$.
Further, define
 $$
 \mathcal{J}(\mathbb{O}(\boldsymbol{e}))=\frac{1}{S} \sum_{s=1}^{S} J\left(\mathcal{O}^{s}(\boldsymbol{e})\right),
 $$
 where
 $$
J\left(\mathcal{O}^{s}(\boldsymbol{e})\right)=\frac{1}{L^2}\left[\sum_{l=1}^{L} \sum_{k=1}^{K}\left(O_{k k}^{[l],s}-\frac{(O_{k}^{[l],s})^{2}}{\sum_{k h} O_{k h}^{[l],s}}\right)+\sum_{l_{1} \neq l_{2}}^{L} \sum_{k=1}^{K}\left(O_{k k}^{{\left[l_{1} l_{2}\right]},s}-\frac{O_{k}^{{\left[l_{1} l_{2}\right]},s} O_{k}^{{\left[l_{2} l_{1}\right]},s}}{\sum_{k h} O_{k h}^{{\left[l_{1} l_{2}\right]},s}}\right)\right].
 $$
Here we suppress the argument $\boldsymbol{e}$ for brevity. Then for convenient, we write
$$
J(\mathcal{O}^s)=\sum_{l=1}^{L} J_{1}\left(O^{{[l]},s}\right)+\sum_{l_{1} \neq l_{2}}^{L} J_{2}\left(O^{{\left[l_{1} l_{2}\right]},s}, O^{{\left[l_{2} l_{1}\right]},s}\right),
$$
where
$$
J_{1}\left(O^{{[l]},s}\right)=\sum_{k=1}^{K}\left(O_{k k}^{{[l]},s}-\frac{(O_{k}^{{[l]},s})^{2}}{\sum_{k h} O_{k h}^{[l],s}}\right),
$$
and
$$
J_{2}\left(O^{\left[l_{1} l_{2}\right],s}, O^{\left[l_{2} l_{1}\right],s}\right)=\sum_{k=1}^{K}\left(O_{k k}^{\left[l_{1} l_{2}\right],s}-\frac{O_{k}^{\left[l_{1} l_{2}\right],s} O_{k}^{\left[l_{2} l_{1}\right],s}}{\sum_{k h} O_{k h}^{\left[l_{1} l_{2}\right],s}}\right).
$$
 Showing the $\hat{\boldsymbol{e}}$ that maximizes the $Q(\boldsymbol{e}, \{\mG(t_s)\}_{s\in[S]})$ is consistent is equivalent to showing the $\hat{\boldsymbol{e}}$ that maximizes the $\mathcal{J}(\mathbb{O}(\boldsymbol{e}))$ is consistent. We show consistency by showing that there exists $\delta_{n,S} \rightarrow 0$, such that
 $$P\left(\max_{\boldsymbol{e}: \eta(\boldsymbol{e}, \boldsymbol{c}) \geq \delta_{n,S}} \mathcal{J}\left(\frac{\mathbb{O}(\boldsymbol{e})}{\mu_{n,S}}\right) < \mathcal{J}\left(\frac{\mathbb{O}(\boldsymbol{c})}{\mu_{n,S}}\right)\right) \rightarrow 1 \text{ as } nS \rightarrow \infty,$$
 where $\eta(\boldsymbol{e}, \boldsymbol{c})=\sum_{l=1}^{L} \sum_{a b}\left|V_{a b}^{[l]}(\boldsymbol{e})-V_{a b}^{[l]}(\boldsymbol{c})\right|$.

 Since $\mathcal{J}(.)$ is Lipschitz in all its arguments, we have
$$
\begin{aligned}
\left|\mathcal{J}\left(\frac{\mathbb{O}(\boldsymbol{e})}{\mu_{n,S}}\right)-\mathcal{J}(\mathcal{H}(\mathcal{R}))\right| &\leq M_1\max _{l}\left\|\sum_{s=1}^S\frac{O^{[l],s}(\mathcal{E})}{\mu_{n,S}}-\sum_{s=1}^SH^{[l],s}(\mathcal{R})\right\|_{\infty}\\
&+M_1\max _{l_1 \neq l_2}\left\|\sum_{s=1}^S\frac{O^{\left[l_1 l_2\right],s}(\mathcal{E})}{\mu_{n,S}}-\sum_{s=1}^SH^{\left[l_1 l_2\right],s}(\mathcal{R})\right\|_{\infty} .
\end{aligned}
$$
Here $\|X\|_{\infty}=\max _{k h}\left|X_{k h}\right|$. To continue with the proof, we need to use the Bernstein's inequality, Lemma A.1 of \cite{zhao2012consistency}.\\
\textbf{Bernstein's inequality} \textit{Let $X_1,\dots,X_n$ be independent variables. Suppose that $|X_i| \leq M$ for all $i$. Then, for all positive $t$}
$$P\left(\left|\sum_{i=1}^n X_i-\sum_{i=1}^n E\left(X_i\right)\right|>t\right) \leq 2 \exp \left(-\frac{t^2 / 2}{\sum \operatorname{var}\left(X_i\right)+M t / 3}\right).$$


Define $\tau=\max _{i j,s} \operatorname{var}\left(A_{i j}^{[l]}(t_s)\right)$. For any $\epsilon<3 \tau$, if we write $\omega=\epsilon n^2S \rho_{n,S}$, we have
$$
\begin{aligned}
P\left(\left|\sum_{s=1}^S\frac{O_{k h}^{[l],s}(\mathcal{E})}{\mu_{n,S}}-\sum_{s=1}^SH_{k h}^{[l],s}(\mathcal{R})\right|>\epsilon\right) & \leq 2 \exp \left(-\frac{\omega^2 / 2}{\operatorname{var}\left(\sum_{s=1}^SO_{k h}^{[l],s}(\mathcal{E})\right)+2 \omega / 3}\right) \\
& \leq 2 \exp \left(-\frac{\epsilon^2 n^4S^2 \rho_n^2}{8 n^2S \rho_n \tau}\right) \\
&=2 \exp \left(-\frac{\epsilon^2 \mu_{n,S}}{8 \tau}\right)
\end{aligned}
$$
Notice that $\operatorname{var}\left(\sum_{s=1}^SO_{k h}^{[l],s}(\mathcal{E})\right) \leq 2 n^2S \max _{i j} \operatorname{var}\left(A_{i j}^{[l]}(t_s)\right)$.

The left hand side of the inequality converges to $0$ in probability uniformly over $\boldsymbol{e}$ as $nS\rho_{n,S} \rightarrow \infty$. Following similar arguments, we can show that
$$
P\left(\left|\sum_{s=1}^S\frac{O_{k h}^{\left[l_1 l_2\right],s}(\mathcal{E})}{\mu_{n,S}}-\sum_{s=1}^SH_{k h}^{\left[l_1 l_2\right],s}(\mathcal{R})\right|>\epsilon\right) \rightarrow 0 \text { as } nS \rightarrow \infty .
$$

Therefore $\mathcal{J}\left(\frac{\mathbb{O}(\boldsymbol{e})}{\mu_{n,S}}\right)$ is uniformly close to $\mathcal{J}(\mathcal{H}(\mathcal{R}(\boldsymbol{e})))$, i.e., there exists $\epsilon_{n,S} \rightarrow 0$ such that
\begin{equation}\label{S2}
P\left(\max _{\boldsymbol{e}}|\mathcal{J}\left(\frac{\mathbb{O}(\boldsymbol{e})}{\mu_{n,S}}\right)-\mathcal{J}(\mathcal{H}(\mathcal{R}(\boldsymbol{e})))|<\epsilon_{n,S}\right) \rightarrow 1 \text { as } nS \rightarrow \infty.
\end{equation}
To show that there exists $\delta_{n,S} \rightarrow 0$, such that
$$P\left(\max_{\boldsymbol{e}: \eta(\boldsymbol{e}, \boldsymbol{c}) \geq \delta_{n,S}} \mathcal{J}\left(\frac{\mathbb{O}(\boldsymbol{e})}{\mu_{n,S}}\right) < \mathcal{J}\left(\frac{\mathbb{O}(\boldsymbol{c})}{\mu_{n,S}}\right)\right) \rightarrow 1 \text{ as } nS \rightarrow \infty.$$

Next we show that $\mathcal{J}(\mathcal{H}(\mathcal{R}(\boldsymbol{e})))$ is uniquely maximized over $\left\{\mathcal{R}: R^{[l]} \geq 0, R^{[l]^{\prime}} \mathbf{1}=\pi^{[l]}, l=\right.$ $1, \ldots, L\}$ at $\mathcal{S}=\mathcal{R}(\boldsymbol{c})$. Since $\mathcal{J}(\mathcal{H}(\mathcal{R}))$ is the population version of $\mathcal{J}\left(\frac{\mathbb{O}(\boldsymbol{e})}{\mu_{n,S}}\right)$, if $\mathcal{J}\left(\frac{\mathbb{O}(\boldsymbol{e})}{\mu_{n,S}}\right)$ is maximized by the true community label $\boldsymbol{c}, J(\mathcal{H}(\mathcal{R}))$ should also be maximized by the true assignment $\mathcal{S}$.
Define
$$
\triangle_{k h}=\left\{\begin{array}{rll}
	1 & \text { for } & k=h \\
	-1 & \text { for } & k \neq h
\end{array}\right.
$$
Using the equalities
$$
\sum_{k}\left(H_{k k}^{[l],s}-\frac{(H_{k}^{[l],s})^{2}}{\sum_{k h} H_{k h}^{[l],s}}\right)+\sum_{k \neq h}\left(H_{k h}^{[l],s}-\frac{H_{k}^{[l],s} H_{h}^{[l],s}}{\sum_{k h} H_{k h}^{[l],s}}\right)=0, \quad l\in[L], s=1, \dots ,S,
$$
and
$$
\sum_{k}\left(H_{k k}^{\left[l_{1} l_{2}\right],s}-\frac{H_{k}^{\left[l_{1} l_{2}\right],s} H_{k}^{\left[l_{2} l_{1}\right],s}}{\sum_{k h} H_{k h}^{\left[l_{1} l_{2}\right],s}}\right)+\sum_{k \neq h}\left(H_{k h}^{\left[l_{1} l_{2}\right],s}-\frac{H_{k}^{\left[l_{1} l_{2}\right],s} H_{h}^{\left[l_{2} l_{1}\right],s}}{\sum_{k h} H_{k h}^{\left[l_{1} l_{2}\right],s}}\right)=0, \quad 1 \leq l_{1} \neq l_{2} \leq L,s=1,\dots,S.
$$
We have
$$
\begin{aligned}
 \mathcal{J}(\mathcal{H}(\mathcal{R}))=&\sum_{s=1}^S\sum_{l=1}^{L} J_{1}\left(H^{[l],s}(\mathcal{R})\right)+\sum_{s=1}^S\sum_{l_{1} \neq l_{2}}^{L} J_{2}\left(H^{\left[l_{1} l_{2}\right],s}(\mathcal{R}), H^{\left[l_{2} l_{1}\right],s}(\mathcal{R})\right) \\
	=& \frac{1}{2}\sum_{s=1}^S \sum_{l=1}^{L} \sum_{k h} \triangle_{k h}\left(H_{k h}^{[l],s}(\mathcal{R})-\frac{H_{k}^{[l],s}(\mathcal{R}) H_{h}^{[l],s}(\mathcal{R})}{\sum_{k h} H_{k h}^{[l],s}(\mathcal{R})}\right)+\\
	\end{aligned}
	$$
	$$
\begin{aligned}
	&\frac{1}{2}\sum_{s=1}^S \sum_{l_{1} \neq l_{2}}^{L} \sum_{k h} \triangle_{k h}\left(H_{k h}^{\left[l_{1} l_{2}\right],s}(\mathcal{R})-\frac{H_{k}^{\left[l_{1} l_{2}\right],s}(\mathcal{R}) H_{h}^{\left[l_{2} l_{1}\right],s}(\mathcal{R})}{\sum_{k h} H_{k h}^{\left[l_{1} l_{2}\right],s}(\mathcal{R})}\right)\\
	=& \frac{1}{2S} \sum_{s=1}^S\sum_{l=1}^{L} \sum_{k h} \triangle_{k h}\left(\sum_{a b} \theta_{a b}^{[l]}(t_s) R_{k a}^{[l]}(\boldsymbol{e}) R_{h b}^{[l]}(\boldsymbol{e})-\frac{\left(\sum_{a q} \theta_{a q}^{[l]}(t_s) R_{k a}^{[l]}(\boldsymbol{e}) \pi_{q}^{[l]}\right)\left(\sum_{b r} \theta_{b r}^{[l]}(t_s) R_{h b}^{[l]}(\boldsymbol{e}) \pi_{r}^{[l]}\right)}{\sum_{k h} H_{k h}^{[l],s}(\mathcal{R})}\right) \\
	+&\frac{1}{2S} \sum_{s=1}^S\sum_{l_{1} \neq l_{2}}^{L} \sum_{k h} \triangle_{k h}\left(\sum_{a b} \theta_{a b}^{\left[l_{1} l_{2}\right]}(t_s) R_{k a}^{\left[l_{l}\right]}(\boldsymbol{e}) R_{h b}^{\left[l_{2}\right]}(\boldsymbol{e})\right.\\
	&\hspace{2.5in}-\left.\frac{\left(\sum_{a q} \theta_{a q}^{\left[l_{1} l_{2}\right]}(t_s) R_{k a}^{\left[l_{1}\right]}(\boldsymbol{e}) \pi_{q}^{\left[l_{2}\right]}\right)\left(\sum_{b r} \theta_{b r}^{\left[l_{2} l_{1}\right]}(t_s) R_{h b}^{\left[l_{2}\right]}(\boldsymbol{e}) \pi_{r}^{\left[l_{1}\right]}\right)}{\sum_{k h} H_{k h}^{\left[l_{1} l_{2}\right],s}(\mathcal{R})}\right)\\
	=& \frac{1}{2S} \sum_{s=1}^S \sum_{l=1}^{L} \sum_{k h} \sum_{a b} \triangle_{k h} R_{k a}^{[l]}(\boldsymbol{e}) R_{h b}^{[l]}(\boldsymbol{e})\left(\theta_{a b}^{[l]}(t_s)-\frac{\left(\sum_{q} \theta_{a q}^{[l]} (t_s)\pi_{q}^{[l]}\right)\left(\sum_{r} \theta_{b r}^{[l]}(t_s) \pi_{r}^{[l]}\right)}{\sum_{k h} H_{k h}^{[l]}(\mathcal{R})}\right) \\
	+&\frac{1}{2S} \sum_{s=1}^S \sum_{l_{1} \neq l_{2}}^{L} \sum_{k h} \sum_{a b} \triangle_{k h} R_{k a}^{\left[l_{1}\right]}(\boldsymbol{e}) R_{h b}^{\left[l_{2}\right]}(\boldsymbol{e})\left(\theta_{a b}^{\left[l_{l} l_{2}\right]}(t_s)-\frac{\left(\sum_{q} \theta_{a q}^{\left[l_{1} l_{2}\right]}(t_s) \pi_{q}^{\left[l_{2}\right]}\right)\left(\sum_{r} \theta_{b r}^{\left[l_{2} l_{1}\right]}(t_s)\pi_{r}^{\left[l_{1}\right]}\right)}{\sum_{k h} H_{k h}^{\left[l_{l} l_{2}\right],s}(\mathcal{R})}\right)\\
		\end{aligned}$$
	$$\begin{aligned}
 \leq & \frac{1}{2S} \sum_{s=1}^S \sum_{l=1}^{L} \sum_{k h} \sum_{a b} \triangle_{a b} R_{k a}^{[l]}(\boldsymbol{e}) R_{h b}^{[l]}(\boldsymbol{e})\left(\theta_{a b}^{[l]}(t_s)-\frac{\left(\sum_{q} \theta_{a q}^{[l]} (t_s)\pi_{q}^{[l]}\right)\left(\sum_{r} \theta_{b r}^{[l]}(t_s) \pi_{r}^{[l]}\right)}{\sum_{k h} H_{k h}^{[l]}(\mathcal{R})}\right) \\
 +&\frac{1}{2S} \sum_{s=1}^S \sum_{l_{1} \neq l_{2}}^{L} \sum_{k h} \sum_{a b} \triangle_{ab} R_{k a}^{\left[l_{1}\right]}(\boldsymbol{e}) R_{h b}^{\left[l_{2}\right]}(\boldsymbol{e})\left(\theta_{a b}^{\left[l_{l} l_{2}\right]}(t_s)-\frac{\left(\sum_{q} \theta_{a q}^{\left[l_{1} l_{2}\right]}(t_s) \pi_{q}^{\left[l_{2}\right]}\right)\left(\sum_{r} \theta_{b r}^{\left[l_{2} l_{1}\right]}(t_s)\pi_{r}^{\left[l_{1}\right]}\right)}{\sum_{k h} H_{k h}^{\left[l_{l} l_{2}\right],s}(\mathcal{R})}\right)\\
 =& \frac{1}{2S} \sum_{s=1}^S\sum_{l=1}^{L} \sum_{a b} \triangle_{a b} \pi_{a}^{[l]} \pi_{b}^{[l]}\left(\theta_{a b}^{[l]}(t_s)-\frac{\left(\sum_{q} \theta_{a q}^{[l]}(t_s) \pi_{q}^{[l]}\right)\left(\sum_{r} \theta_{b r}^{[l]}(t_s) \pi_{r}^{[l]}\right)}{\sum_{k h} H_{k h}^{[l],s}(\mathcal{S})}\right) \\ 
 +&\frac{1}{2S}\sum_{s=1}^S \sum_{l_{1} \neq l_{2}}^{L} \sum_{a b} \triangle_{a b} \pi_{a}^{\left[l_{1}\right]} \pi_{b}^{\left[l_{2}\right]}\left(\theta_{a b}^{\left[l_{1} l_{2}\right]}(t_s)-\frac{\left(\sum_{q} \theta_{a q}^{\left[l_{2} l_{1}\right]}(t_s) \pi_{q}^{\left[l_{2}\right]}\right)\left(\sum_{r} \theta_{b r}^{\left[l_{1} l_{2}\right]}(t_s) \pi_{r}^{\left[l_{1}\right]}\right)}{\sum_{k h} H_{k h}^{\left[l_{1} l_{2}\right],s}(\mathcal{S})}\right) \\
 =& \sum_{s=1}^S\sum_{l=1}^{L} J_{1}\left(H^{[l],s}(\mathcal{S})\right)+\sum_{s=1}^S\sum_{l_{1} \neq l_{2}} J_{2}\left(H^{\left[l_{1} l_{2}\right],s}(\mathcal{S}), H^{\left[l_{2} l_{1}\right],s}(\mathcal{S})\right)=\mathcal{J}(\mathcal{H}(\mathcal{S})) 	
\end{aligned}.
$$
Here we used the conditions in Theorem 1 for the inequality, and the relationship that
$$
\sum_{k h} H_{k h}^{[l],s}(\mathcal{R})=\frac{1}{S}\sum_{k h} \sum_{a b} \theta_{a b}^{[l]}(t_s) R_{k a}^{[l]}(\boldsymbol{e}) R_{h b}^{[l]}(\boldsymbol{e})=\frac{1}{S}\sum_{a b} \theta_{a b}^{[l]}(t_s) \pi_{a}^{[l]} \pi_{b}^{[l]}=\sum_{k h} H_{k h}^{[l],s}(\mathcal{S}),
$$
and
$$
\sum_{k h} H_{k h}^{\left[l_{1} l_{2}\right],s}(\mathcal{R})=\frac{1}{S}\sum_{k h} \sum_{a b} \theta_{a b}^{\left[l_{1} l_{2}\right]}(t_s)R_{k a}^{\left[l_{1}\right]}(\boldsymbol{e}) R_{h b}^{\left[l_{2}\right]}(\boldsymbol{e})=\frac{1}{S}\sum_{a b} \theta_{a b}^{\left[l_{l} l_{2}\right]}(t_s) \pi_{a}^{\left[l_{1}\right]} \pi_{b}^{\left[l_{2}\right]}=\sum_{k h} H_{k h}^{\left[l_{1} l_{2}\right],s}(\mathcal{S}).
$$
We have shown that $\mathcal{S}$ is a maximizer of $\mathcal{J}(\mathcal{H}(\mathcal{R}))$.

Next we need to show that $\mathcal{S}$ is the unique maximizer of $\mathcal{J}(\mathcal{H}(\mathcal{R}))$. This can be shown using Lemma $3.2$ in \cite{bickel2009nonparametric}. Since the inequality $\mathcal{J}(\mathcal{H}(\mathcal{R})) \leq \mathcal{J}(\mathcal{H}(\mathcal{S}))$ holds only if $\triangle_{k h}=\triangle_{a b}$ whenever $R_{k a}^{[l]}(\boldsymbol{e}) R_{h b}^{[l]}(\boldsymbol{e})>0, l\in[L]$, and $\triangle$ does not have two identical columns, using the results in Lemma 3.2, we have $\mathcal{S}$ uniquely maximizes $J(\mathcal{H}(\mathcal{R}))$.
Now that we have shown that $\mathcal{J}(\mathcal{H}(\mathcal{R}))$ is uniquely maximized by $\mathcal{S}$. By the continuity of $\mathcal{J}(.)$ in the neighborhood of $\mathcal{S}$, there exists $\delta_{n,S} \rightarrow \infty$, such that
$$
J(\mathcal{H}(\mathcal{R}))-J(\mathcal{H}(\mathcal{S})) \geq 2 \epsilon_{n,S} \quad \text { for } \quad \eta(\boldsymbol{e}, \boldsymbol{c}) \geq \delta_{n,S}.
$$
Here we used the fact that
$$
\begin{aligned}
	\eta(\mathcal{R}(\boldsymbol{e}), \mathcal{S}) &=\sum_{l=1}^{L} \sum_{a b}\left|\pi_{b}^{[l]} V_{a b}^{[l]}(\boldsymbol{e})-\pi_{b}^{[l]} V_{a b}^{[l]}(\boldsymbol{c})\right| \\
	& \geq\left(\min _{l, b} \pi_{b}^{[l]}\right) \times \sum_{l=1}^{L} \sum_{a b}\left|V_{a b}^{[l]}(\boldsymbol{e})-V_{a b}^{[l]}(\boldsymbol{c})\right|=\left(\min _{l, b} \pi_{b}^{[l]}\right) \times \eta(\boldsymbol{e}, \boldsymbol{c}).
\end{aligned}
$$
Thus, with (\ref{S2}), we have that
$$
\begin{aligned}
	&P\left(\max _{\boldsymbol{e}: \eta(\boldsymbol{e}, \boldsymbol{c}) \geq \delta_{n,s}} \mathcal{J}\left(\frac{\mathbb{O}(\boldsymbol{e})}{\mu_{n,S}}\right) < \mathcal{J}\left(\frac{\mathbb{O}(\boldsymbol{c})}{\mu_{n,S}}\right)\right) \\
	&\geq P\left(\left|\max _{\boldsymbol{e}: \eta(\boldsymbol{e}, \boldsymbol{c}) \geq \delta_{n,S}} \mathcal{J}\left(\frac{\mathbb{O}(\boldsymbol{e})}{\mu_{n,S}}\right)-\max _{\boldsymbol{e}: \eta(\boldsymbol{e}, \boldsymbol{c}) \geq \delta_{n,S}} \mathcal{J}(\mathcal{H}(\mathcal{R}))\right|<\epsilon_{n,S},\left|\mathcal{J}\left(\frac{\mathbb{O}(\boldsymbol{c})}{\mu_{n,S}}\right)-\mathcal{J}(\mathcal{H}(\mathcal{S}))\right| \leq \epsilon_{n,S}\right) \rightarrow 1,
\end{aligned}
$$
and this implies that
$$
P\left(\eta(\hat{\boldsymbol{c}}, \boldsymbol{c}) \leq \delta_{n,S}\right) \rightarrow 1,
$$
where
$$
\hat{\boldsymbol{c}}=\arg \max _{\boldsymbol{e}} \mathcal{J}\left(\frac{\mathbb{O}(\boldsymbol{e})}{\mu_{n,S}}\right),
$$
since
$$
\begin{aligned}
	\frac{1}{n} \sum_{l=1}^{L} \sum_{i=1}^{n_{l}} I\left(\hat{c}_{i}^{[l]} \neq c_{i}^{[l]}\right)=\sum_{l=1}^{L} \sum_{k} \pi_{k}^{[l]}\left(1-V_{k k}^{[l]}(\hat{\boldsymbol{c}})\right) & \leq \sum_{l}^{L} \sum_{k}\left(1-V_{k k}^{[l]}(\hat{\boldsymbol{c}})\right) \\
	&=\frac{1}{2} \sum_{l=1}^{L}\left(\sum_{k}\left(1-V_{k k}^{[l]}(\hat{\boldsymbol{c}})\right)+\sum_{k \neq h} V_{k h}^{[l]}(\hat{\boldsymbol{c}})\right) \\
	&=\eta(\hat{\boldsymbol{c}}, \boldsymbol{c}) / 2.
\end{aligned}
$$
We have thus established the consistency property of $\hat{\boldsymbol{c}}$.

\newpage
\section{Additional simulation results}\label{sim}
In this section, we provide some additional simulation results for simulation settings 1-3,  where the network generation is the same as in the simulation section of the text, except that we have also considered the case where $G^{[1]}$ has a weak community structure while $G^{[2]}$ has no community structure, leading to $r_{1}=0.05$ in \textbf{Scenarios S1} and \textbf{S2}.\\
\textbf{Scenario S1}: 
$\theta_{1}=0.5$, $\theta_{2}=0.6$, $\theta_{3}=0.3$, $r_{1}=0.05$, $r_{2}=0$, \\
\textbf{Scenario S2}: $\theta_{1}=0.1$, $\theta_{2}=0.2$, $\theta_{3}=0.05$, $r_{1}=0.05$, $r_{2}=0$.

\begin{figure}[!h]
	\centering
	\subfigure[Type 1 nodes, Scenario S1]{
		\centering
		\includegraphics[trim=0 0 0 5mm, width=0.3\linewidth]{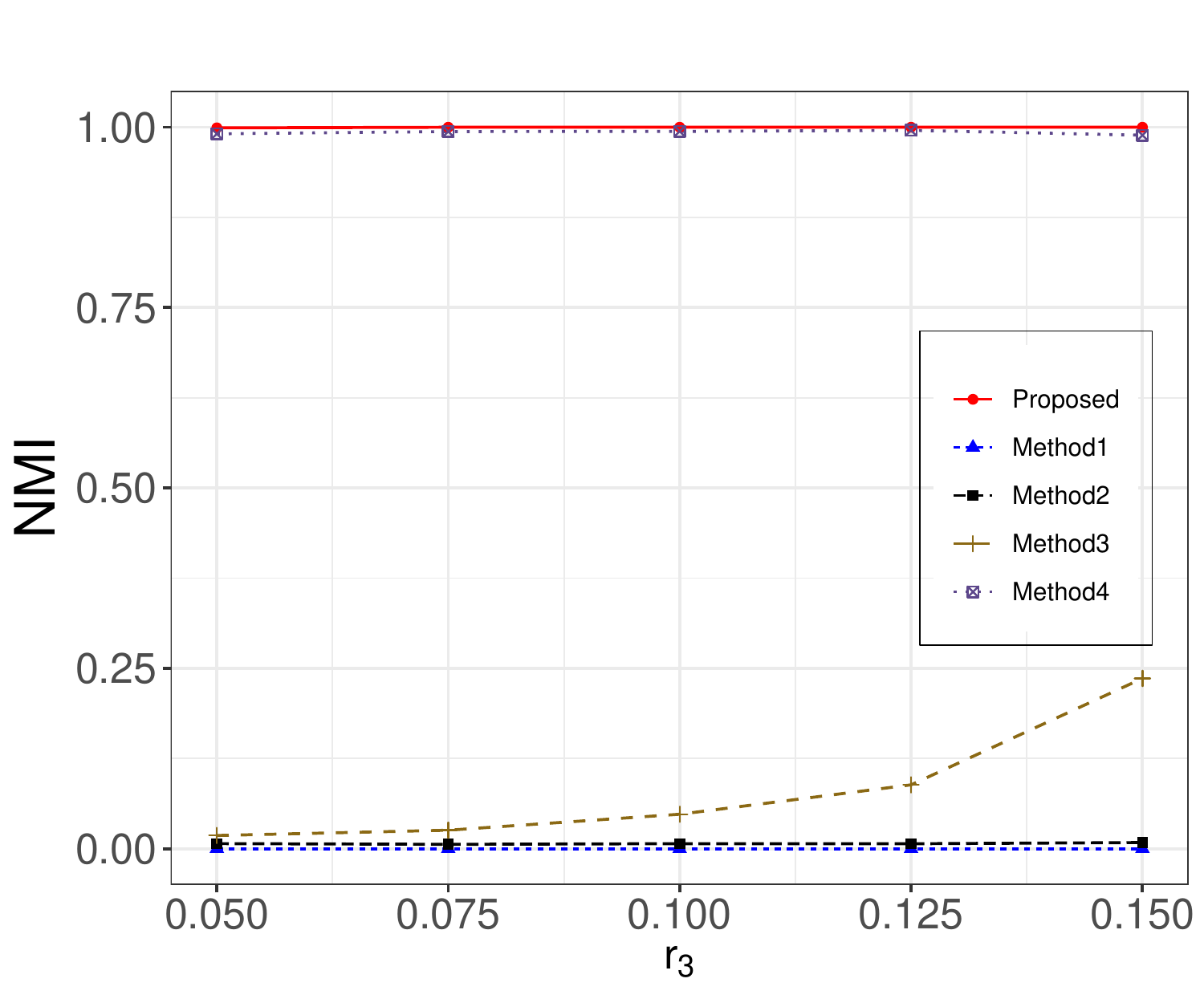}}
	\subfigure[Type 1 nodes, Scenario S2]{
		\centering
		\includegraphics[trim=0 0 0 5mm, width=0.3\linewidth]{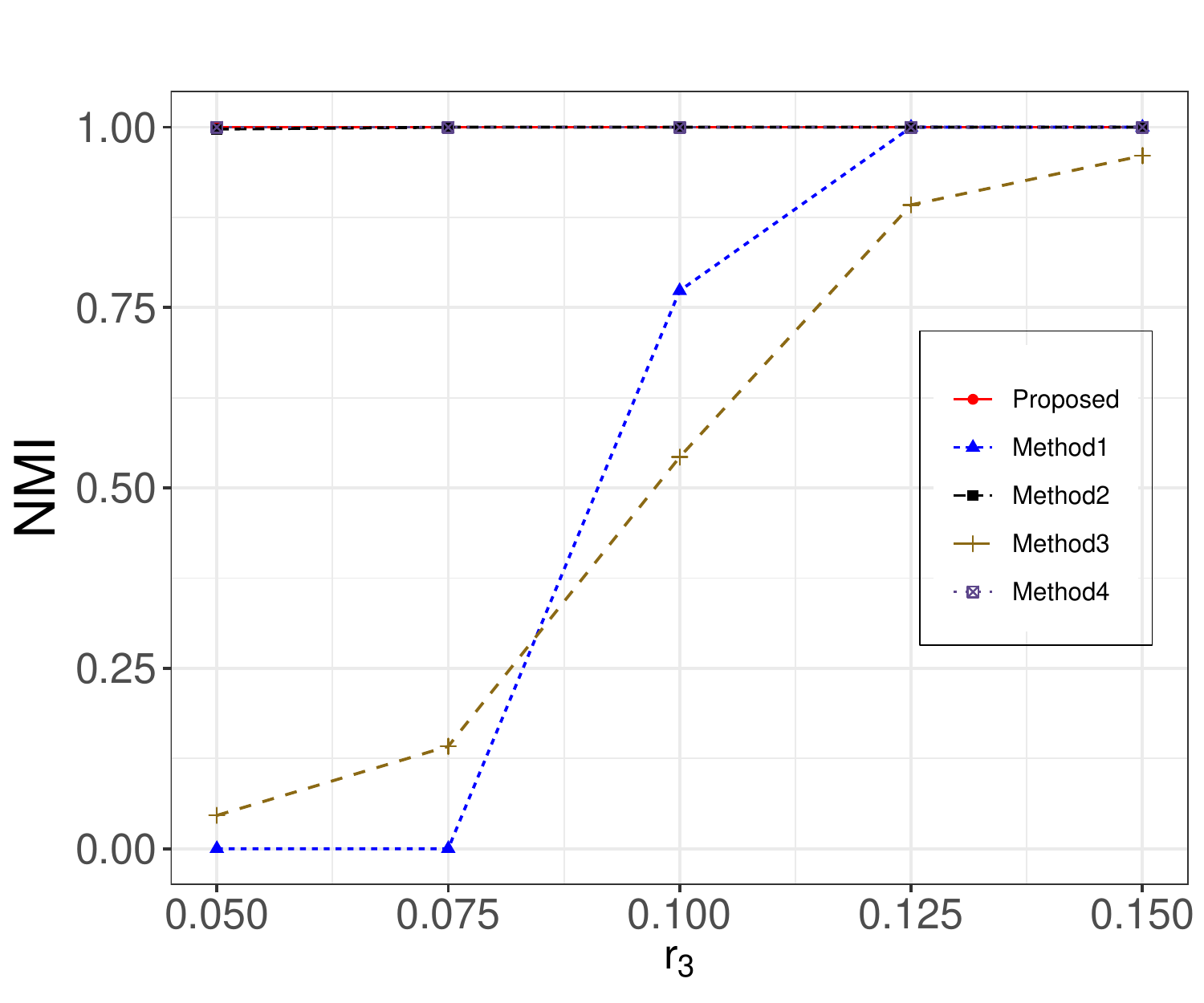}}
	\centering\\
	\subfigure[Type 2 nodes, Scenario S1]{
		\centering
		\includegraphics[trim=0 0 0 5mm, width=0.3\linewidth]{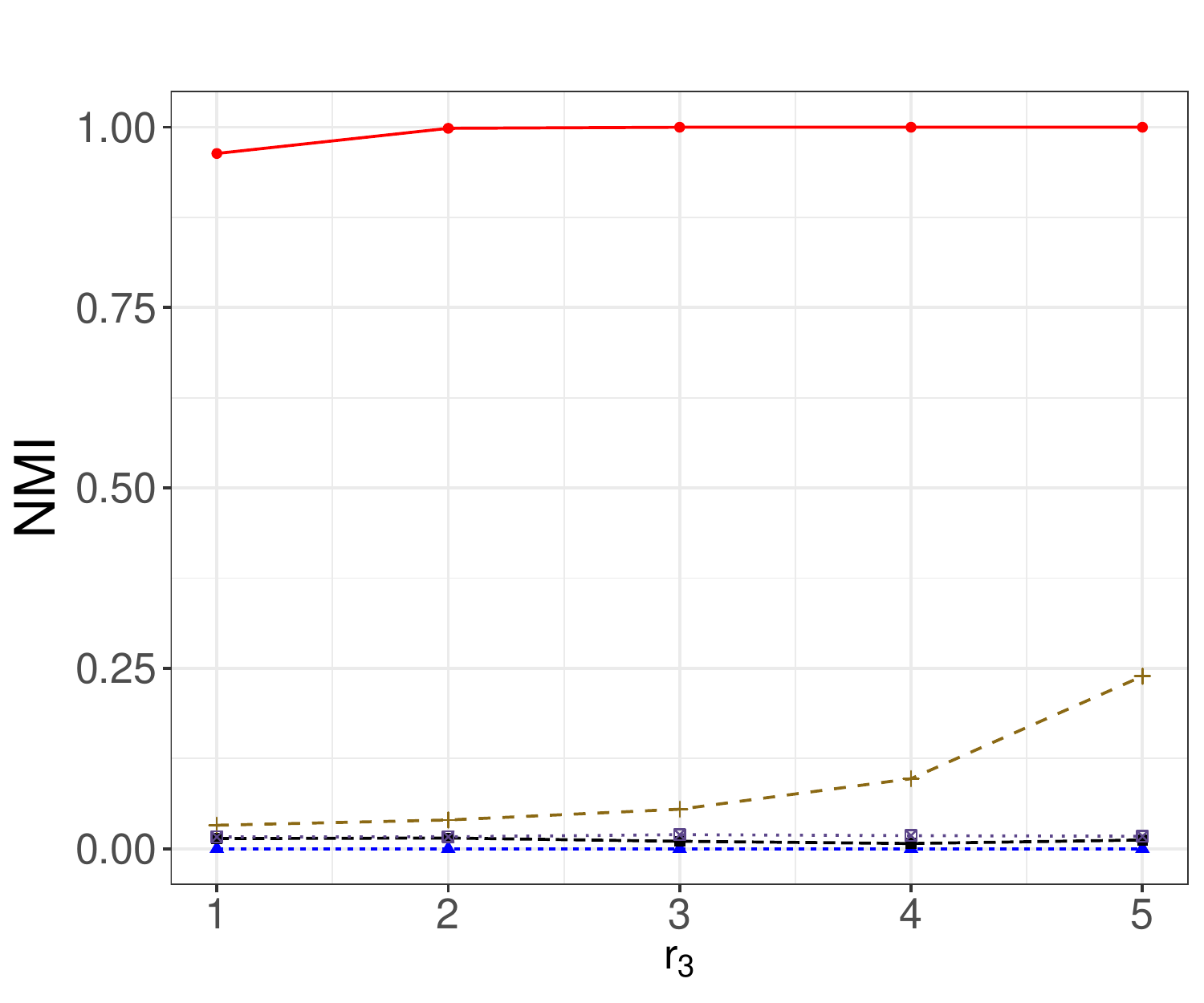}}
	\subfigure[Type 2 nodes, Scenario S2]{
		\centering
		\includegraphics[trim=0 0 0 5mm, width=0.3\linewidth]{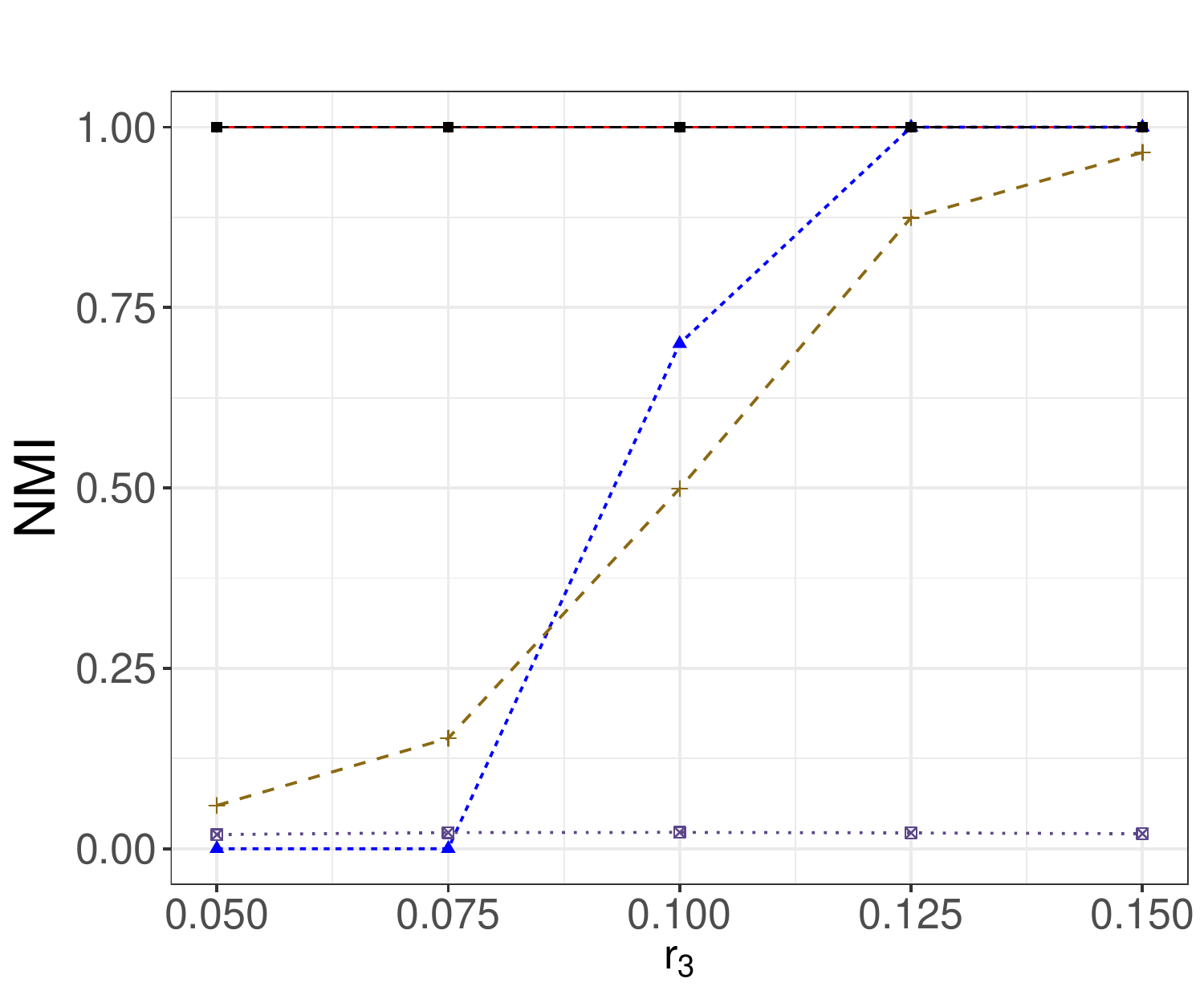}}
	\caption{Average NMIs against the value of $r_3$ for different methods in Setting 1 under Scenarios S1-S2.} \label{set1}
\end{figure}

\begin{figure}[!t]
	\centering
	\subfigure[Type 1 nodes, Scenario S1]{
		\centering
		\includegraphics[trim=0 0 0 5mm, width=0.3\linewidth]{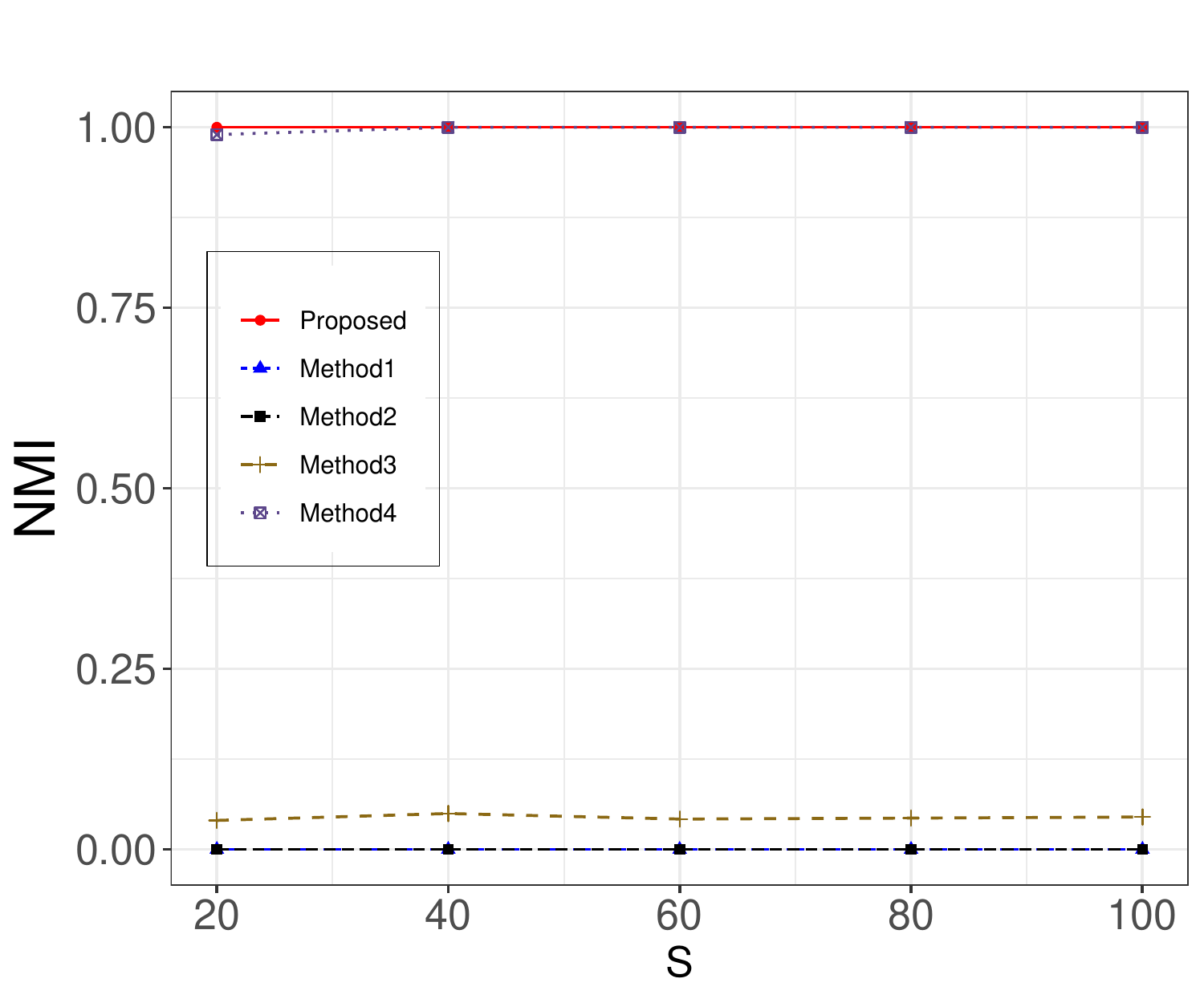}}
	\subfigure[Type 1 nodes, Scenario S2]{
		\centering
		\includegraphics[trim=0 0 0 5mm, width=0.3\linewidth]{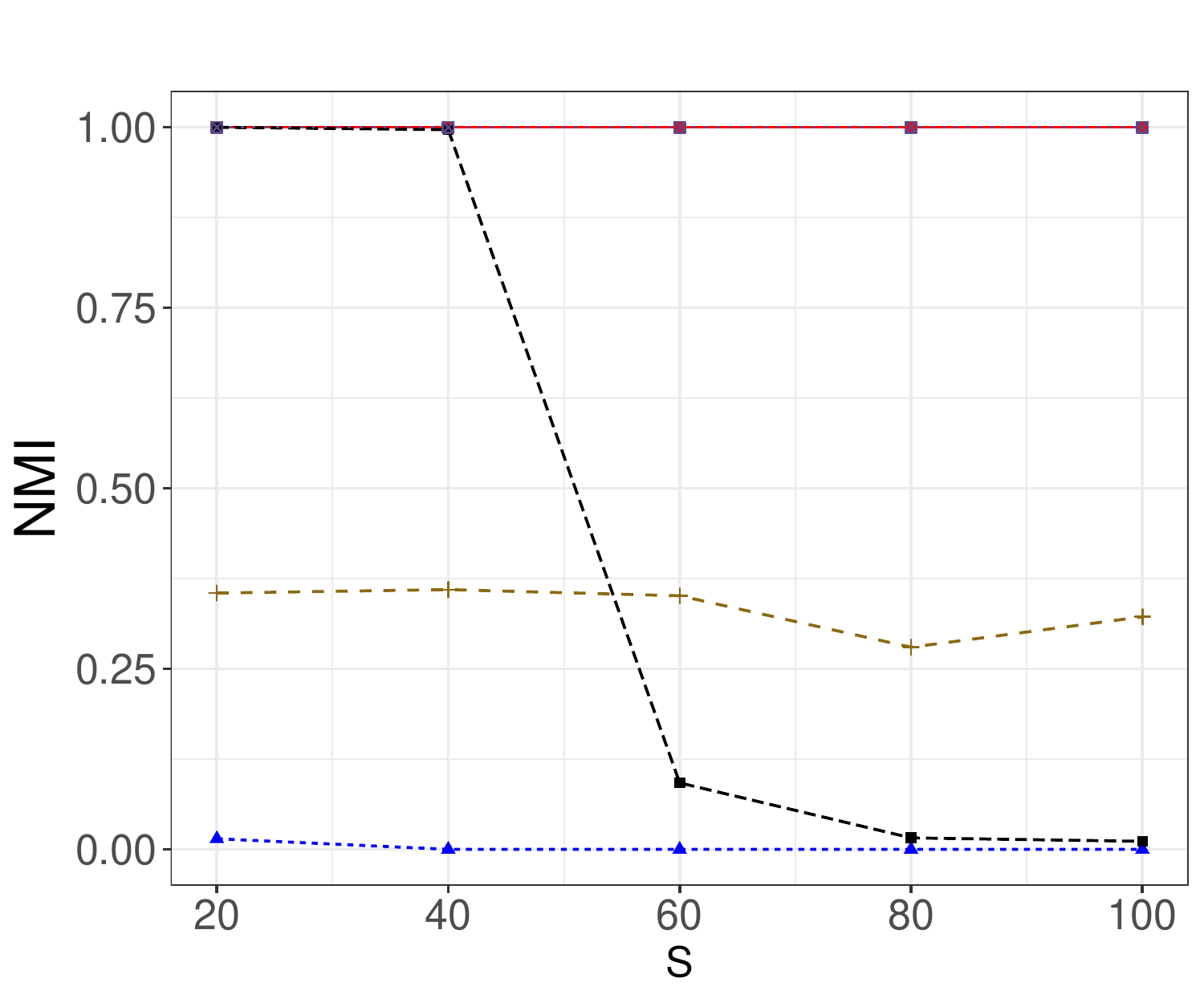}}
	\centering\\
	\subfigure[Type 2 nodes, Scenario S1]{
		\centering
		\includegraphics[trim=0 0 0 5mm, width=0.3\linewidth]{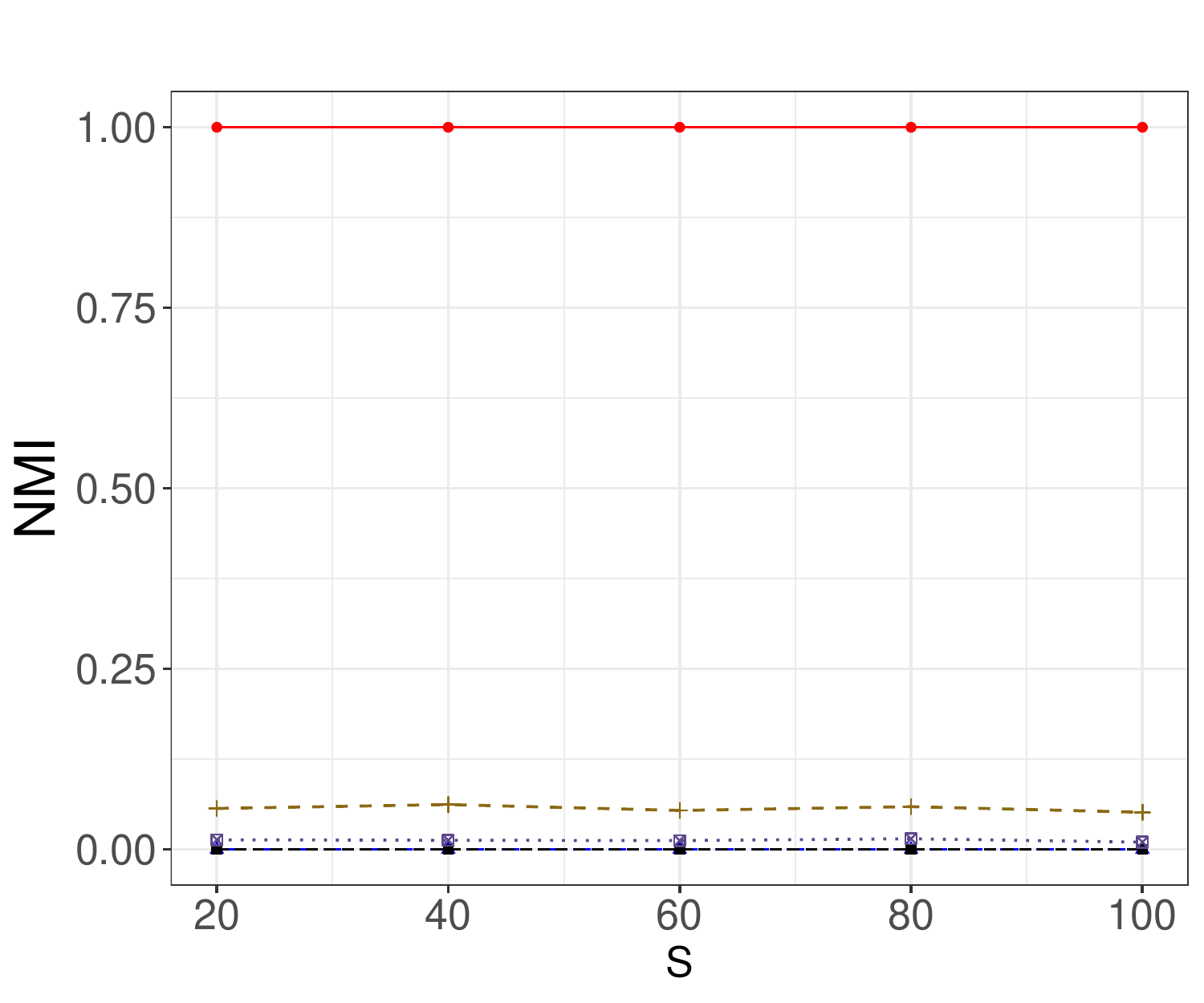}}
	\subfigure[Type 2 nodes, Scenario S2]{
		\centering
		\includegraphics[trim=0 0 0 5mm, width=0.3\linewidth]{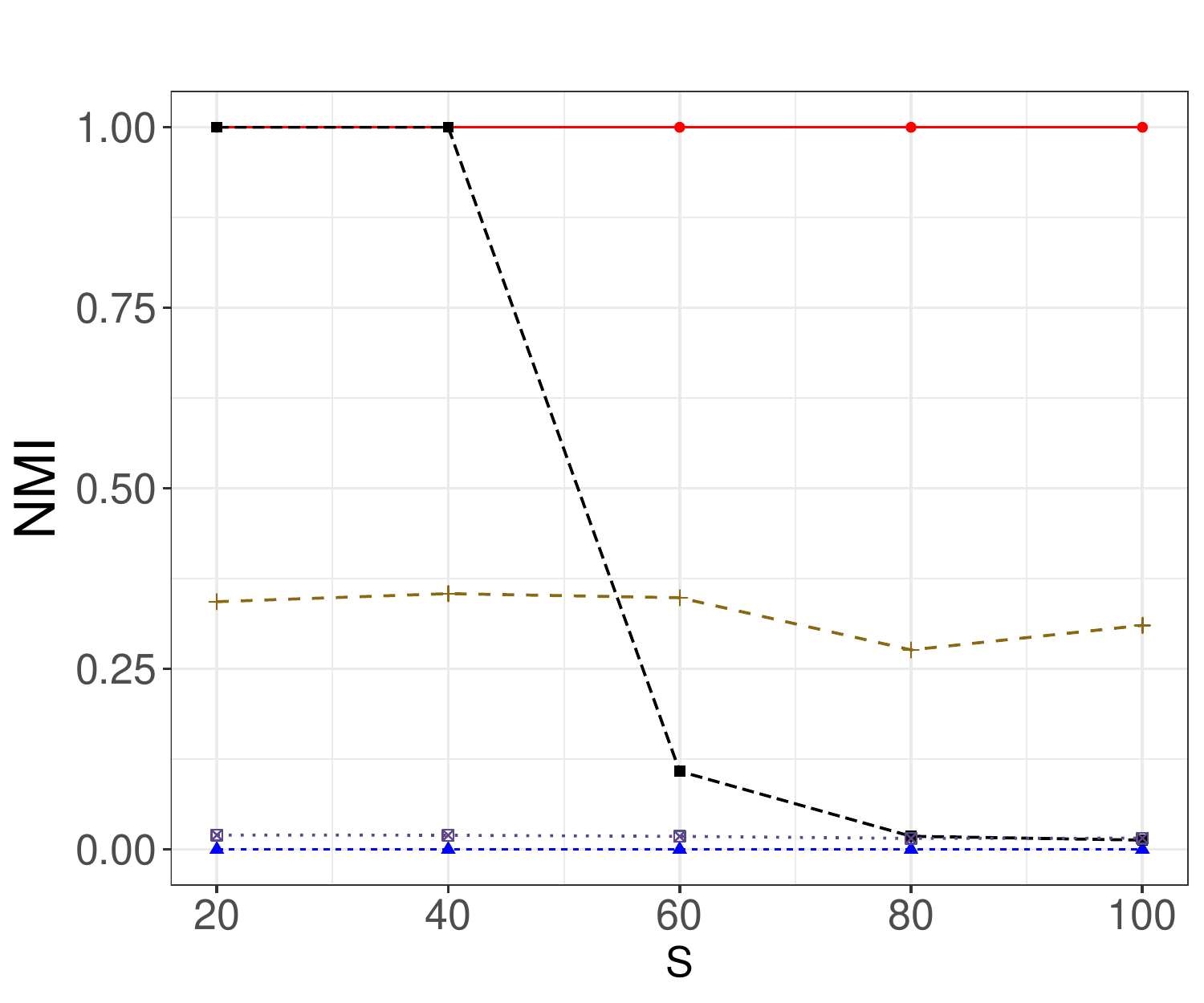}}
	\caption{Average NMIs against the value of $S$ for different methods in Setting 2 under Scenarios S1-S2.} \label{set2_dense}
\end{figure}


\begin{figure}[!t]
	\centering
	\subfigure[Type 1 nodes, Scenario 1]{
		\centering
		\includegraphics[trim=0 0 0 5mm, width=0.3\linewidth]{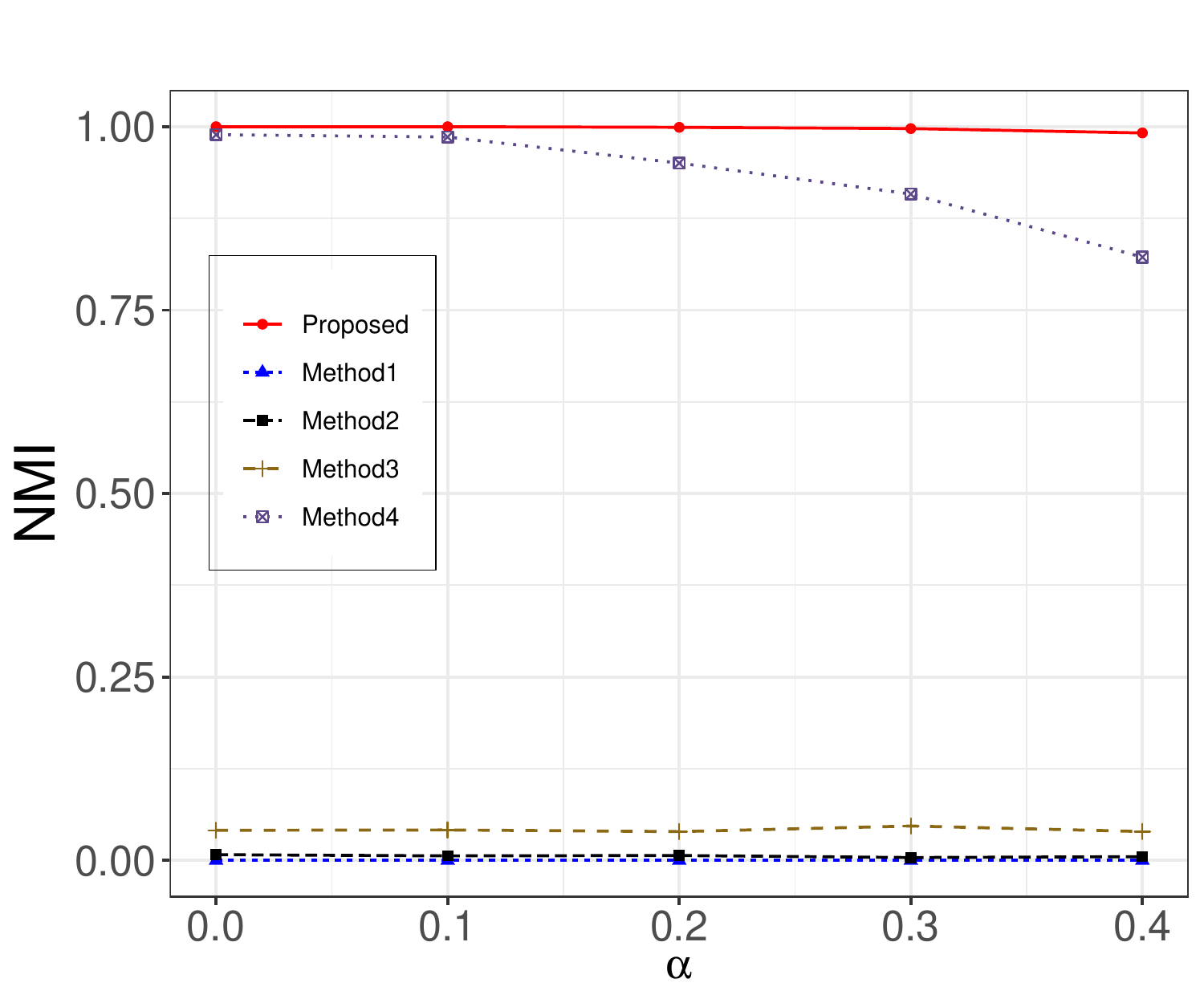}}
	\subfigure[Type 1 nodes, Scenario 2]{
		\centering
		\includegraphics[trim=0 0 0 5mm, width=0.3\linewidth]{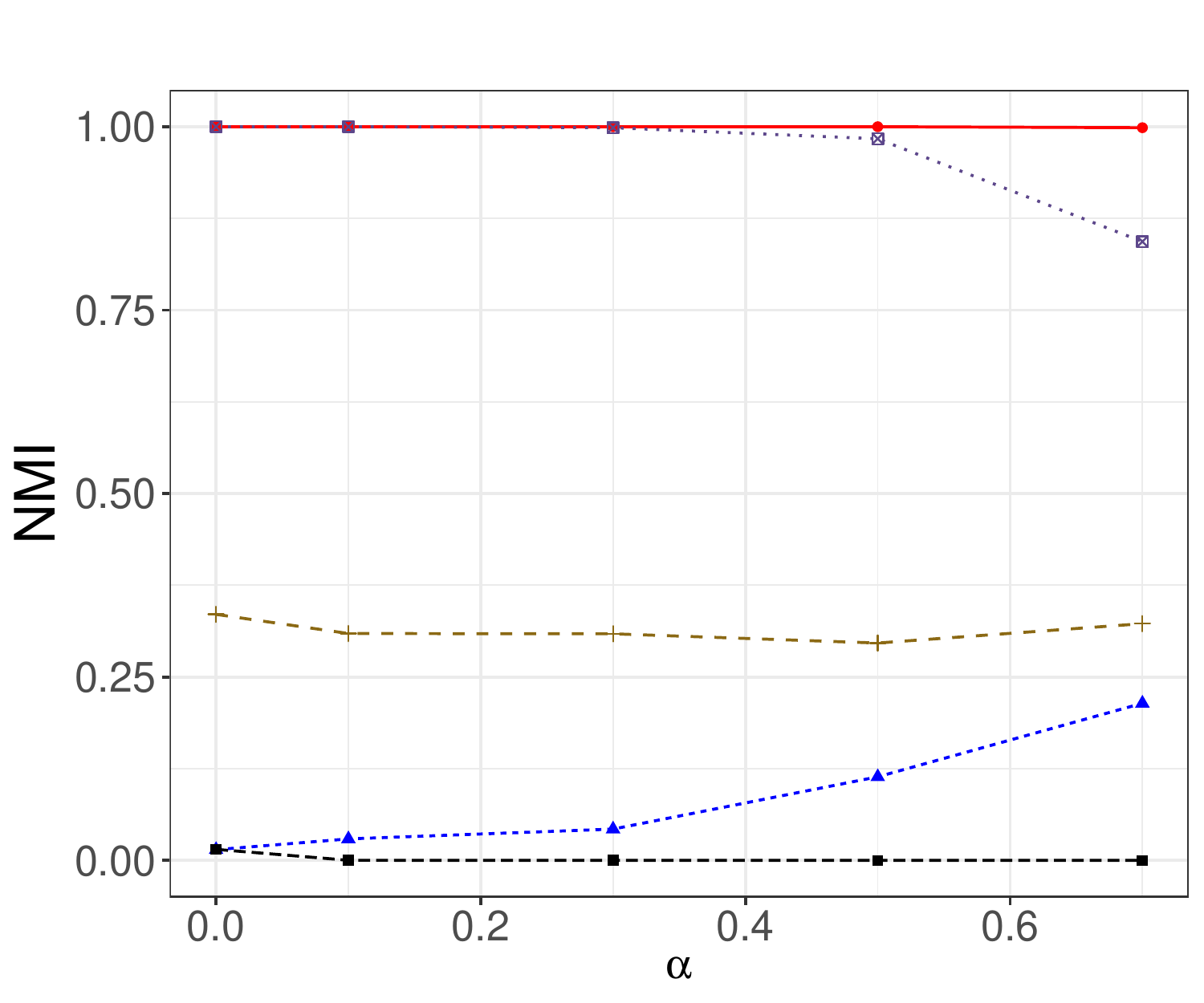}}
	\centering\\
	\subfigure[Type 2 nodes, Scenario 1]{
		\centering
		\includegraphics[trim=0 0 0 5mm, width=0.3\linewidth]{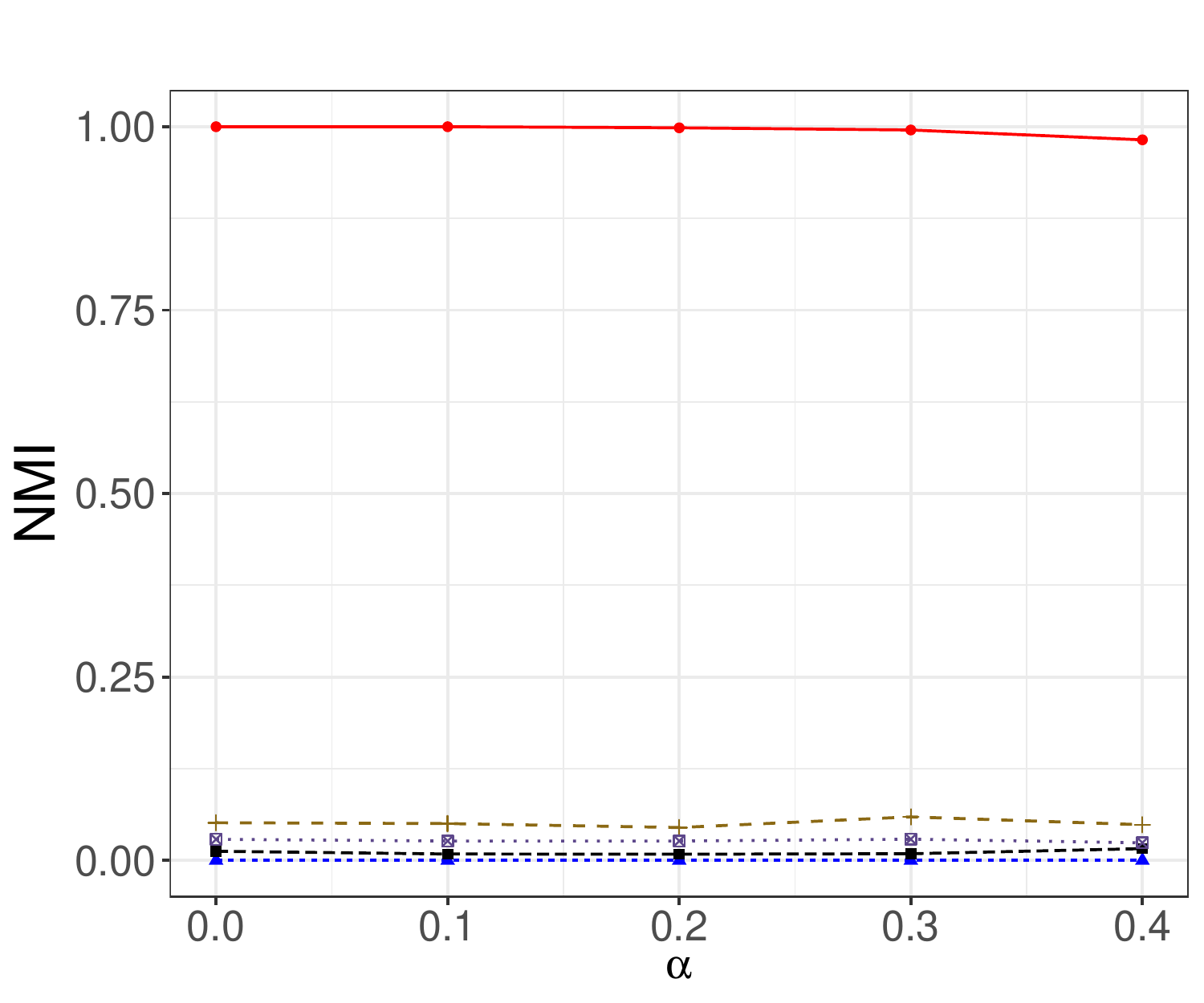}}
	\subfigure[Type 2 nodes, Scenario 2]{
		\centering
		\includegraphics[trim=0 0 0 5mm, width=0.3\linewidth]{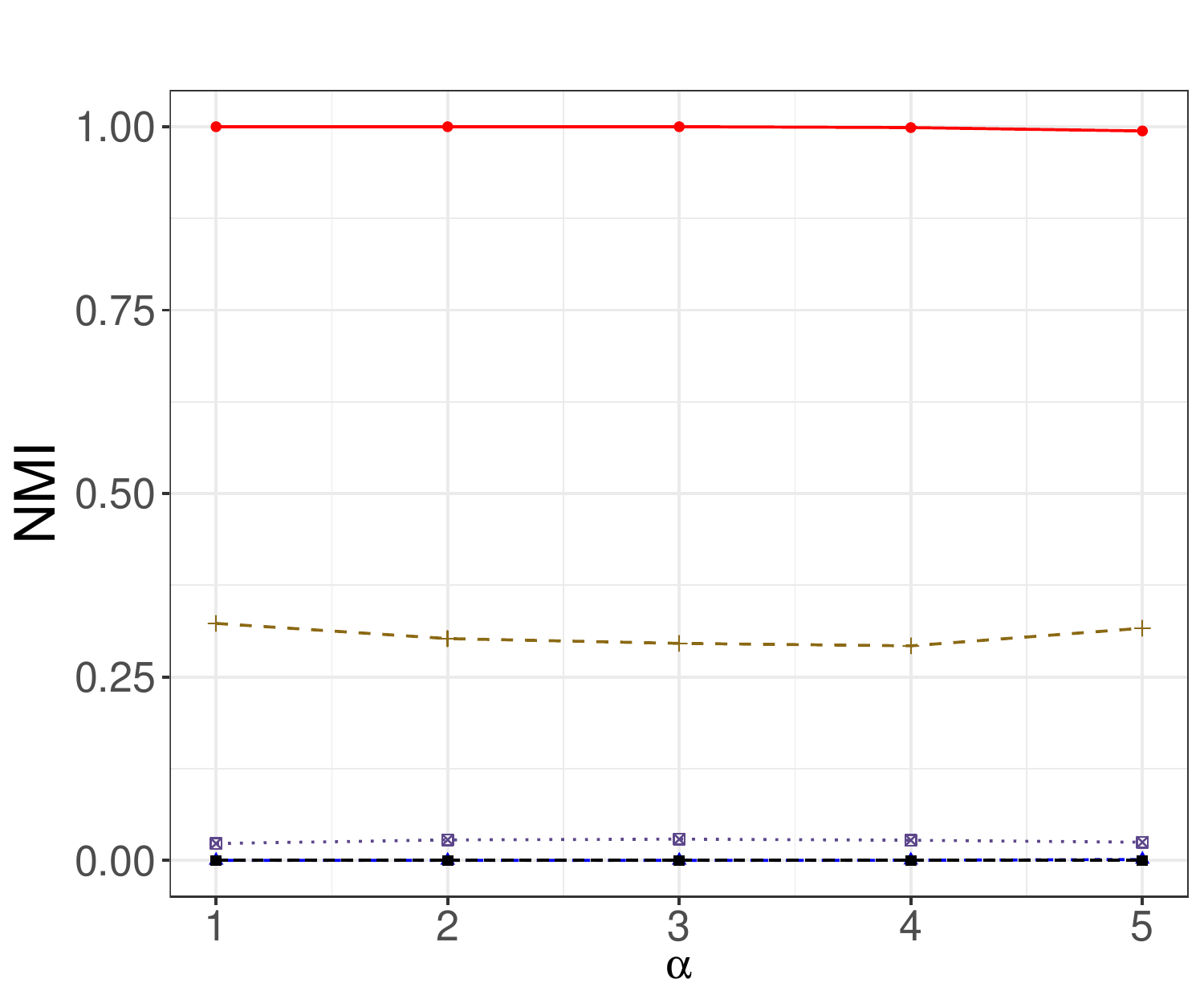}}
	\caption{Average NMIs against the value of $\alpha$ for different methods in Setting 3 under Scenarios 1-2.} \label{set3_dense}
\end{figure}

\section{Additional real data results}\label{real}
\begin{table}[H]\footnotesize
	\caption{\footnotesize The detected communities by the proposed method.}
	\label{Tab:2}
	\centering
	\begin{tabular}{lp{12cm}p{1.3cm}}
		\toprule
		community &category&number\\
		\midrule
		1&"Animal Shelters","Community Service/Non-Profit", "Greek", "Mediterranean", "Middle Eastern", "Pet Groomers", "Pet Services", "Pet Stores", "Pets", "Turkish", "Veterinarians" &11\\
	\hline
	2&  "Barbeque", "Cajun/Creole" "Southern", "Tex-Mex" &4\\
	\hline
	3& "Bagels" "Bakeries" "Breakfast \& Brunch" "Bubble Tea"  "Burgers" "Cafes" "Cantonese" "Caribbean" "Caterers" "Chicken Wings" "Chinese" "Coffee \& Tea" "Delis", "Desserts", "Dim Sum" "Diners" "Donuts"  "Fast Food" "Food Delivery Services" "Gluten-Free" "Hot Dogs" "Ice Cream \& Frozen Yogurt" "Italian"
	"Juice Bars \& Smoothies"  "Kosher" "Latin American" "Mexican" "Noodles" "Pizza" "Restaurants" "Salad" "Sandwiches" "Soup" "Tea Rooms" "Thai" "Vegan"       "Vegetarian" "Vietnamese"  &38 \\
	\hline
	4&"Custom Cakes" "French"  "Pasta Shops"  "Professional Services" "Seafood" "Steakhouses" "Taiwanese" &7 \\
	\hline
	5&"Beer" "Candy Stores" "Cheese Shops" "Chocolatiers \& Shops" "Convenience Stores"   "Drugstores" "Ethnic Food" "Farmers Market" "Florists" "Food" "Fruits \& Veggies" "Grocery" "Health Markets" "Meat Shops" "Modern European" "Organic Stores" "Seafood Markets" "Specialty Food" "Wine \& Spirits" & 19 \\
	\hline
	6&"Adult Entertainment" "Bars" "Beer Bar" "Breweries" "Brewpubs" "British"  "Cocktail Bars" "Comfort Food" "Dance Clubs" "Dive Bars" "Gay Bars" "German" "Irish" "Irish Pub" "Jazz \& Blues"  "Karaoke" "Lounges" "Music Venues" "Nightlife" "Pool Halls" "Pubs" "Spanish" "Sports Bars" "Tapas Bars" "Tapas/Small Plates" "Wine Bars"&26    \\
	\hline
	7& "Barbers" "Beauty \& Spas" "Cosmetics \& Beauty Supply" "Day Spas"  "Doctors" "Fitness \& Instruction", "Gyms", "Hair Removal" "Hair Salons" "Health \& Medical" "Makeup Artists" "Massage" "Nail Salons", "Optometrists", "Skin Care" "Trainers", "Waxing", "Yoga" & 18 \\
	\hline
	8&"Accessories" "Antiques" "Appliances" "Art Galleries" "Art Supplies" "Arts \& Crafts"  "Bike Rentals"  "Bike Repair/Maintenance" "Bikes" "Books" "Bookstores" "Cards \& Stationery" "Children's Clothing" "Colleges \& Universities" "Computers" "Department Stores" "Dry Cleaning \& Laundry" "Education" "Electronics" "Fashion" "Flowers \& Gifts", "Furniture Stores" "Gift Shops"   "Hardware Stores" "Hobby Shops" "Home \& Garden" "Home Decor" "Home Services" "Jewelry" "Kitchen \& Bath" "Laundry Services" "Local Services" "Mags"  "Men's Clothing" "Music \& DVDs" "Music \& Video" "Nurseries \& Gardening" "Outdoor Gear" "Real Estate" "Shoe Stores" "Shopping" "Shopping Centers" "Specialty Schools" "Sporting Goods" "Sports Wear" "Thrift Stores" "Toy Stores" "Used" "Vintage \& Consignment" "Vinyl Records" "Women's Clothing" & 51   \\
	\hline
	9&"Active Life" "Amusement Parks" "Arcades" "Arts Entertainment" "Botanical Gardens" "Bowling" "Cinema" "Event Planning \& Services" "Hotels" "Hotels \& Travel" "Landmarks \& Historical Buildings" "Local Flavor" "Museums" "Parks"  "Party \& Event Planning" "Public Services \& Government" "Public Transportation" "Tours" "Transportation" "Venues \& Event Spaces" & 21 \\
	\hline
	10& "Asian Fusion", "Buffets", "Indian", "Japanese", "Korean", "Pakistani", "Sushi Bars" &7 \\
	\hline
	11&"Auto Parts \& Supplies" "Auto Repair" "Automotive" "Gas Stations"  "Tires"&5 \\
		\bottomrule
	\end{tabular}
\end{table}

\begin{table}[H]\small
	\caption{\footnotesize The detected communities by Method 2.}
	\label{Tab:bookRWCal}
	\centering
	\begin{tabular}{lp{13cm}}
		\toprule
		community &category\\
		\midrule
			1&"Animal Shelters"   "Community Service/Non-Profit", "Greek"                        "Mediterranean", "Middle Eastern",  "Pet Groomers", "Pet Services", "Pet Stores"   "Pets"  "Turkish", "Vegan"  "Vegetarian"  "Veterinarians" \\
			\hline
		2& "Barbeque", "Dim Sum", "Southern", "Tapas/Small Plates", "Tex-Mex" \\
			\hline
		3& "Bagels"  "Bakeries" "Breakfast \& Brunch" "Bubble Tea"   "Burgers"   "Cafes"   "Cajun/Creole"   "Cantonese" "Caribbean" "Caterers" "Chicken Wings"             "Chinese" "Coffee \& Tea"  "Delis"   "Desserts"  "Diners"   "Donuts" "Fast Food"  "Food Delivery Services" "German"   "Hot Dogs" "Ice Cream \& Frozen Yogurt" "Italian"  "Kosher"   "Latin American" "Mexican"  "Noodles""Pizza"  "Restaurants"   "Salad"   "Sandwiches"  "Seafood"  "Soup"   "Spanish"   "Steakhouses"               "Tea Rooms"  "Thai"   \\
			\hline
		4&"Asian Fusion" "Custom Cakes" "French" "Hobby Shops" "Japanese"     "Korean" "Sushi Bars"   "Taiwanese"   "Vietnamese"   \\
			\hline
		5& ""Beer"  "Candy Stores" "Cheese Shops" "Chocolatiers \& Shops"   "Convenience Stores" "Drugstores"   "Ethnic Food" "Farmers Market"        "Florists"   "Flowers \& Gifts"    "Food" Fruits \& Veggies"   "Gluten-Free"            "Grocery"   "Health Markets"  "Juice Bars \& Smoothies" "Meat Shops"             "Organic Stores"  "Pasta Shops"  "Seafood Markets"   "Specialty Food"   "Wine \& Spirits"   \\
			\hline
		6&"Adult Entertainment"  "Arts \& Entertainment" "Bars" "Beer Bar"             "Breweries" "Brewpubs"  "British" "Cinema" "Cocktail Bars"  "Comfort Food"         "Dance Clubs" "Dive Bars" "Gay Bars" "Irish"  "Irish Pub"n"Jazz \& Blues"         "Karaoke"  "Lounges"  "Modern European"   "Music Venues" "Nightlife"           "Performing Arts"  "Pool Halls"   "Pubs"   "Sports Bars" "Tapas Bars"           "Wine Bars"   \\
			\hline
		7& "Barbers"   "Beauty  \& Spas"  "Cosmetics \& Beauty Supply" "Day Spas"  "Doctors"  "Hair Removal"  "Hair Salons"    "Health \& Medical"  "Makeup Artists"   "Massage"   "Nail Salons"  "Optometrists"  "Skin Care"     "Waxing"  \\
			\hline 
		8&"Accessories" Antiques"  "Appliances"  "Art Galleries"  "Art Supplies"            "Arts \& Crafts"   "Bike Rentals" "Bike Repair/Maintenance" "Bikes"                   "Books" "Bookstores" "Cards \& Stationery"  "Children's Clothing"  "Computers"  "Department Stores"   "Dry Cleaning \& Laundry" "Electronics"             "Fashion"  "Furniture Stores"  "Gift Shops"   "Hardware Stores"  "Home \& Garden"           "Home Decor"  "Home Services"   "Jewelry"   "Kitchen \& Bath"  "Laundry Services"  "Local Services" "Mags" "Men's Clothing"  "Music \& DVDs" "Music \& Video" "Nurseries \& Gardening"  "Outdoor Gear"   "Professional Services"   "Real Estate"  "Shoe Stores"  "Shopping" "Shopping Centers"  "Specialty Schools" "Sporting Goods"  "Sports Wear"  "Thrift Stores" "Toy Stores"  "Used" "Vintage \& Consignment" "Vinyl Records"   "Women's Clothing"  \\
			\hline
		9&"Active Life" "Amusement Parks"  "Arcades"    "Botanical Gardens"    "Bowling"  "Colleges \& Universities"   "Education"  "Event Planning \& Services"  "Fitness \& Instruction"  "Gyms"  "Hotels"  "Hotels \& Travel"  "Landmarks \& Historical Buildings" "Local Flavor"   "Museums"  "Parks" "Party \& Event Planning"           "Public Services \& Government"  "Public Transportation"  "Tours"    "Trainers"  "Transportation"  "Venues \& Event Spaces"    "Yoga"   \\
			\hline
		10&"Buffets" "Indian"  "Pakistani"  \\
			\hline
		11&"Auto Parts \& Supplies" "Auto Repair" "Automotive" "Gas Stations"  "Tires" \\
		\bottomrule
	\end{tabular}
\end{table}

\end{document}